\documentclass[aps,pre,preprint,groupedaddress,showpacs]{revtex4-2}
\usepackage{graphicx,amsmath,amssymb}
\begin{document}

\title{CORRELATIONS IN NON-EQUILIBRIUM  DIFFUSIVE SYSTEMS}


\author{P.L. Garrido}
\email[]{garrido@onsager.ugr.es}
\affiliation{Instituto Carlos I de F{\'\i}sica Te{\'o}rica y Computacional. Universidad de Granada. E-18071 Granada. Spain }

\date{\today}
\begin{abstract}
We study the behavior of stationary non-equilibrium two-body correlation functions for Diffusive Systems with equilibrium reference states (DSe). We describe a DSe at the mesoscopic level by $M$ locally conserved continuum fields that evolve through coupled Langevin equations with white noises. The dynamic is designed such that the system may reach equilibrium states for a set of boundary conditions. In this form, we make the system driven to a non-equilibrium stationary state by changing the equilibrium boundary conditions. We decompose the correlations in a known local equilibrium part and another one that contains the non-equilibrium behavior and that we call {\it correlation's excess} $\bar C(x,z)$. We formally derive the differential equations for $\bar C$. To solve them order by order, we define a perturbative expansion around the equilibrium state. We show that the $\bar C$'s first-order expansion, $\bar C^{(1)}$, is always zero for the unique field case, $M=1$. Moreover $\bar C^{(1)}$  is always long-range or zero when $M>1$. Surprisingly we show that their associated fluctuations,   the space integrals of $\bar C^{(1)}$, are always zero. Therefore, fluctuations are dominated by local equilibrium up to second-order in the perturbative expansion around the equilibrium. We derive the behaviors of $\bar C^{(1)}$ in real space for dimensions $d=1$ and $2$ explicitly. Finally, we derive the two first perturbative orders of the correlation's excess for a generic $M=2$ case and a hydrodynamic model.
\end{abstract}
\pacs{18-3e}
\maketitle

\section{Introduction}

Particle systems are characterized by the dynamics they follow: classical or quantum for material particles, stochastic rules for models in ecology, biology,... 
Moreover, boundary conditions are an essential part of dynamics because they  determine the values that some variables must take in some spatial regions. 
Today, we can establish from first principles, theories, and observations the dynamical rules of a given system with reasonable precision.
However, to extract valuable information for our understanding, we need to solve coupled ordinary nonlinear differential equations, partial differential equations, or stochastic equations with many degrees of freedom. Our mathematical tools are minimal for this enormous task.

However, we managed to get an idea of the properties of the system by simplifying the original dynamics by focusing on the aspects that we consider relevant to studying a particular observed phenomenon. The most common strategy is to adapt the modeling of the system to our mathematical knowledge. That allows us to use the tools that we master to extract some answers from those complex equations. This natural-looking scheme has some drawbacks. In our opinion, the most relevant is the robustness of the chosen model. Whether or not small changes in the dynamic rule imply proportionally small changes in the observed behavior. This problem is far from trivial, but it is generally neglected because we know many relevant physical situations in which the models we build are robust by construction. 
For example, we know that most details of the structure of a molecule and the interaction between them are irrelevant to describing equilibrium macroscopic properties of a system as the equation of state. Moreover, for equilibrium system we have Thermodynamics and the Ensemble Theory that help us to design simple microscopic models that contain the phenomena we want to characterize with detail.
 In conclusion, we are in a controlled environment in many developped theories where we have models that are reasonably simple and typically robust. Let us mention as a counterexample that there are very relevant equilibrium systems, such as water, where we do not know a simple model that contains all the rich set of properties and phases that \cite{water} has been observed. 

Nature is far from an equilibrium state. There are currents and flows of particles and energy, unbalanced chemical reactions, births, and deaths. Dynamic details and boundary conditions frequently determine the system's overall qualitative behavior. Therefore, the modeling of these systems becomes a very subtle issue, and robustness is always under deep scrutiny. Fortunately, there are cases in which we have successfully managed all those issues.
For instance, after centuries of observation, experiments, and theories, we derived a successful macroscopic theory as the Navier-Stokes equations for fluids. They have been the starting point to understanding many exciting phenomena associated with them as turbulence, convection,... \cite{Bat}. In other relevant cases, as in ecology, many efforts have been made to determine the basic principles and to build the most resilient models \cite{eco}.

In recent years, we have been interested in looking for a common theoretical framework that permits us to model different systems from diverse disciplines, each with its own particular dynamic rules. That lets us look for generic properties that can be of common interest. The first step in this direction was the Onsager-Machlup's theory for irreversible processes \cite{Onsager} where a Markovian mechanism is proposed to explain how the thermodynamic variables relax and fluctuate towards and around their equilibrium value. They assume that the macroscopic variables evolve by a Langevin equation where its deterministic evolution is proportional to the ''causes'' that provoke it and call them thermodynamic forces. For instance, for fluids, the local heat current is proportional to the local temperature gradient (Fourier's law), or the local particle current is proportional to the chemical potential gradient (Fick's law).
Moreover, the stochastic process is a white noise design, so the model fulfills the fluctuations at equilibrium. 

The Onsager and Machlup idea was recently extended by Bertini et al. \cite{Bertini2} to nonequilibrium systems where the time-reversibility is lost, which is typical in equilibrium. They developed the Macrocopic Fluctuating Theory (MFT) in the context of Diffusive Systems because we know rigorous results about hydrodynamic limits and large deviation properties. Many MFT ideas were already developed for systems with a discreet number of degrees of freedom \cite{seminal}, and they are easily generalized to different models with or without local conserved quantities \cite{Garrido0}. It is, in our opinion, the natural context to develop theoretical tools that help us understand the complex behavior of nonequilibrium systems.  

One exciting object in MFT is the quasipotential that defines the stationary measure in the weak noise limit. It is the non-equilibrium equivalent to the thermodynamic potential for systems at equilibrium. The quasipotential has been derived for some one-dimensional system \cite{Derrida,Bertini}. Also, there is some algebraic method that may help in getting them as the solution from a Hamilton-Jacobi equation \cite{Garrido1}. The quasipotentials have a highly complex structure with a non-local behavior that strongly depends on the boundary conditions.   That makes it very difficult to find regularities and generic behaviors to build, if possible, complete non-equilibrium thermodynamics beyond the one based on local-equilibrium assumptions \cite{Groot}. Therefore, it is convenient to get some more insight into the system's behavior by studying the correlations. We know that the correlations are just the inverse of the kernel coming from the second-order expansion of the quasipotential around the stationary state. Moreover, they contain precious physical information about the system's physical structure. Correlations have been extensively studied in fluids by experiments and theories. Here, we can mention the Fluctuating Hidrodyamics that is an MFT \cite{Fox}. Fluctuating Hydrodynamics is built following an Onsager-Machlup's type of assumption by adding a local equilibrium white noise to the deterministic Navier Stokes equations. In this context, we may highlight a couple of classical works by Tremblay et al.  \cite{Tremblay} and by Mansour et al. \cite{Mansour} where they deeply study the correlations for fluid by linearizing the Navier-Stokes equations in different situations and approximations. 

Inspired by the classical works in fluids, we study the two-body correlations in a generic non-equilibrium model with three main properties: (1) the system is described by $M$ fields that are locally conserved by the dynamics, (2) the local currents are proportional to the local field's gradients,  and (3) the equilibrium state may be reached by the system for a given set of external parameters. We call these systems DSe: Diffusive Systems with reference equilibrium states. We will use the last property to have a reasonable definition for the noise term and, later, to make a perturbative expansion around the equilibrium to get precise results. 

Section II presents the model definition through the Langevin equations and its connection with the reference equilibrium state. We also point out the properties we will assume in the paper, for instance, an unique locally stable stationary state. Section III obtains the partial differential equations for the two-body equal time correlation functions from the Hamilton-Jacobi equation for the quasi-potential. We also decompose the correlations in a local equilibrium contribution and a {\it correlation excess} that carries the non-equilibrium structure because it is equal to zero at equilibrium. Section III is devoted to extracting some general property by doing a perturbative expansion of the correlation excess around the equilibrium. For example, we find that all DSe systems with only one field have a zero first-order correction. We also find that, in general, all DSe with parallel plates as boundary conditions have their field fluctuations (integrals over the space of the two-body correlations) equal to zero at first order in the expansion around the equilibrium, although their correlation excess to such order being non-zero. In Section V, we focus on studying the basic correlation function $F$, which is the common part for the correlation excess at first order in the perturbation for any model. We study its behavior numerically in one dimension in real space after a non-trivial transformation. We also look at dimensions greater than one in the thermodynamic limit but near a system's boundary. We see the rich power-law behaviors depending on how we do the long-distance limits.
In Section VI, we show the first-order perturbation correlation excess in the case of two fields in dimensions one and two. Finally, Section VII is devoted to getting the correlation excess up to second order in the perturbation expansion for a two-dimensional particle model whose hydrodynamic equations have been derived recently \cite{Garr}. Some comments, most of the detailed computations, and the math relations we have derived to get the results shown in the central part of the paper have been left in six appendices.

\section{The model}
Let us define a mesoscopic system defined by $M$ conserved real fields $\phi_{\alpha}(x,t)$, $\alpha=1,\ldots,M$  in a $d$-dimensional region $x\in\Lambda\subset R^d$. The fields evolve by the Langevin equation:
\begin{equation}
\partial_t\phi_\alpha(x,t)+\nabla J_\alpha(x,t)=0\label{Lan}
\end{equation}
 $J_\alpha$ is the local vector current associated to the $\phi_\alpha$ field that  it is composed by a deterministic part, $J_\alpha^D$  and a fluctuating one, $J_\alpha^R$:
 \begin{equation}
 J_\alpha(x,t)=J_\alpha^D(x,t)+J_\alpha^R(x,t)
 \end{equation}
 
 We study in this paper Diffusive Systems with equilibrium reference states (DSe). That is, we impose two conditions on the form of the currents: (1) $J^D$ should be linear combinations of the field's gradients (Diffusive System), and (2) it should describe an equilibrium system with the appropriate boundary conditions (equilibrium reference state).  
 We will assume in this paper only spatially uniform equilibrium reference states, and we won't consider the action of external fields, like gravity, on the system.
 We may think of this model as the linear approximation around a given stationary state of a much more complex non-equilibrium conserved model as, for instance, the fluctuating hydrodynamics \cite{Fox} that, as we know, contains equilibrium states as a part of its description. In this class of models, non-equilibrium stationary states are built by changing the boundary conditions without introducing any other external effect. Therefore DSe's currents have the form:
 \begin{equation}
J_{\alpha}^D(x)=\sum_{\beta}g_{\alpha\beta}(\phi(x))\nabla\phi_\beta(x)\label{dif}
\end{equation}
 and
 \begin{equation}
J_\alpha^R(x,t)= \sum_{j=1}^d\sum_{\beta=1}^2\sigma_{\alpha,i;\beta,j}(\phi(x,t))\psi_{\beta,j}(x,t)\quad i=1,\ldots,d
 \end{equation}
All the sums over greek symbols run from $1$ to $M$ (the number of fields), and the ones with Latin symbols from $1$ to $d$ (the spatial dimension).
 $\phi(x,t)\equiv\{\phi_{\alpha}(x,t)\}_{\alpha=1}^M$, $\psi_{\alpha,i}(x,t)$ is an uncorrelated white noise:
\begin{equation}
\langle \psi_{\alpha,i}(x,t)\psi_{\beta,j}(x',t') \rangle=\Omega^{-1}\delta_{\alpha,\beta}\delta_{i,j}\delta(x-x')\delta(t-t')
\end{equation}
and $\Omega>>1$ is a large parameter that characterizes the separation between the microscopic and macroscopic scales.

The condition of having an equilibrium reference state implies a relation between $g$ and $\sigma$. We know from MFT 
\cite{Bertini2,Garrido0} that the deterministic current that describe a system at equilibrium should be of the form:
\begin{equation}
J_{\alpha,i}^D(x)=-\frac{1}{2}\sum_{\gamma}\sum_{k}\chi_{\alpha, i;\gamma,j}(\phi(x))\partial_k\frac{\delta V_{eq}[\phi]}{\delta\phi_{\gamma}(x)}\label{MFT}
\end{equation}
where
\begin{equation}
\chi_{\alpha,i;\beta,j}[\phi;x]=\sum_{\gamma}\sum_{k}\sigma_{\alpha,i;\gamma,k}(\phi(x))\sigma_{\beta,j;\gamma,k}(\phi(x))\label{chi}
\end{equation}
and $V_{eq}$ is the equilibrium mesoscopic potential that defines the equilibrium probability distribution:
 \begin{equation}
 P_{eq}[\phi]\simeq\exp[-\Omega V_{eq}[\phi]]\quad,\quad \Omega\rightarrow\infty
 \end{equation}
In order to get the linear form (\ref{dif}) from (\ref{MFT}) we need to assume:
\begin{equation}
\chi_{\alpha,i;\beta,k}(\phi)=2 L_{\alpha,\beta}(\phi)\delta_{i,k}\quad,\quad \frac{\delta V_{eq}[\phi]}{\delta\phi_\alpha(x)}=-\frac{\partial\tilde s(\phi)}{\partial\phi_\alpha}\biggr\vert_{\phi=\phi(x)}+cte\label{noi}
\end{equation}
where $L$ is a  positive defined symmetric matrix by construction and $\tilde s(\phi)$ is a function of $M$ variables.
 Finally, we find that 
  \begin{equation}
g= L S\label{g}
\end{equation}
 $S_{\alpha,\beta}(\phi)=\partial^2\tilde s(\phi)/\partial\phi_\alpha\partial\phi_\beta$. In conclusion, the Langevin equation for DSe models is completely determined by giving the $L$ symmetric matrix and the function $\tilde s(\phi)$. 

We see that from eq.(\ref{noi}), we may deduce, by a simple integration, the particular form of the equilibrium potentials that give rise to this linear set of currents:
\begin{equation}
V_{eq}[\phi]=-\int_\Lambda dx \left[\tilde s(\phi(x))-\tilde s(\phi_{eq})-\sum_{\alpha}(\phi_\alpha(x)-\phi_{eq,\alpha})\frac{\partial \tilde s(\phi)}{\partial\phi_\alpha}\biggr\vert_{\phi=\phi_{eq}}\right]\label{Veq0}
\end{equation}
 where we have made use of the known properties $V_{eq}[\phi_{eq}]=0$ and $\delta V[\phi]/\delta\phi_\alpha\vert_{\phi=\phi_{eq}}=0$. $\phi_{eq}$ are the equilibrium values of the fields.

The deterministic current structure (\ref{dif}) reminds us of the macroscopic linear laws we observe in Nature as Fick's law for diffusion or the Fourier's law for heat conduction. We know that there is a set of theories that describe how a system characterized by its mesoscopic variables fluctuates around its equilibrium state (Einstein theory of fluctuations \cite{Einstein} and Appendix I) or how it relaxes towards the equilibrium (Onsager's theory \cite{Onsager}). Both of them are contained in the so-called {\it Non-equilibrium Thermodynamics} \cite{Groot}.
Let us connect our description above with this classic point of view. et $s(\phi)$ be the entropy per unit volume of a system in equilibrium with macroscopic observables $\phi=(\phi_1,\phi_2,\ldots,\phi_M)$. A macroscopic system relaxing to the equilibrium state from a nearby initial state is reasonable to think it is locally at equilibrium with an entropy $s(\phi(x))$ at each macroscopic point $x\in\Lambda$ in the system. It is again assumed that $s(\phi)$ does not contain an explicit dependence on position $x$ due to the action of an external field such, for instance, gravitation.  
 In this situation, the macroscopic currents associated to the conserved fields, $\phi$,  are found to have the form
\begin{equation}
J_{\alpha,i}^D(x)=\sum_{\beta}\sum_{k}\tilde L_{\alpha,i;\beta,k}(\phi(x))X_{\beta,k}(\phi(x))\label{on}
\end{equation}
where $\tilde L$ (the Onsager's coefficients) is a symmetric matrix on $(\alpha,i)$-index and $X$'s are the so-called {\it thermodynamic forces} that are defined from the local entropy $s(\phi)$:
\begin{equation}
X_{\alpha,i}(\phi)=\partial_i \tilde y_{\alpha}(\phi)\quad ,\quad \tilde y_\alpha(\phi)=\frac{\partial s(\phi)}{\partial\phi_\alpha}
\end{equation}
Observe that this classical description coincides with ours above if we identify $\tilde s(\phi)=s(\phi)$, the thermodynamic entropy, and $\tilde L_{\alpha,i;\beta,k}=L_{\alpha,\beta}\delta_{i,k}$. Moreover, the expression (\ref{Veq0})  can be derived from  Equilibrium Statistical Mechanics (see Appendix I). Our model includes the classical description of how the macroscopic variables of systems perturbed from their equilibrium state evolve effectively towards it by assuming that local equilibrium is fulfilled.  

A final property is asked for $J^D$: the equilibrium state is stable under small perturbations at the deterministic level. That is, the determistics evolution starting from any initial set of fields near the equilibrium should relax towards it. The deterministic evolution equation is:
\begin{equation}
\partial_t\phi_\alpha^D(x,t)=-\nabla\left[\sum_{\beta}g_{\alpha\beta}(\phi^D)\nabla\phi_\beta^D\right]
\end{equation}
Let us assume that $\phi_\alpha(x,t)=\phi_{eq,\alpha}+\theta(x,t)$ with $\theta$ small and we expand the deterministic equation up to first order in $\theta$:
\begin{equation}
\partial_t\theta(x,t)=-\sum_\beta g_{\alpha\beta}(\phi_{eq})\nabla\theta_\beta(x,t)+O(\theta^2)
\end{equation}
We see that the evolution is characterized by the $g$ matrix evaluated at equilibrium. We rewrite this evolution equation for the Fourier Transform of $\theta$, $\hat\theta$:  
\begin{equation}
\partial_t\hat\theta_\alpha(k,t)=k^2\sum_{\beta}g_{\alpha\beta}(\phi_{eq})\hat\theta_\beta(k,t)
\end{equation}
and then expand $\hat\theta$ in the eigenvector basis of $g$:
\begin{equation}
\hat\theta_{\alpha}(k,t)=\sum_{n}a_n(k,t)v_{n,\alpha}\quad,\quad gv_n=\lambda_nv_n
\end{equation}
and the resulting evolution equation for $a$'s is given by:
\begin{equation}
\partial_t a_n(k,t)=k^2\lambda_n a_n(k,t)
\end{equation}
whose solution is:
\begin{equation}
a_n(k,t)=a_n(k,0)\exp[\lambda_n k^2 t]\label{eigen}
\end{equation}
The evolution of $a_n$ goes to zero and then the equilibrium state is stable if and only if all the eigenvalues of $g$ have their real part negative:
\begin{equation}
Re(\lambda_n)<0\quad \forall n
\end{equation}

These equations describe the dynamics of a DSe relaxing to the equilibrium state whenever the boundary conditions are compatible with such state, for instance, $\phi(x)=\phi_{eq}\quad \forall x\in\partial\Lambda$. If we change such boundaries, the stationary distribution is no longer the equilibrium one defined by $V_{eq}[\phi]$. Moreover,  the local equilibrium property is lost, and the stationary state's quasi-potential has a non-local structure that implies long-range correlations (see, for instance, Ref. \cite{Derrida,Bertini,Garrido1}). This model permits us to answer some interesting and refined questions: What happens near the equilibrium? How is local equilibrium lost? What are the fluctuations of the observables? 

Once we have constructed the model, let us study its behavior in a generic non-equilibrium stationary state. First, the deterministic stationary solution would depend on $x$, $\phi^*(x)$, and it is solution of: 
\begin{equation}
\nabla J_{\alpha}^D(x)=0\Rightarrow \nabla\left[\sum_{\beta}g_{\alpha\beta}(\phi^*(x))\nabla\phi_\beta^*(x)\right]=0\label{stat}
\end{equation}
The boundary conditions affects dramatically the system behavior. Typically they are assumed to be of Diritchlet type, $\phi(x)=\phi_0(x)\,\forall x\in\partial\Lambda$. As a helpful example, we discuss in Appendix II the conditions on the system's boundaries and/or in the system's dynamics when we require to have constant currents $\vec j_\alpha=\sum_\beta g_{\alpha\beta}(\phi^*)\vec\nabla\phi_\beta^*$. Moreover, there could also be global conservation laws of a field, $\int_\Lambda dx \phi_\alpha^*(x)=cte$ that we also discuss their effect in the correlations in Appendix III.
 
The fluctuating properties of such non-equilibrium stationary states are studied by using the Fokker-Planck equation associated to the above Langevin equation:
\begin{eqnarray}
\partial_tP[\phi;t]&=&\sum_{\alpha}\sum_{i}\int_\Lambda dx\,\left(\partial_i\frac{\delta}{\delta\phi_\alpha(x)}\right)\biggl[-J_{\alpha,i}^D(x)P[\phi;t]\\\nonumber
&+&\frac{1}{\Omega}\sum_{\beta}\left(\partial_i\frac{\delta}{\delta\phi_\beta(x)}\right)\left(L_{\alpha,\beta}(\phi)P[\phi;t]\right)\biggr]\label{ME}
\end{eqnarray}

The stationary distribution when $\Omega\rightarrow\infty$ is of the form
\begin{equation}
P_{st}[\phi]\simeq \exp[-\Omega V_0[\phi]]
\end{equation}
where $V_0[\phi]$ is called the {\it quasi-potential} that it is solution of the Hamilton-Jacobi equation:
\begin{equation}
0=\sum_{\alpha}\sum_{i}\int_\Lambda dx\,\left(\partial_i\frac{\delta V_0[\phi]}{\delta\phi_\alpha(x)}\right)\biggl[J_{\alpha,i}^D(x)+\sum_{\beta}L_{\alpha,\beta}(\phi)\partial_i\frac{\delta V_0[\phi]}{\delta\phi_\beta(x)}\biggr]\label{HJ}
\end{equation}
This simple derivation of the Hamilton-Jacobi equation hides a set of important quasi-potential properties that we do not address here. We ask the reader to look at refs. \cite{seminal,Bertini2} for a complete description of them.
The quasi-potential contains all the relevant behavior about the system's stationary state, but it isn't easy to get explicit solutions from the Hamilton-Jacobi equation for generic cases \cite{Garrido1}. However, let us show that from the Hamilton-Jacobi equation, we can derive a set of closed equations for the equal-time correlation functions of the stationary state. We know that these capture the essential features of the system's spatial structure and are closely related to the quasi-potential shape around the stationary state.

\section{Equal-time Correlation Functions}

The correlations for our $M$-field model are defined as
\begin{equation}
\tilde C_{\alpha_1\alpha_2\ldots\alpha_n}(x_1,x_2\ldots x_n)\equiv\langle(\phi_{\alpha_1}(x_1)-\langle\phi_{\alpha_1}(x_1)\rangle_{st})\ldots(\phi_{\alpha_n}(x_n)-\langle\phi_{\alpha_n}(x_n)\rangle_{st})\rangle_{st}
\end{equation}
where $\langle\cdot\rangle_{st}=\int D\phi \,\cdot P_{st}[\phi]$. 

In the weak noise limite (large values of $\Omega$) we can use the quasipotentical $V_0$ to compute the correlations. It is a matter of algebra to show that
\begin{eqnarray}
 C_{\alpha_1\alpha_2\ldots\alpha_n}(x_1,x_2\ldots x_n)&\equiv&\lim_{\Omega\rightarrow\infty}\Omega^{n-1}\tilde C_{\alpha_1\alpha_2\ldots\alpha_n}(x_1,x_2\ldots x_n)\nonumber\\
&=&-\frac{\delta F[\phi^*[b],b]}{\delta b_{\alpha_1}(x_1)\ldots\delta b_{\alpha_n}(x_n)}\biggr\vert_{b=0}
\end{eqnarray} 
where
\begin{equation}
F[\phi,b]=V_0[\phi]-\sum_{\alpha=1}^M\int_\Lambda dx\,b_\alpha(x)\phi_\alpha(x)
\end{equation}
and $\phi^*[b]$ is solution of
\begin{equation}
\frac{\delta F[\phi,b]}{\delta\phi_\alpha(x)}\biggr\vert_{\phi=\phi^*[b]}=0\quad \Leftrightarrow\quad \frac{\delta V_0[\phi]}{\delta\phi_\alpha(x)}\biggr\vert_{\phi=\phi^*[b]}=b_\alpha(x)
\end{equation}
$\phi^*[0]=\phi^*$ is the stationary solution of the Langevin equation without noise given by eq.(\ref{stat}).
We construct a set of closed equations for the correlations by using the Hamilton-Jacobi equation (\ref{HJ})  with $\phi\rightarrow\phi^*[b]$ and then expanding the equation in powers of $b$'s (see for instance ref.\cite{Garrido0}  a detailed computation for the $M=1$ case). We get, at order $b^2$, the general equations for the two body correlations:
\begin{eqnarray}
\sum_{\beta}\int_{\Lambda}&dy&\,\left[K_{\alpha\beta}(x,y)C_{\beta\gamma}(y,z)+K_{\gamma\beta}(z,y)C_{\beta\alpha}(y,x)\right]\nonumber\\
&=&2\sum_{i}\partial_{x_i}\partial_{z_i}\left[L_{\alpha,\gamma}(\phi^*(x))\delta(x-z)\right]
\end{eqnarray}
where
\begin{equation}
K_{\alpha\beta}(x,y)=\frac{\delta\nabla J_{\alpha}^D(x) }{\delta\phi_\beta(y)}\biggr\vert_{\phi=\phi^*}
\end{equation}

We can now substitute the $J^D$ corresponding to the DSe (\ref{dif}) and we get:
\begin{eqnarray}
&&\sum_\beta\nabla_x\left[\vec{b}_{\alpha\beta}(x)C_{\beta\gamma}(x,z)+g_{\alpha\beta}(\phi^*(x))\nabla_x C_{\beta\gamma}(x,z)\right]+\nonumber\\
&&\sum_\beta\nabla_z\left[\vec{b}_{\gamma\beta}(z)C_{\beta\alpha}(z,x)+g_{\gamma\beta}(\phi^*(z))\nabla_z C_{\beta\alpha}(z,x)\right]\nonumber\\
&&=2\partial_{x_i}\partial_{z_i}\left[L_{\alpha\gamma}(\phi^*(x))\delta(x-z)\right]\label{corr}
\end{eqnarray}
where
\begin{equation}
\vec{b}_{\alpha\beta}(x)=\sum_\gamma \frac{\partial g_{\alpha\gamma}(\phi)}{\partial\phi_\beta}\biggr\vert_{\phi=\phi^*}\nabla_x\phi_\gamma^*(x)
\end{equation}
In the equilibrium case $\phi^*(x)=\phi_{eq}$. Therefore $\vec{b}_{\alpha\beta}(x)=0$ and the equations (\ref{corr}) become:
\begin{eqnarray}
&&\sum_\beta g_{\alpha\beta}(\phi_{eq})\nabla_x^2 C_{\beta\gamma}^{(0)}(x,z)+
\sum_\beta g_{\gamma\beta}(\phi_{eq})\nabla_z^2 C_{\beta\alpha}^{(0)}(z,x)\nonumber\\
&&=-2L_{\alpha\gamma}(\phi_{eq})\nabla_x^2\delta(x-z)\label{corre}
\end{eqnarray}
whose solution for boundary conditions such that $\phi(x)=\phi_{eq}\,\forall x\in\partial\Lambda$ is
\begin{equation}
C_{\alpha\beta}^{(0)}(x,y)=-(S^{-1})_{\alpha\beta}(\phi_{eq})\delta(x-y)\label{correq}
\end{equation}
This result could be  obtained directly from the equilibrium quasipotential $V_{eq}[\phi]$:
\begin{equation}
C_{\alpha\beta}^{(0)}(x,y)=\left(\frac{\delta^2V_{eq}[\phi]}{\delta\phi_\alpha(x)\delta\phi_\beta(y)}\right)^{-1}(\phi_{eq})\delta(x-y)
\end{equation}

In this paper we are going to consider only the case of equilibrium fluctuating boundary conditions: $C_{\alpha\beta}(x,z)=C_{\alpha\beta}^{(0)}(x,z)$ $\forall\, x$ or $z\in\partial\Lambda$. 

At this point, it is convenient to decompose the correlations in two terms, one that represents the local-equilibrium contributions (equilibrium correlations  evaluated at each macroscopic point with the corresponding field values of the stationary state) and the rest that contains the strong nonequilibrium behavior:
\begin{equation}
C_{\alpha\beta}(x,y)=-(S^{-1})_{\alpha\beta}(\phi^*(x))\delta(x-y)+\bar C_{\alpha\beta}(x,y)\label{dec}
\end{equation} 
After substituting eq.(\ref{dec}) into eq.(\ref{corre}) we obtain the central equation for the two body correlations:
\begin{eqnarray}
&&\sum_\beta\nabla_x\left[\vec{b}_{\alpha\beta}(x)\bar C_{\beta\gamma}(x,z)+g_{\alpha\beta}(\phi^*(x))\nabla_x \bar C_{\beta\gamma}(x,z)\right]+\nonumber\\
&&\sum_\beta\nabla_z\left[\vec{b}_{\gamma\beta}(z)\bar C_{\beta\alpha}(z,x)+g_{\gamma\beta}(\phi^*(z))\nabla_z \bar C_{\beta\alpha}(z,x)\right]\nonumber\\
&&=(\nabla_x\vec{A}_{\alpha\gamma}(x))\delta(x-z)+(\vec{A}_{\alpha\gamma}(x)-\vec{A}_{\gamma\alpha}(x))\nabla_x \delta(x-z)\label{exc}
\end{eqnarray}
where 
\begin{equation}
\vec{A}_{\alpha\gamma}(x)=\sum_{\beta}\left[\vec{b}_{\alpha\beta}(x)\left(S^{-1}\right)_{\beta\gamma}+g_{\alpha\beta}(\phi^*(x))\nabla_x\left(S^{-1}\right)_{\beta\gamma}\right]\label{AA}
\end{equation}
with boundary conditions: $\bar C_{\alpha\beta}(x,z)=0\, \forall\, x$ or $z\in\partial\Lambda$.
This equation has the symmetry $(\alpha,x)\leftrightarrow(\gamma,z)$ and also that the $\nabla\delta$-term doesn't exist in the one-field case. We see that these coupled equations for the {\it correlation's excess}  are highly nonlinear because it depends on the non-equilibrium stationary state $\phi^*(x)$, on the equilibrium reference state represented by the entropy hessian $S$ and on the diffusive model $g$. We are interested in studying the role of the local equilibrium at the level of correlations. Therefore, we will expand these correlations near the equilibrium state to get some generic results on their properties.

We should mention that our system is typically open because of the boundary conditions. However, we could think of models where some fields have global conservation constraints. For instance, in a system of particles enclosed in a container where only energy is exchanged at the boundaries. The field corresponding to the density is precisely conserved at any time, but, in contrast, the field associated with the energy is not strictly conserved.
In the Appendix III we study the effect in the correlations of the existence of global conservation in some fields. We show there that the correlations $C_{\alpha\beta}^{SC}(x,y)$ when a set of global conserved fields, $\tilde M$, can be expressed as combinations of the correlations corresponding to the non-conserved case, $C_{\alpha\beta}^{OB}(x,y)$:
\begin{equation}
C_{\alpha\beta}^{SC}(x,y)=C_{\alpha\beta}^{OB}(x,y)-\sum_{\bar\alpha\bar\beta\in\tilde M}\int_\Lambda dz_1\int_\Lambda dz_2\, C_{\alpha\bar\alpha}^{OB}(x,z_1)\left(A^{-1}\right)_{\bar\alpha\bar\beta}C_{\bar\beta\beta}^{OB}(z_2,y)
\end{equation}
where
\begin{equation}
 A_{\alpha\beta}=\int_\Lambda dx\int_\Lambda dy\, C_{\alpha\beta}^{OB}(x,y)\quad \alpha,\beta\in\tilde M
\end{equation}
Therefore global conservation do not introduce new complexities at this level and we just focus in cases where all the fields are globally non-conserved.

\section{Nonequilibrium correlations near the equilibrium: two theorems}

The DSe are driven from an equilibrium state to a non-equilibrium stationary state by changing the boundary conditions. Let us assume that the system's stationary state is near the equilibrium. Therefore, a parameter $0\leq\epsilon\ll 1$ represents the distance of the values of its boundaries to their corresponding equilibrium ones. Then, let us assume that the deterministic stationary state, $\phi^*(x)$, can be analytically expanded:
\begin{equation}
\phi_\alpha(x)^*=\phi_{eq,\alpha}+\epsilon h_\alpha^{(1)}(x)+\epsilon^2 h_\alpha^{(2)}(x)+O(\epsilon^3)\label{per1}
\end{equation}
Then, from eq.(\ref{stat}) we find that $h_\alpha^{(1,2)}$ are solution of:
\begin{equation}
\sum_{\beta}g_{\alpha\beta}(\phi_{eq})\nabla_x^2h_\beta^{(1)}(x)=0
\end{equation}
\begin{equation}
\sum_{\beta}g_{\alpha\beta}(\phi_{eq})\nabla_x^2h_\beta^{(2)}(x)=-\sum_\beta\nabla_x\left(g_{\alpha\beta}^{(1)}(x)\nabla_xh_\beta^{(1)}(x)\right)
\end{equation}
where
\begin{equation}
g_{\alpha\beta}^{(1)}(x)=\sum_{\gamma}\frac{\partial g_{\alpha\gamma}}{\partial\phi_{\beta}}\biggr\vert_{\phi_{eq}}h_\gamma^{(1)}(x)
\end{equation}
with given boundary conditions. For instance, in a one dimension system in a unit box $[0,1]$, when $\phi_\alpha(0)=\phi_{eq,\alpha}+\epsilon$ and $\phi_\alpha(1)=\phi_{eq,\alpha}$ then $h_\alpha^{(1,2)}(1)=0$ and $h_\alpha^{(1)}(0)=1$,  $h_\alpha^{(2)}(0)=0$.

When $\epsilon\rightarrow 0$ the correlations tend to their equilibrium value $C_{\alpha\beta}(x,y)\rightarrow C_{\alpha\beta}^{(0)}(x,y)$ and therefore, $\bar C_{\alpha\beta}(x,y)\rightarrow 0$  in such limit. Thus, we can assume the existence of an analytic $\epsilon$-expansion for the correlation's excess, $\bar C$:
\begin{equation}
\bar C_{\alpha\beta}(x,y)=\epsilon \bar C_{\alpha\beta}^{(1)}(x,y)+\epsilon^2 \bar C_{\alpha\beta}^{(2)}(x,y)+O(\epsilon^3)\label{per2}
\end{equation}
We substitute eqs.(\ref{per1}) and (\ref{per2}) into (\ref{exc}) and we get a hierarchy of closed equations that for  $\bar C_{\alpha\beta}^{(1)}(x,y)$ and $\bar C_{\alpha\beta}^{(2)}(x,y)$ are:
\begin{equation}
\sum_{\beta}g_{\alpha\beta}\nabla_x^2\bar C_{\beta\gamma}^{(1)}(x,z)+\sum_{\beta}g_{\gamma\beta}\nabla_z^2\bar C_{\alpha\beta}^{(1)}(x,z)=\vec a_{\alpha\gamma}^{(1)}\cdot\nabla_x\delta(x-z)\label{C1}
\end{equation}
\begin{eqnarray}
&&\sum_{\beta}g_{\alpha\beta}\nabla_x^2\bar C_{\beta\gamma}^{(2)}(x,z)+\sum_{\beta}g_{\gamma\beta}\nabla_z^2\bar C_{\alpha\beta}^{(2)}(x,z)=\nabla_x\vec A_{\alpha\gamma}^{(2)}(x)\delta(x-z)\nonumber\\
&&+\vec a_{\alpha\gamma}^{(2)}(x)\cdot\nabla_x\delta(x-z)-\sum_{\beta}g_{\alpha\beta}^{(1)}(x)\nabla_x^2\bar C_{\beta\gamma}^{(1)}(x,z)-\sum_{\beta}g_{\gamma\beta}^{(1)}(z)\nabla_z^2\bar C_{\alpha\beta}^{(1)}(x,z)\nonumber\\
&&-\sum_\beta\left(\vec b_{\alpha\beta}^{(1)}(x)+\nabla_xg_{\alpha\beta}^{(1)}(x)\right)\cdot\nabla_x\bar C_{\beta\gamma}^{(1)}(x,z)-\sum_\beta\left(\vec b_{\gamma\beta}^{(1)}(z)+\nabla_zg_{\gamma\beta}^{(1)}(z)\right)\cdot\nabla_z\bar C_{\alpha\beta}^{(1)}(x,z)
\label{C2}
\end{eqnarray}
where 
\begin{equation}
\vec a_{\alpha\gamma}^{(i)}(x)=\vec A_{\alpha\gamma}^{(i)}(x)-\vec A_{\gamma\alpha}^{(i)}(x)\label{aa0}
\end{equation}
and
\begin{equation}
\vec A_{\alpha\gamma}^{(1)}=\sum_{\bar\gamma}\left(\nabla_x h_{\bar\gamma}^{(1)}\right)\sum_\beta\left(\frac{\partial g_{\alpha\bar\gamma}}{\partial\phi_\beta}-\frac{\partial g_{\alpha\beta}}{\partial\phi_{\bar\gamma}}\right)(S^{-1})_{\beta\gamma}\label{aa}
\end{equation}
\begin{equation}
\vec A_{\alpha\gamma}^{(2)}(x)=\biggl[\sum_\sigma\nabla_xh_\sigma^{(2)}(x)+\sum_{\sigma\eta'}(\nabla_xh_\sigma^{(1)}(x))h_{\eta'}^{(1)}(x)\frac{\partial}{\partial\phi_{\eta'}}\biggr]\sum_{\beta}(S^{-1})_{\beta\gamma}\sum_\eta\frac{\partial L_{\alpha\eta}}{\partial\phi_\beta}S_{\eta\sigma}
\end{equation}
\begin{equation}
\vec b_{\alpha\gamma}^{(1)}(x)=\nabla_x g_{\alpha\gamma}^{(1)}(x)
\end{equation}
We have simplified the notation: $g_{\alpha\beta}\equiv g_{\alpha\beta}(\phi_{eq})$ and  $\partial g_{\alpha\beta}/\partial\phi_{\bar\gamma}\equiv \partial g_{\alpha\beta}(\phi)/\partial\phi_{\bar\gamma} \vert_{\phi=\phi_{eq}}$. In general, after any operation, a functional that depend on $\phi$ is considered to be evaluated at $\phi_{eq}$.

At this point we find the first general result:
\begin{itemize}
\item{\bf Therorem 1:} {\bf All DSe systems with one field, $M=1$, have $\bar C^{(1)}=0$. }
\end{itemize}
That is, the excess of correlations is, at most, of order $\epsilon^2$. That is due because $\vec A^{(1)}=0$ and the solution of eq.(\ref{C1}) is an harmonic function whose maximum or minimum should be at the boundary that in our case is always zero: $\bar C^{(1)}(x,z)=0$ $\forall x$ or $z \in\partial\Lambda$ and therefore $\bar C^{(1)}(x,z)=0 \, \forall\, x,z $. This property was already observed in two specific one-dimensional models, the Symmetric Simple Exclusion process (SSEP) \cite{Derrida} and the Kipnis, Marchioro, Presutti Model (KMP) \cite{Bertini}. In these works it is shown that  $\bar C=\epsilon^2 F(x,z)$ for any $\epsilon$. 

In order to go forward we need to give specific boundary conditions. Our natural choice is to place our system between two parallel plates placed at $x_1=0$ and $L$ where the values of the fields are given: 
\begin{equation}
\phi_\alpha(0,x_\perp)=\phi_{eq,\alpha}\quad,\quad \phi_\alpha(L,x_\perp)=\phi_{eq,\alpha}+\Delta\phi_\alpha
\end{equation}
where $\Delta\phi_\alpha$ are given constants.
Therefore:
\begin{equation}
\bar C_{\alpha,\beta}^{(1)}(x,z)=0\quad,\quad x\,\text{and/or}\,z\in\partial\Lambda=\{(0,w_\perp)\}\cup\{(L,w_\perp\}
\end{equation}
We also assume periodic boundary conditions in the perpendicular $d-1$ directions:  $\bar C_{\alpha\beta}(x_1,x_\perp\pm\eta La_{j})=\bar C_{\alpha\beta}(x_1,x_\perp)\, ,\forall j=2,\ldots d$. Where  $x=(x_1,x_\perp)$ and $a_{i}$ are the unit vectors on the principal directions. $\eta>0$ is a form factor.  

These boundary conditions have the advantage to give us a simple stationary state around the equilibrium (see Appendix II). In particular $\vec{a}_{\alpha\gamma}^{(1)}=a_{\alpha\gamma}^{(1)}\hat\i$ in eq.(\ref{C1}), with $a_{\alpha\gamma}^{(1)}$  constant. We apply to the $\bar C_{\alpha\beta}^{(1)}(x,z)$ functions the Sinus Fourier's Transform to the $x_1$, $z_1$ coordinates because they incorporate the boundary conditions  and a normal Fourier's Transform to the perpendicular coordinates $x_\perp$, $z_\perp$ in  eq.(\ref{C1}) (see details about properties of the Fourier's Transform in Appendix IV). Then
\begin{eqnarray}
\bar C_{\alpha\beta}^{(1)}(x,z)&=&\frac{1}{(L')^{d-1}}\sum_{n_\perp\in\mathbb{Z}^{d-1}}e^{i\frac{2\pi}{L'}n_{\perp}\cdot(x_\perp-z_{\perp})}\nonumber\\
&&\sum_{n=1}^{\infty} \sum_{m=1}^{\infty}\sin\left(\frac{n\pi x_1}{L}\right)\sin\left(\frac{m\pi z_1}{L}\right)\hat C_{\alpha\beta}^{(1)}(n,m;n_\perp)\label{C1real}
\end{eqnarray}
where $L'\equiv\eta L$ and $\hat C_{\alpha\beta}$-functions are solution of the equations:
\begin{eqnarray}
(n^2+\frac{4}{\eta^2}n_\perp^2)\sum_\beta g_{\alpha\beta}\hat C_{\beta\gamma}^{(1)}(n,m;n_\perp)+(m^2&+&\frac{4}{\eta^2}n_\perp^2)\sum_\beta g_{\gamma\beta}\hat C_{\beta\alpha}^{(1)}(n,m;n_\perp)\nonumber\\
&=&\Lambda(n,m)a_{\alpha\gamma}^{(1)}
\end{eqnarray}
where 
\begin{eqnarray}
\Lambda(n,m)&=&-\frac{4}{\pi^2}\left(1-(-1)^{n+m}\right)\frac{nm}{n^2-m^2}\quad (n\neq m)\nonumber\\
&=&0\quad (n=m)\label{lambda}
\end{eqnarray}
We observe that for a given set of values $(n,m,n_\perp)$, we have an ensemble of equations with the unknowns linearly related. We can express them in  matrix notation:
\begin{equation}
d(n)g\,\hat C^{(1)}+d(m)\hat C^{(1)}g^T=\Lambda(n,m)a\label{mat1}
\end{equation}
where $d(n)=n^2+4n_\perp^2/\eta^2$ and we only show the arguments that change to simplify the notation. To solve these equations, let us assume that the matrix $g$ can be diagonalized or, in other words, there is an eigenvector basis that spans the $M$-dimensional space. We define:
\begin{equation}
g^Tw(s)=\lambda(s)w(s)\quad,\quad Re(\lambda(s))<0\quad\forall\,s
\end{equation}
we multiply by $w(s')^T$ the left of equation (\ref{mat1})  and by $w(s)$ its right and we can isolate the $\hat C^{(1)}$ matrix components:
\begin{equation}
w(s')^T\hat C^{(1)}w(s)=\bar G_{s's}=\frac{\Lambda(n,m)w(s')^Taw(s)}{d(n)\lambda(s')+d(m)\lambda(s)}
\end{equation}
or, in cartesian coordinates
\begin{equation}
\hat C^{(1)}=\left(P^{-1}\right)^T\bar GP^{-1}\label{Pmat}
\end{equation}
where $P$ is the matrix where its $s$-column is  the components of $w(s)$: $Pe(s)=w(s)$ with $e(s)_i=\delta_{s,i}$ being the canonical orthonormal basis. Finally, we can write $\hat C^{(1)}$ in components:
\begin{equation}
\hat C^{(1)}_{\alpha\beta}(n,m;n_\perp)=\sum_{\sigma\sigma'}G_{\sigma\sigma';\alpha\beta}\hat F(n,m;n_{\perp};\sigma,\sigma')\label{hatC1}
\end{equation}
where
\begin{equation}
G_{\sigma\sigma';\alpha\beta}=\left(P^{-1}\right)_{\sigma\alpha}\left(P^{-1}\right)_{\sigma'\beta}\sum_{s}\sum_{s'}a_{ss'}^{(1)}P_{s\sigma}P_{s'\sigma'}\label{hatC2}
\end{equation}
and
\begin{equation}
\hat F(n,m;n_{\perp};\sigma,\sigma')=\frac{\Lambda(n,m)}{\left(n^2+4n_\perp^2/\eta^2\right)\lambda(\sigma)+\left(m^2+4n_\perp^2/\eta^2\right)\lambda(\sigma')}\label{hatC3}
\end{equation}
where $\vec a_{ss'}^{(1)}=a_{ss'}^{(1)}\hat\i$ is given by eq.(\ref{aa0}). Observe that  the property  $a_{\alpha\gamma}^{(1)}=-a_{\gamma\alpha}^{(1)}$ 
 imply $G_{\sigma\sigma';\alpha\beta}=-G_{\sigma'\sigma;\beta\alpha}$.

\begin{figure}[h!]
\begin{center}
\includegraphics[height=7cm]{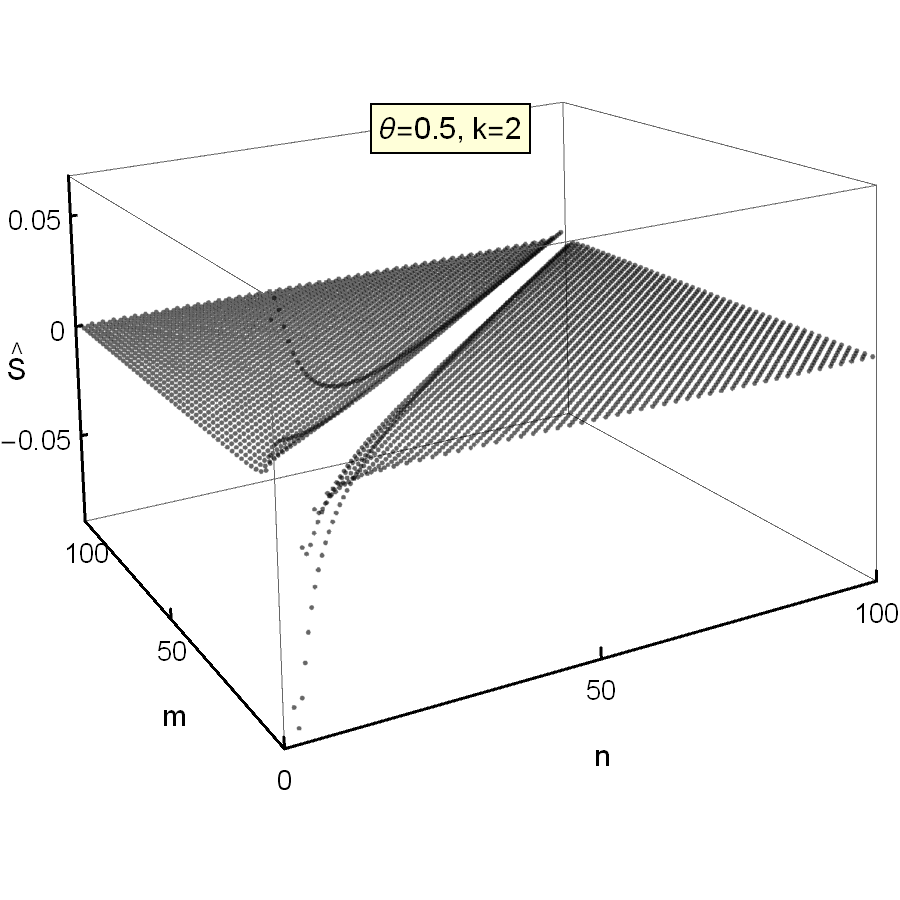}  
\includegraphics[height=7cm]{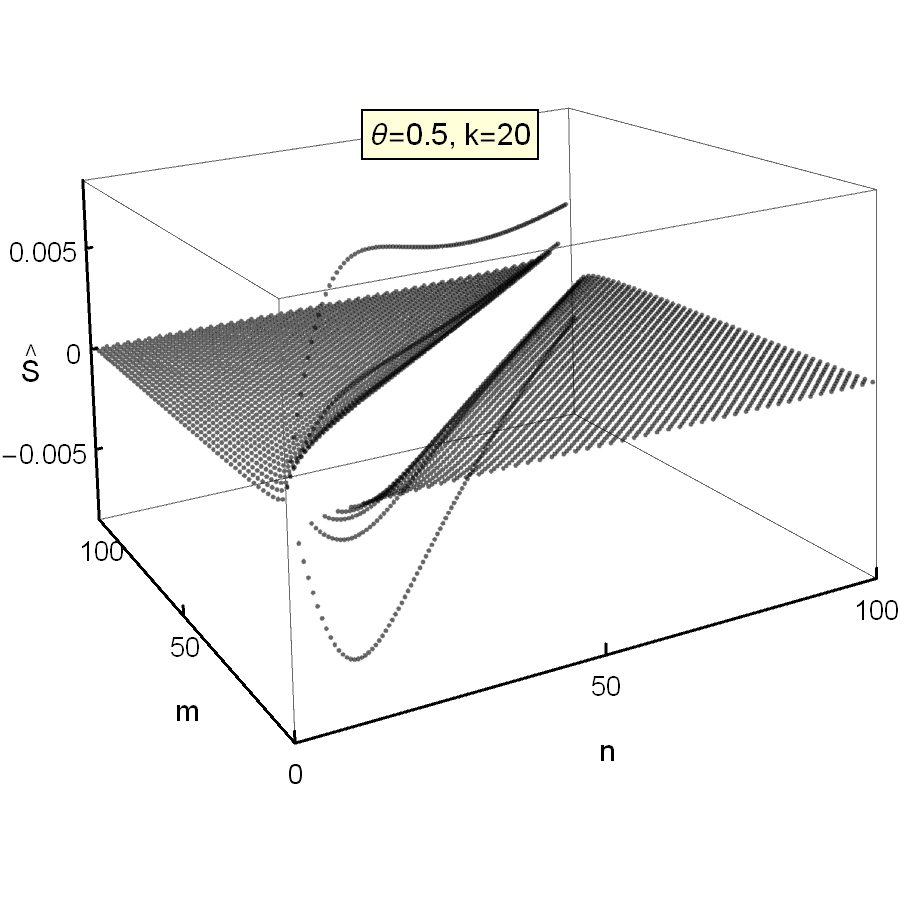}            
\end{center}
\kern -1.5cm
\caption{The basic Structure Function $\hat S(n,m;k,\theta)$. It is shown only the $(n,m)$ values where $\hat S$ is non-zero. The sub-lattices with equal parity, i.e. both $n$ and $m$ even or odd, have $\hat S=0$.\label{struc} }
\end{figure}

Please, observe that $\hat C_{\alpha\beta}^{(1)}(n,m;n_\perp)$ is a linear combination of the {\it basic structure function} $\hat F(n,m;n_\perp;\sigma,\sigma')$. We show in Figure \ref{struc} the behavior of a related function that only depends on the relation between eigenvalues:
\begin{equation}
\hat S(n,m;k,\theta_{\sigma\sigma'})\equiv\lambda(\sigma')\hat F(n,m;n_\perp;\sigma,\sigma')
\end{equation}
where $k^2=4n_\perp^2/\eta^2$ and $\theta_{\sigma\sigma'}^2=\lambda(\sigma)/\lambda(\sigma')$. We only need to plot values with $\theta_{\sigma\sigma'}<1$ because the relation:
\begin{equation}
\hat S(m,n;k,\frac{1}{\theta_{\sigma\sigma'}})=-\theta_{\sigma\sigma'}^2\hat S(n,m;k,\theta_{\sigma\sigma'})
\end{equation}

Finally, the correlations in real space given by eq. (\ref{C1real}) can be written:
\begin{equation}
\bar C_{\alpha\beta}^{(1)}(x,z)=\sum_{\sigma\sigma'}G_{\sigma\sigma';\alpha\beta}F(x,z;\sigma,\sigma')\label{CR1}
\end{equation}
where  we call $F$ {\it the basic correlation function} and it is written:
\begin{equation}
F(x,z;\sigma,\sigma')=\tilde F(x,z;\sigma,\sigma')-\tilde F(z,x;\sigma',\sigma)\equiv \lambda(\sigma')^{-1}S(x,z;\theta_{\sigma\sigma'})\label{CR2}
\end{equation}
and
\begin{eqnarray}
&&\tilde F(x,z;\sigma,\sigma')=-\frac{8}{\pi^2}\frac{1}{(L')^{d-1}}\sum_{n_\perp\in\mathbb{Z}^{d-1}}e^{i\frac{2\pi}{L'}n_{\perp}\cdot(x_\perp-z_{\perp})}\sum_{n=1}^{\infty}\sin\left(\frac{(2n-1)\pi x_1}{L}\right)\sum_{m=1}^{\infty}\sin\left(\frac{2m\pi z_1}{L}\right)\nonumber\\
&&\frac{(2n-1)2m}{(2n-1)^2-(2m)^2}\frac{1}{\left((2n-1)^2+4n_\perp^2/\eta^2\right)\lambda(\sigma)+\left((2m)^2+4n_\perp^2/\eta^2\right)\lambda(\sigma')}\label{CR3}
\end{eqnarray}

One interesting observable associated with the two-body correlation is the field's spatial average fluctuation. At equilibrium, these magnitudes are related to other characteristics of the system. For instance, the Einstein relation between the system's overall energy fluctuations and its specific heat. For a system composed of $M$ fields, we can define the fluctuations between the fields $\alpha$ and $\beta$ at the stationary state as:
\begin{equation}
\Delta_{\alpha\beta}=\langle(e_\alpha-e_\alpha^*)(e_\beta-e_\beta^*)\rangle_{ss}
\end{equation}
where $e_\alpha$ is the spatially averaged field $\phi_\alpha$:
\begin{equation}
e_\alpha=\frac{1}{\vert\Lambda\vert}\int_\Lambda dx\,\phi_\alpha(x)
\end{equation}
and $e_\alpha^*$ is its average value at the stationary state. Fluctuations can be written as the sum of correlations:
\begin{equation}
\Delta_{\alpha\beta}=\frac{1}{\vert\Lambda\vert^2}\int_\Lambda dx\int_\Lambda dz \,C_{\alpha\beta}(x,z)
\end{equation}
This expression  for  the DSe is writen as the sum of two contributions:
\begin{equation}
\Delta_{\alpha\beta}=\Delta_{\alpha\beta}^{leq}+\Delta_{\alpha\beta}^{neq}
\end{equation}
where the local equilibrium contribution is
\begin{equation}
\Delta_{\alpha\beta}^{leq}=\frac{1}{\vert\Lambda\vert^2}\int_\Lambda dx\,\left(-S^{-1}\right)_{\alpha\beta}(\phi^*(x))
\end{equation}
and the remaining, the nonequilibrium part, is
\begin{equation}
\Delta_{\alpha\beta}^{neq}=\frac{1}{\vert\Lambda\vert^2}\int_\Lambda dx\int_\Lambda dz \,\bar C_{\alpha\beta}(x,z)\label{fluct}
\end{equation}
When we $\epsilon$-expand $\Delta_{\alpha\beta}^{neq}$ through the correlation expansion we get our second general result: 
\begin{itemize}
\item{\bf Theorem 2:} {\bf $\Delta_{\alpha\beta}^{neq}=\mathcal{O}(\epsilon^2)$ for all DSe with parallel plates as boundary conditions. }
\end{itemize}
In other words, the fluctuations for DSe systems with parallel plates as boundary conditions are, near to the equilibrium, at most of the order $\epsilon^2$. The field's global averaged values are very well described by the local equilibrium approximation whenever the stationary state is at the linear regime (order $\epsilon$). The nonequilibrium corrections appear at order $\epsilon^2$ despite their correlations that deviate from local equilibrium already at order $\epsilon$, and they are long-range.

The proof of this theorem is straightforward. We compute explicitly  $\Delta_{\alpha\beta}^{neq}$ near to the equilibrium at first order in $\epsilon$ and in the case of parallel plates (see section III above). We just substitute  expression (\ref{hatC1}) into (\ref{C1real}):
\begin{equation}
\Delta_{\alpha\beta}^{neq,(1)}=\frac{1}{\pi^2(L')^{d-1}}\sum_{n=1}^\infty\sum_{m=1}^\infty\frac{1}{nm}\left(1-(-1)^n\right)\left(1-(-1)^m\right)\hat C_{\alpha\beta}^{(1)}(n,m;0)\label{zero}
\end{equation}
We observe that the sums over $n$ and $m$ in eq.(\ref{zero}) runs over odd values due to the factors in front of $\hat C^{(1)}$. Moreover $\hat C_{\alpha\beta}^{(1)}$ includes the factor $\Lambda(n,m)$ given by eq.(\ref{lambda}) that is different from zero whenever $n$ and $m$ have different parities and therefore the overall result is zero. 

\section{The behavior of the basic correlation function}

We observe that the correlations at the first order in the $\epsilon$ expansion are given by eq. (\ref{CR3}) which is a linear combination of $F$-functions ({\it basic correlation function}). Therefore, $F$  contains the structural part of the nonequilibrium correlations in real space, and it is interesting to get some insight into it.

Let us begin the study of $F$ with the one-dimensional case. We see that we can get some idea of its behavior by doing numerically the sums in eq.(\ref{CR3}) for given values of the ratio $\theta_{\sigma\sigma'}=\lambda(\sigma)/\lambda(\sigma')$. However,  the sums converge very poorly due to the sinus functions. Therefore we had to transform it to a new one with a better numerical convergence behavior. After some algebra (see details in Appendix V) we transform eq.(\ref{CR2}) into:
\begin{eqnarray}
S(x,z;\theta_{\sigma\sigma'})\equiv \lambda(\sigma')F(x,z;\sigma,\sigma')&=&\frac{1}{\pi}\frac{1}{1+\theta_{\sigma\sigma'}^2}\biggr[\sum_{m=1}^{\infty}\arctan A_{\sigma\sigma'}(m;x,z)\nonumber\\&
-&\sum_{m=1}^{\infty}\arctan A_{\sigma'\sigma}(m;z,x)+\frac{\pi}{2}\text{sgn}(x-z)
\biggl]\label{F}
\end{eqnarray}
where
\begin{equation}
A_{\sigma\sigma'}(m;x,z)=4\cos\left(\frac{\pi}{2}\bar x\right)\frac{\cosh\left(\frac{\pi}{2}\theta_{\sigma\sigma'}(2m-1)\right)\sinh\left(\frac{\pi}{2}\theta_{\sigma\sigma'}\bar z\right)}{\cosh\left(\pi\theta_{\sigma\sigma'}(2m-1)\right)-\cosh\left(\pi\theta_{\sigma\sigma'}\bar z\right)+2\cos^2\left(\frac{\pi}{2}\bar x\right)}
\end{equation}
with $\theta_{\sigma\sigma'}^2=\lambda(\sigma)/\lambda(\sigma')$, $\bar x=2x/L-1$ and $\bar z=2z/L-1$ and $\text{sign}(x)=x/\vert x\vert$ when $x\neq 0$ and $\text{sign}(0)=0$. 

\begin{figure}[h!]
\begin{center}
\includegraphics[height=9cm]{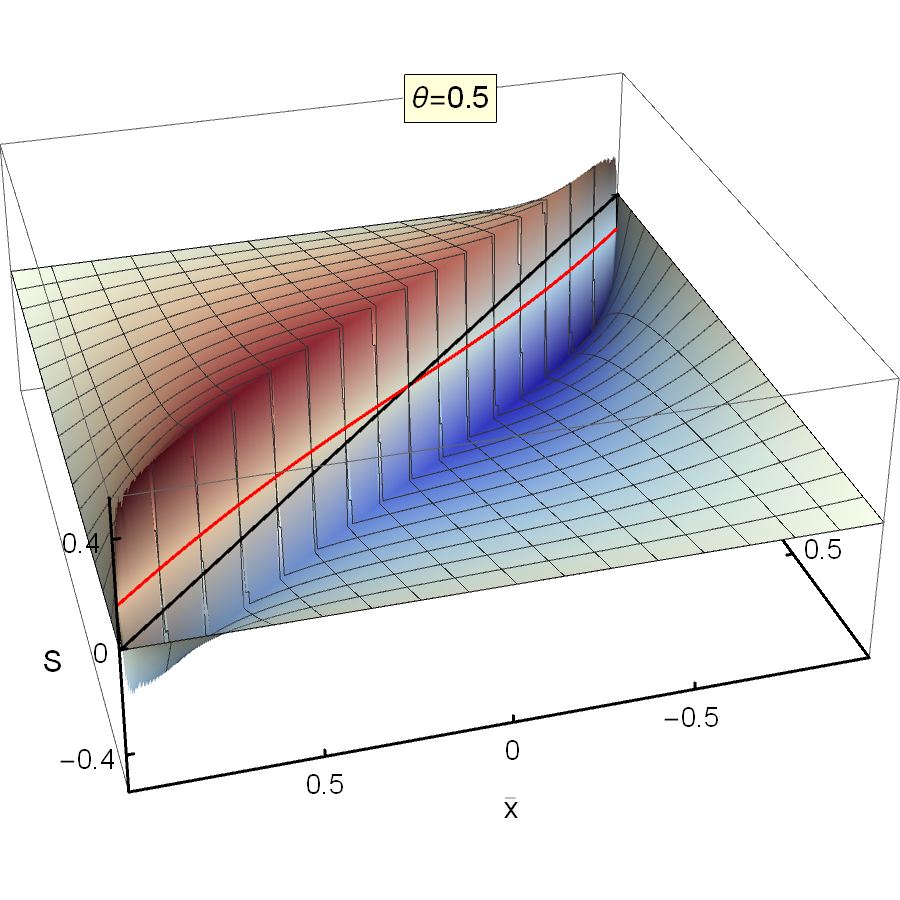}  
\end{center}
\kern -1.5cm
\caption{The $S(x,z;\theta)$ vs. $(\bar x=2x/L-1,\bar z=2z/L-1)$ for $\theta=0.5$. The red line shows the function $S(x,x;\theta)$. The black line is a reference $(\bar x,\bar x,0)$.\label{Ffig} }
\end{figure}

We show in figure \ref{Ffig} the behavior of the $S(x,z;\sigma,\sigma')$ vs. $(\bar x=2x/L-1,\bar z=2z/L-1)$ for $\theta_{\sigma\sigma'}=0.5$. We obtain the figure by computing $S$ numerically with eq. (\ref{F}). There are several points to remark. First, we see how $S(x,z;\theta)$ is zero for values $(x,z)$ located at the boundaries. Moreover, let us observe a defined discontinuity along the line $\bar x=\bar z$ where two anti-symmetric halves meet, forming a well-defined gap. There are some apparent rounding effects near the points $(\bar x,\bar z)=(\pm 1,\pm 1)$ but are just due to numerical computation difficulties. $S(x, x;\theta)$ is shown separately in figure \ref{Ffig} by a red line located at the middle of the gap.

We can analytically compute the magnitude of the gap along the line $(\bar x,\bar x)$ for a given $\theta$ from eq.(\ref{F}) and we find:
\begin{equation}
\Delta S=\vert\lim_{\epsilon\rightarrow 0}\left[S(x_0+\epsilon,x_0-\epsilon;\theta)-S(x_0-\epsilon,x_0+\epsilon;\theta)\right]\vert=\frac{1}{1+\theta^2}
\end{equation}
It seems remarkable that the size of the gap is independent of $x_0$. Similarly, we also find the limiting value at $\bar x=\pm 1$:
\begin{equation}
S(L,L;\theta)=-S(0,0;\theta)=\frac{1}{\pi}\frac{1}{1+\theta^2}\left[\arctan\theta^{-1}-\arctan\theta\right]
\end{equation} 
All these exact results are, of course, consistent with the numerical behavior obtained in Fig.  \ref{Ffig}.

In the $d=1$ case, we got the qualitative behavior of $S$ by computing numerically part of the infinite sum (\ref{F}) and/or extracting some analytical results from it.  There we were lucky because we could obtain a fast converging expression that made possible its overall description. However, we have been unable to find a similar expression to eq. (\ref{F}) when $d>1$. We circumvent this difficulty by studying analytically the limit $L\rightarrow\infty$ where we can use the Riemann summation formula to substitute the sums by integrals. As we will see below, we pay the price of only describing the correlations near the system's boundaries.  

Let us write from eq.(\ref{CR2}):
\begin{equation}
S(x,z;\theta)=\tilde S(x,z;\theta)-\frac{1}{\theta^2}\tilde S(z,x;\frac{1}{\theta})
\end{equation}
where
\begin{equation}
\tilde S(x,z;\theta_{\sigma\sigma'})=\lambda(\sigma')\tilde F(x,z;\sigma,\sigma')
\end{equation}
and  $\tilde F$ is given by eq.(\ref{CR3}). 
We can do the summation over $m$  similarly as  we did for the $d=1$ case (see Appendix V).  We get
\begin{eqnarray}
\tilde S(x,z;\theta)&=&\frac{2}{\pi}\frac{1}{1+\theta^2}\frac{1}{(L')^{d-1}}\sum_{n_\perp\in\mathbb{Z}}\exp\left[i\frac{2\pi}{L'}n_\perp\cdot (x_\perp-z_\perp)\right]\nonumber\\
&&\sum_{n=1}^\infty\sin\left[(2n-1)\pi\frac{x_1}{L}\right]\frac{2n-1}{(2n-1)^2+4n_\perp^2/\eta^2}\biggl\{\cos\left[(2n-1)\pi\frac{z_1}{L}\right]\nonumber\\
&-&\frac{\displaystyle\sinh\left[\alpha(n,n_\perp;\theta)\frac{\pi}{2}\left(1-\frac{2z_1}{L}\right)\right]}{\displaystyle\sinh\left[\alpha(n,n_\perp;\theta)\frac{\pi}{2}\right]}\biggr\}\label{Ft2}
\end{eqnarray}
with
\begin{equation}
\alpha(n,n_\perp;\theta)=\left[\theta^2(2n-1)^2+4(1+\theta^2)n_\perp^2/\eta^2\right]^{1/2}
\end{equation}
We can do explicitly the first sum by using some of the Fourier sums that we derived in Appendix IV and we obtain:
\begin{eqnarray}
&&\sum_{n=1}^\infty \sin\left[(2n-1)\pi\frac{x_1}{L}\right]\frac{2n-1}{(2n-1)^2+4n_\perp^2/\eta^2}\cos\left[(2n-1)\pi\frac{z_1}{L}\right]\nonumber\\
&=&\frac{\pi}{8}(\cosh\left(\frac{\vert n_\perp\vert\pi}{\eta}\right))^{-1}\biggl[
\Theta(1-\frac{1}{L}(x_1+z_1))\cosh\left(\frac{\vert n_\perp\vert\pi}{\eta}\left[1-2\frac{x_1+z_1}{L}\right]\right)\nonumber\\
&-&\Theta(\frac{1}{L}(x_1+z_1)-1)\cosh\left(\frac{\vert n_\perp\vert\pi}{\eta}\left[2\frac{x_1+z_1}{L}-3\right]\right)\nonumber\\
&+&\text{sign}(x_1-z_1)\cosh\left(\frac{\vert n_\perp\vert\pi}{\eta}\left[1-2\frac{\vert x_1-z_1\vert}{L}\right]\right)\biggr]
\end{eqnarray}
In order to go beyond these expressions we need to convert those remaining sums into integrals by means of the Riemann summation formula:
\begin{equation}
\lim_{L\rightarrow\infty}\frac{1}{L}\sum_{n=-N}^N f\left(\frac{n}{L}\right)=\int_{-\infty}^\infty dq\,f(q)\quad,\quad N>L^\sigma\quad\sigma>0
\end{equation}
 The formula can be applied under some conditions on $f$ (see for instance ref.\cite{Riem}). Therefore we substitute $n_\perp=L \eta q_\perp/2\pi$ and $n=Lq/2\pi$ in eq. (\ref{Ft2}), we keep $x$ and $z$ fixed and we do the limit $L\rightarrow\infty$. The result is:
 \begin{eqnarray}
&&\tilde S(x,z;\theta)=\frac{2}{(2\pi)^d}\frac{1}{1+\theta^2}\int_{\mathbb{R}^{d-1}}dq_\perp\,e^{ i  q_\perp\cdot (x_\perp-z_\perp)}\nonumber\biggl[ \frac{\pi}{4}e^{-\vert q_\perp\vert (x_1+z_1)}\\
&+&\frac{\pi}{4}\text{sign}(x_1-z_1)e^{-\vert q_\perp\vert \vert  x_1-z_1\vert}
-\int_0^\infty dq\, \frac{q\sin(x_1 q)}{q^2+q_\perp^2}e^{-z_1 \tilde\alpha(q,q_\perp;\theta)}
\biggr]\label{Ft3}
\end{eqnarray}
 where 
 \begin{equation}
 \tilde\alpha(q,q_\perp;\theta)=\sqrt{\theta^2q^2+(1+\theta^2)q_\perp^2}\label{Ft4}
 \end{equation}
 Let us remark that keeping fix $x$ and $z$ and doing $L\rightarrow\infty$ we are describing effectively the $S$ function around the boundary $x_1$ or $z_1=0$ because the other one is at $x_1$ or $z_1=L$ that  is now infinitely far away. Therefore, $\tilde S$ only incorporates the boundary condition $\tilde S=0$ when $x_1$ or $z_1=0$. 
 
 It can be shown that $S(x,z;\theta)=2\tilde S(x,z;\theta)$ by using eq. (\ref{rel1}) in Appendix VI. This property only applies in the $L\rightarrow\infty$ limit as the even/odd modes differences in eq.(\ref{CR3}) disappear.
  
We can compute the integrals explicitly, but their technicalities depend on the dimension. Let's start with $d=1$ to check this limiting case with the overall description we already found numerically. 
\begin{figure}[h!]
\begin{center}
\includegraphics[height=7cm]{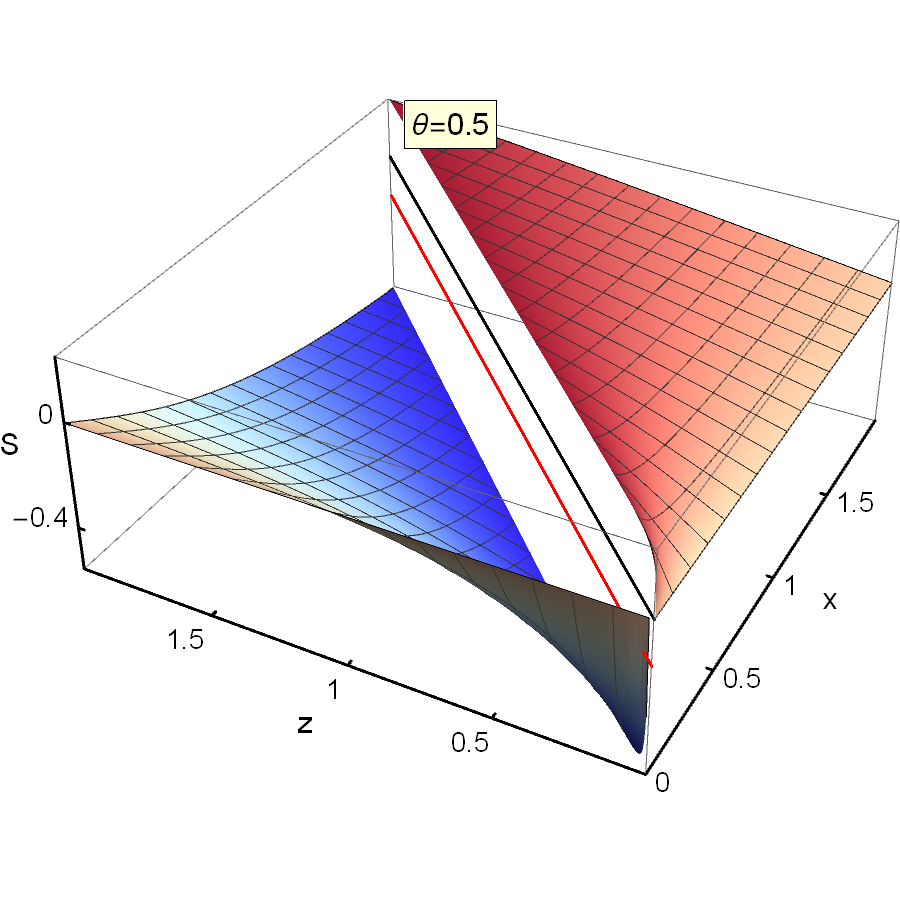}  
\includegraphics[height=7cm]{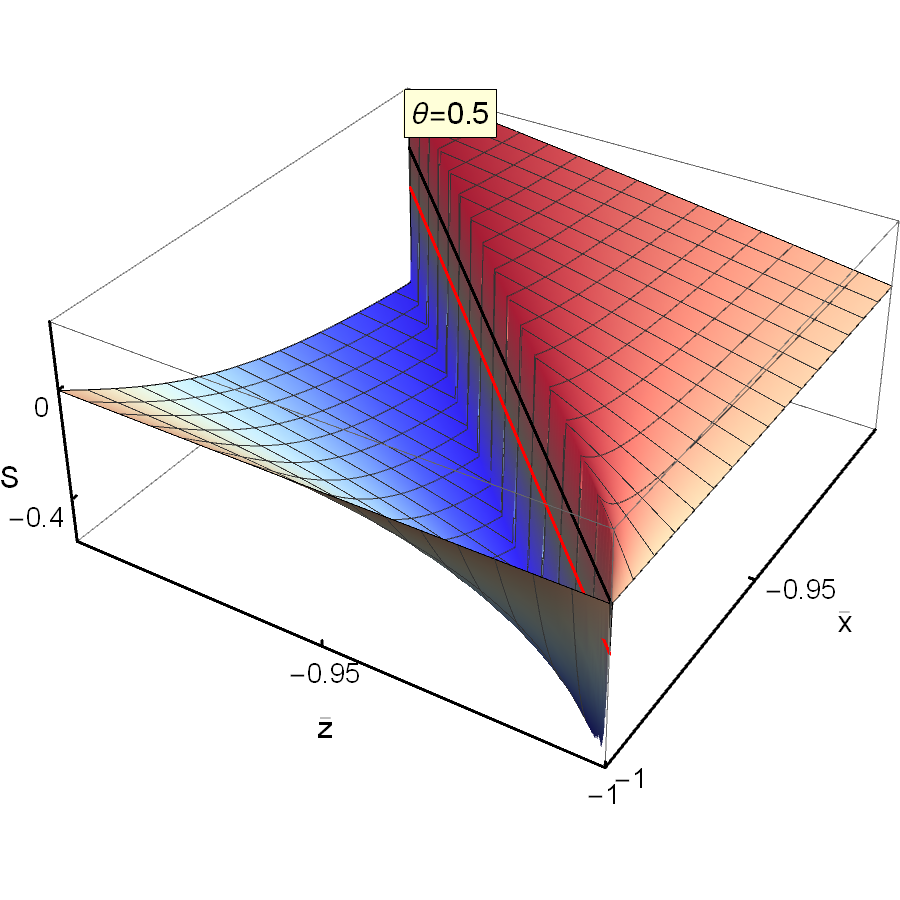}    
\end{center}
\kern -1.5cm
\caption{Left: $S$ vs. $(x,z)$ for $\theta=0.5$ using the expression (\ref{F1d}).  Right: Figure \ref{Ffig} expanded around $(\bar x,\bar z)=(-1,-1)$. The red curve is $(x, x,F(x,x))$. The black line is a reference $(\bar x,\bar z,0)$.\label{Fcont1d} }
\end{figure}
For $d=1$ the corresponding expression (\ref{Ft3}) is equivalent to $q_\perp=0$ and the integral over $q_\perp$ disappears. The unique integral that remains to do is 
 \begin{equation}
 \int_0^\infty dq\, \frac{\sin(x_1 q)}{q}e^{-z_1\theta q}=\arctan\left(\frac{x_1}{\theta z_1}\right)
 \end{equation}
(see GR.3.941.1 in \cite{Grads}) and therefore 
\begin{equation}
\tilde S(x,z;\theta)=\frac{1}{4(1+\theta^2)}\left[1+\text{sign}(x_1-z_1)-\frac{4}{\pi}\arctan\left(\frac{x_1}{\theta z_1}\right)\right]
\end{equation}
and
\begin{equation}
S(x,z;\theta)=\frac{1}{2(1+\theta^2)}\left[\text{sign}(x_1-z_1)+\frac{2}{\pi}\arctan\left(\frac{z_1^2\theta^2-x_1^2}{2\theta x_1z_1}\right)\right]\label{F1d}
\end{equation}
From eq. (\ref{F1d}) we derive that $S$ behaves as a power law for large values of $x_1$:
\begin{equation}
S(x_1,z_1;\theta)\simeq\frac{2\theta}{1+\theta^2}\frac{z_1}{\pi x_1}\quad ,\,x_1\rightarrow\infty
\end{equation}
that is typical for systems at non-equilibrium stationary states.
One can check that  $S$ has all the properties we discussed above for the $d=1$ case numerically solved. Moreover, we compare in figure \ref{Fcont1d} the $S$ obtained from eq. (\ref{F1d}) and the  $S$ from eq. (\ref{F}) computed for values around $(\bar x,\bar z)=(-1,-1)$. We observe that our computation captures with precision the nontrivial behavior near that boundary. 
This limiting approach contains the most relevant part of the description of the non-equilibrium correlations for DSe models, and therefore, it permits us to get analytical results.

\begin{figure}[h!]
\begin{center}
\includegraphics[height=7cm]{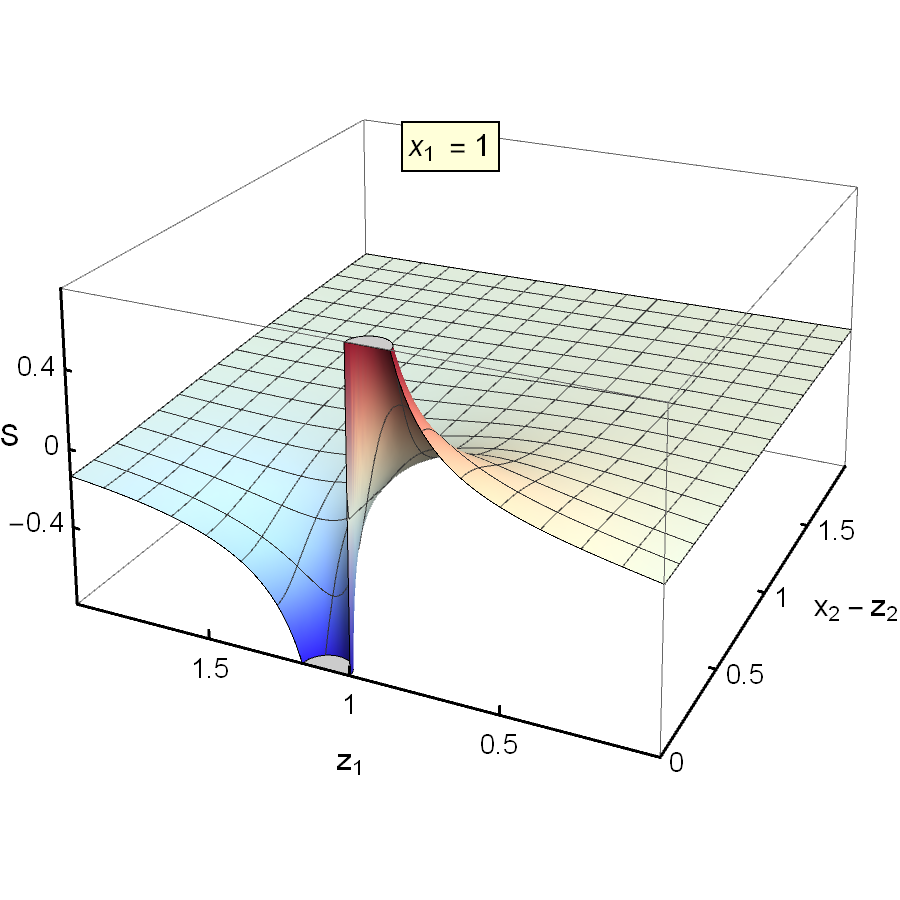}  
\includegraphics[height=7cm]{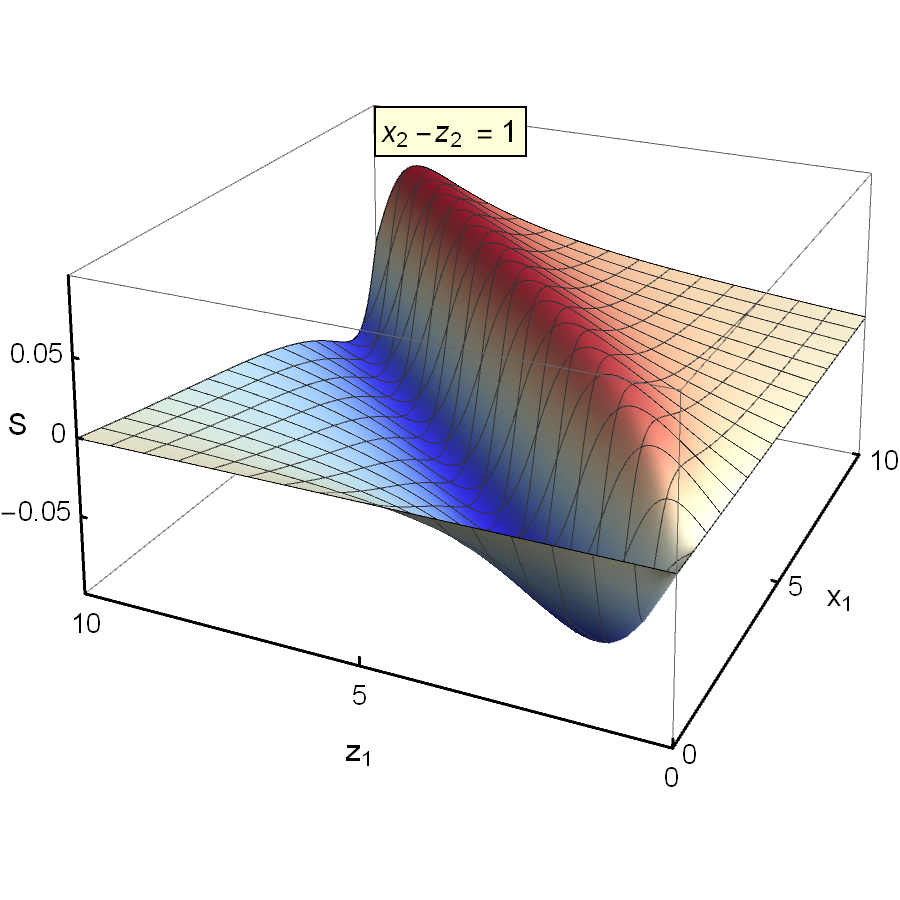}    
\includegraphics[height=7cm]{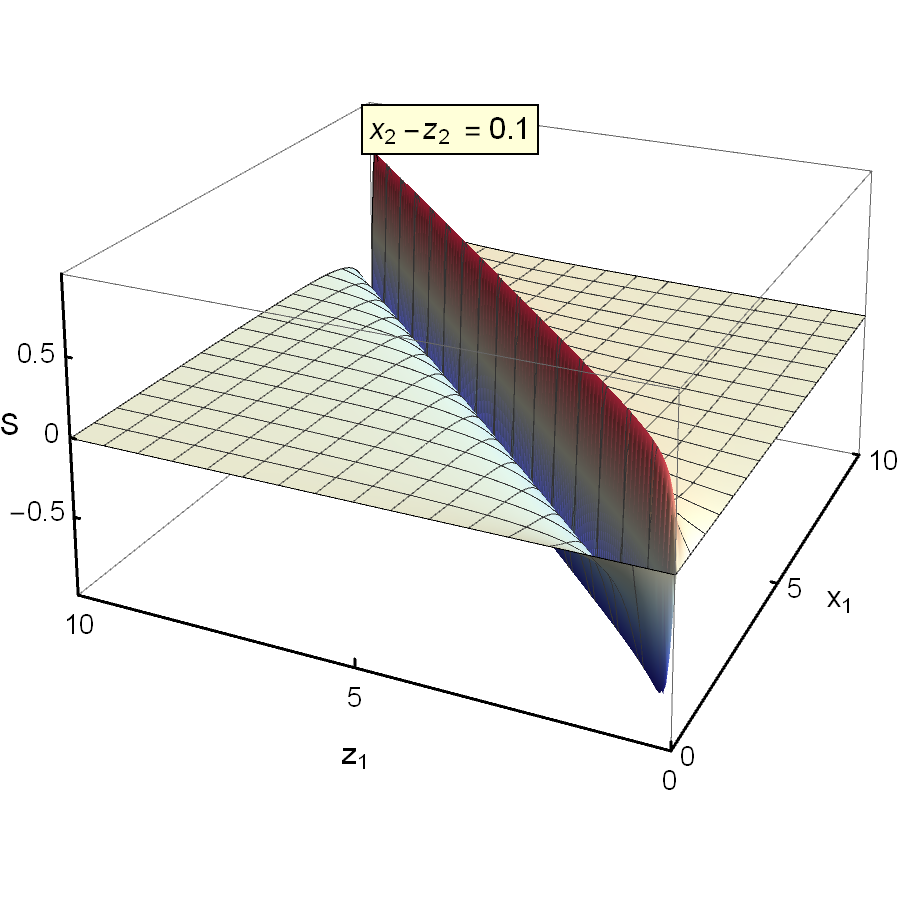} 
\includegraphics[height=7cm]{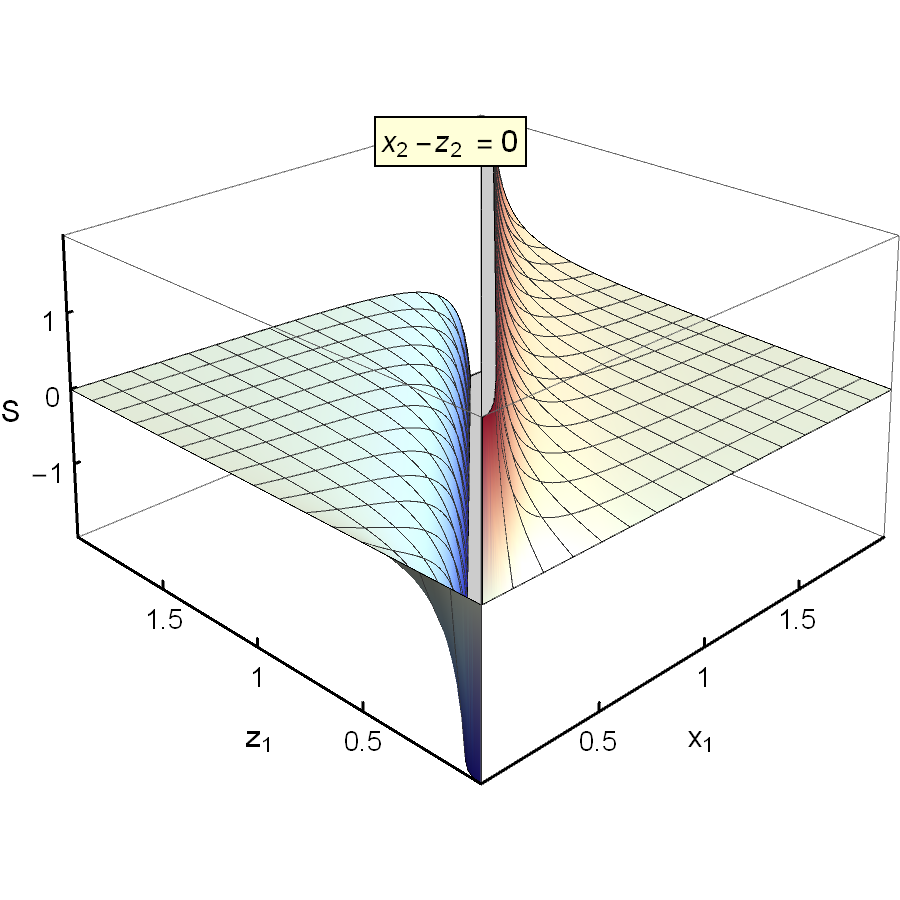}   
\end{center}
\kern -1.5cm
\caption{Several behaviors of $S$ for $\theta=0.5$ from eq. (\ref{F2dtot}). Top-Left: $S$ vs. $(z_1,x_2-z_2)$ for $x_1=1$. We see the singularity at $z_1=1$. $S$ vs. $(z_1,x_1)$ for $x_2-z_2=1$ (Top-Right), $x_2-z_2=0.1$  (Bottom-Left) and $x_2-z_2=0$  (Bottom-Right) respectively.   \label{F2dfig} }
\end{figure}

Having certified that our limiting theory is describing correctly the system near the boundary we can show with some more confidence the results for $d=2$ where no numeric data is, at this moment, available. $S$ is given by eq.(\ref{CR2}) by using eqs. (\ref{Ft3},\ref{Ft4}) restricted to $d=2$:
\begin{eqnarray}
S(x,z;\theta)&=&\frac{4}{(2\pi)^2}\frac{1}{1+\theta^2}\int_{-\infty}^\infty dq_2\, e^{iq_2 (x_2-z_2)}\biggl[\frac{\pi}{4}e^{-\vert q_2\vert (x_1+z_1)}\nonumber\\
&+&\frac{\pi}{4}\text{sign}(x_1-z_1)e^{-\vert q_2\vert \vert  x_1-z_1\vert}
-\int_0^\infty dq\, \frac{q\sin(x_1 q)}{q^2+q_2^2}e^{-z_1 \sqrt{\theta q^2+(1+\theta^2)q_2^2}}
\biggr]\label{Ft5}
\end{eqnarray}
The first two integrals over $q_2$ can be easily done by using GR.3.893.2 from \cite{Grads}. The last integral is explicitly done in Appendix VI (see eq. (\ref{exp7})). The final result is:
\begin{equation}
S(x,z;\theta)=-\frac{1}{\pi}\frac{x_1z_1}{1+\theta^2}\frac{(1+\theta^2)(z_1^2-x_1^2)+(1-\theta^2)(x_2-z_2)^2}{D_1(x,z)D_2(x,z;\theta)}\label{F2dtot}
\end{equation}
with
\begin{eqnarray}
D_1(x,z)&=&[(x_1-z_1)^2+(x_2-z_2)^2][(x_1+z_1)^2+(x_2-z_2)^2]\nonumber\\
D_2(x,z;\theta)&=&[(1+\theta^2)x_1^2+\theta^2((1+\theta^2)z_1^2+(x_2-z_2)^2)]^{1/2}\label{D1D2}
\end{eqnarray}
Observe that $S$ has the limiting behaviors:
\begin{eqnarray}
S(x,z;\theta)&\simeq&\frac{1}{\pi\sqrt{1+\theta^2}}\frac{z_1}{x_1^2}\quad ,\, x_1\rightarrow\infty\nonumber\\
&\simeq&-\frac{1-\theta^2}{\pi\theta(1+\theta^2)}\frac{x_1z_1}{\vert x_2-z_2\vert^3}\quad ,\, \vert x_2-z_2\vert\rightarrow\infty\nonumber\\
&=&\frac{1}{\pi\sqrt{1+\theta^2}}\frac{x_1z_1}{(x_1^2-z_1^2)\sqrt{x_1^2+\theta^2z_1^2}}\quad,\, x_2=z_2
\end{eqnarray}
Please, observe that the decay for large distances in the x-axis is $x_1^{-2}$ while in the y-axis is $\vert x_2-z_2\vert^{-3}$ even though the boundaries have been sent to infinity. Moreover, there is a singularity at $x_1=z_1$ when $x_2=z_2$. We show in figure \ref{F2dfig} some overall examples of the $S$ behavior.

We see that we obtain a rich complex behavior for the $S$ functions that are the basis for the $\bar C^{1}$ correlations. Let us now study  $M=2$ case.

\section{The case of two-locally conserved fields (\boldmath$M=2$)}

We have seen that the correlation's excess is a linear combination of the basic correlation function $F$ whose properties can be derived independently from the DSe model we chose to study. Therefore, most of their geometrical behavior is already formally determined. However, we need to compute $G$ from eq. (\ref{hatC2}) that depends on the model studied to get the correlations' expressions. This section gives a detailed account of the correlations of a DSe with two fields ($M=2$). We'll see that the correlations $\bar C_{11}^{(1)}$ and $\bar C_{22}^{(1)}$ are, except for a model-dependent factor, universal. This fact permits us to have a generic vision of part of the correlations in this case.

Let us assume that  DSe is defined by a $2\times 2$ given matrix $g$. To obtain $G$ we need mainly to get the eigenvalues and eigenvectors of the $g^T$-matrix computed at the equilibrium. The eigenvalues are
\begin{equation}
\lambda(1,2)=\frac{1}{2}\left[g_{11}+g_{22}\pm D^{1/2}\right]\quad,\quad D=(g_{11}-g_{22})^2+4g_{12}g_{21}
\end{equation}
And the corresponding $P$ matrix (see eq. (\ref{Pmat})) is
\begin{equation}
P=
\begin{pmatrix}
 -g_{21}&-g_{21} \\
g_{11}-\lambda(1)&g_{11}-\lambda(2)
 \end{pmatrix}
\end{equation}
Finally, our boundary conditions imply that $\vec a_{\alpha\gamma}=a_{\alpha\gamma}\hat\i$. Where the matrix $a$ is of the form:
\begin{equation}
a=
\begin{pmatrix}
0& a_{12} \\
-a_{12}&0 
 \end{pmatrix}
\end{equation}
With all this items at hand we get $G$ from eq. (\ref{hatC2}): 
\begin{eqnarray}
G_{\sigma\sigma';\alpha\beta}&=&a_{12} g_{21}\left(\lambda(\sigma')-\lambda(\sigma)\right)\left(P^{-1}\right)_{\sigma\alpha}\left(P^{-1}\right)_{\sigma'\beta}\quad,\,\sigma'\neq\sigma\nonumber\\
&=&0\quad,\, \sigma'=\sigma
 \end{eqnarray}
Therefore, from eq. (\ref{CR1}) the nonequilibrium correlations $\bar C_{\alpha\beta}^{(1)}$ are:
\begin{eqnarray}
\bar C_{11}^{(1)}(x,z)&=&\frac{a_{12}g_{12}}{\pi\lambda(2)^2}D_{11}(x,z;\theta)  \label{M2C11}\\
\bar C_{22}^{(1)}(x,z)&=&-\frac{a_{12}g_{21}}{\pi\lambda(2)^2}D_{11}(x,z;\theta)\label{M2C22}\\
\bar C_{12}^{(1)}(x,z)&=&-\frac{a_{12}}{\pi\lambda(2)^2}D_{12}(x,z;\theta)
\label{M2C12}
\end{eqnarray}
and
\begin{eqnarray}
D_{11}(x,z;\theta)&=&\frac{\pi}{ 1-\theta^2}\left[S(x,z;\theta)+S(z,x;\theta)\right]\nonumber\\
D_{12}(x,z;g)&=&\frac{\pi}{ 1-\theta^2}\biggl[(g_{11}-\lambda(2))S(x,z;\theta)+(g_{11}-\lambda(1))S(z,x;\theta)\biggr]
\end{eqnarray}
where $\theta^2=\lambda(1)/\lambda(2)$.
 
Let us remark that the expressions (\ref{M2C11}), (\ref{M2C22}) and (\ref{M2C12}) are generic for the $M=2$ case.
We see that the spatial structure of $\bar C_{11}^{(1)}$ and $C_{22}^{(1)}$ is the same independently on the model studied. Let us see how the correlation excess behaves for $d=1$ and $d=2$.

\subsection{\boldmath$d=1$:}

In this case $F$ is given by eq.(\ref{F}). Therefore:
\begin{eqnarray}
&&D_{11}(x,z;\theta)=\frac{1}{1-\theta^4}\left[I(x,z;\theta)-I(x,z;\frac{1}{\theta})+I(z,x,\theta)-I(z,x,\frac{1}{\theta})\right]\label{D11}\\
&&D_{12}(x,z;g)=\frac{\lambda(2)}{2(1+\theta^2)}\biggl[\pi\text{sgn}(x-z)+I(z,x,\frac{1}{\theta})-I(x,z,\frac{1}{\theta})+I(z,x,\theta)-I(x,z,\theta)\biggr]\nonumber\\
&&
\phantom{12345678912}+\frac{g_{11}-g_{22}}{2(1-\theta^4)}\biggl[I(x,z,\theta)+I(z,x,\theta)
-I(z,x,\frac{1}{\theta})-I(x,z,\frac{1}{\theta})\biggr]
\end{eqnarray}
 and
\begin{equation}
I(x,z;\theta)\equiv\sum_{m=1}^\infty \arctan A_{12}(m;x,z)
\end{equation}

\begin{figure}[h!]
\begin{center}
\includegraphics[height=7cm]{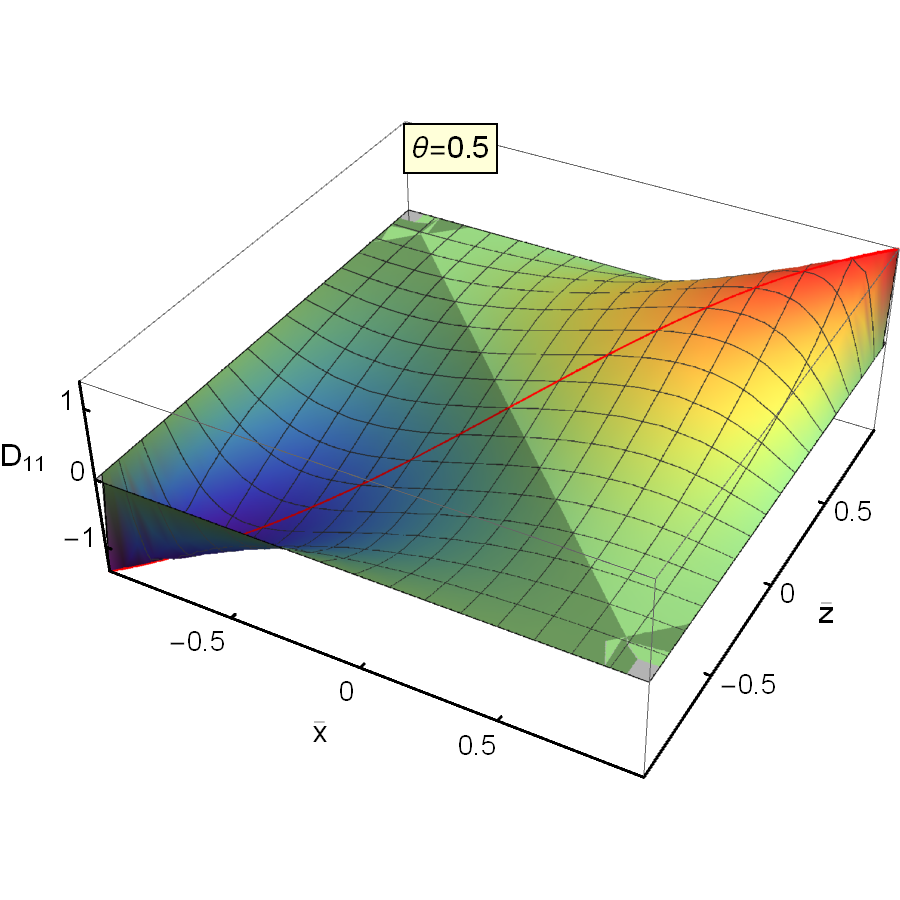}  
\includegraphics[height=7cm]{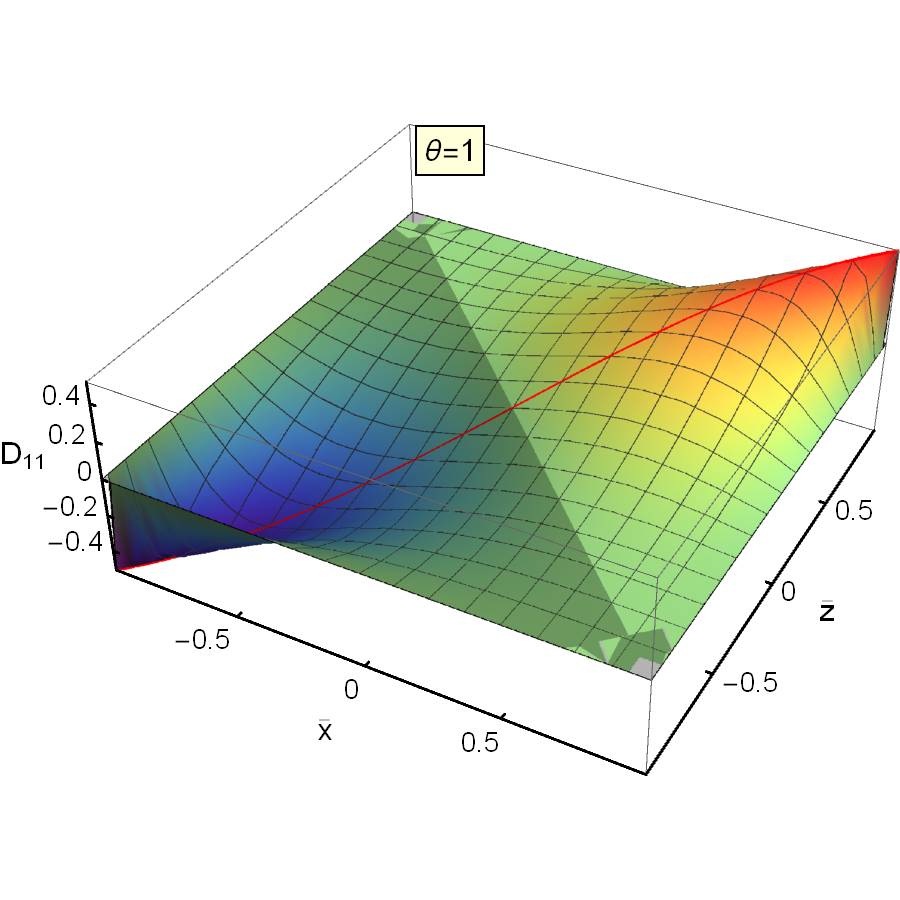} 
\includegraphics[height=7cm]{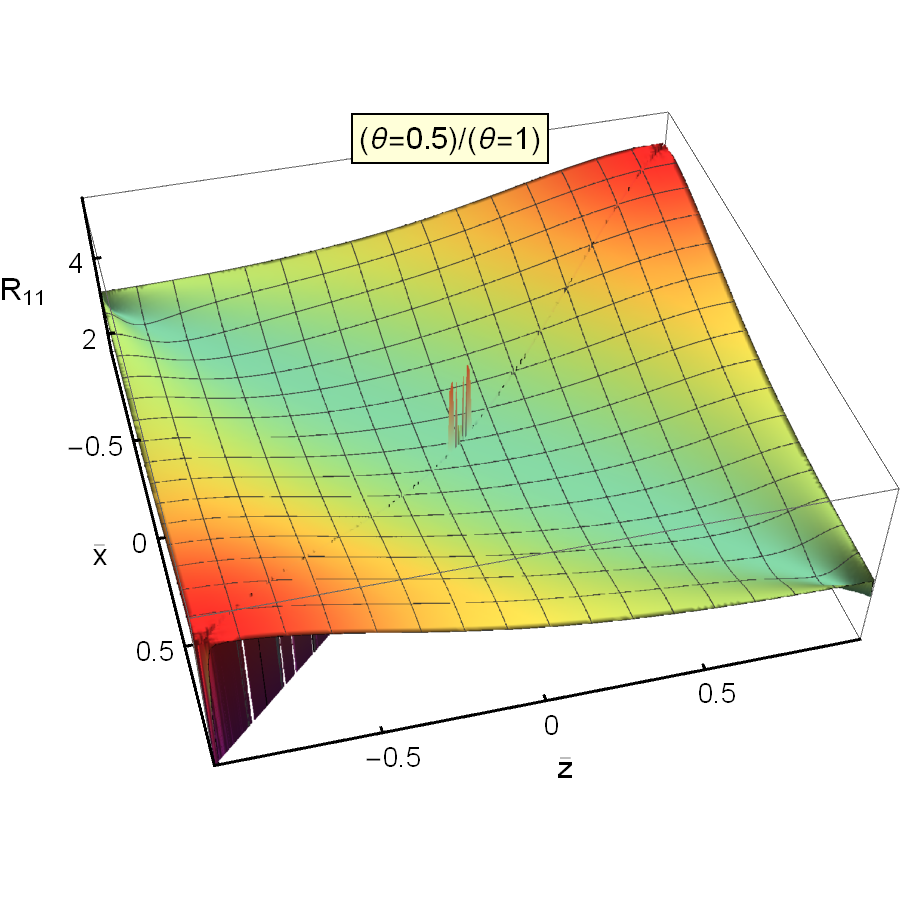}    
\end{center}
\kern -1.5cm
\caption{$D_{11}$ vs. $(\bar x,\bar z)$ numerically computed from eq. (\ref{D11}) for $d=1$. Top Left and Right: $\theta=0.5$ and $1$ respectively . The red curve is $(\bar x,\bar x,D_{11}(\bar x,\bar x))$. The gray plane is the zero-reference $(\bar x,\bar z,0)$. The bottom figure is the ratio $R_{11}=D_{11}(\theta=0.5)/D_{11}(\theta=1)$  to show that they lack of proportionality.  \label{C11} }
\end{figure}

In figure \ref{C11} we show $D_{11}(x,z)$ for $\theta=1/2$ and $1$. We see that  the shape is very similar in both cases but it is not proportional as it is shown in  the same figure  where we plot the  ratio $R_{11}\equiv D_{11}(x,z;\theta=0.5)/D_{11}(x,z;\theta=1)$. Moreover we see that the discontinuity that appeared in $F$  at the $x=z$ line dissapears in $D_{11}$ by symmetry.  We also see that $D_{11}(x,z)>0$  whenever $\bar x+\bar z>1$ and negative otherwise. The maximum  and the minimum of the $\bar C_{11}^{(1)}$ correlation are at $(x,z)=(L,L)$ and  $(0,0)$ respectively and their values are:
\begin{equation}
D_{11}(L,L)=-D_{11}(0,0)=\frac{2}{1-\theta^4}\left[\arctan\theta^{-1}-\arctan\theta\right]
\end{equation}
they range from $1/2$ for $\theta=1$ to $\pi$  when $\theta=0$.

We can get analytic expressions for $D_11$ by using the $F$'s functions given by eq.  (\ref{F1d})  that we obtained in the limit $L\rightarrow\infty$. Therefore:
\begin{equation}
D_{11}(x,z)=\frac{2}{1-\theta^4}\left[\arctan\left(\frac{\theta(x^2+z^2)}{(1-\theta^2)xz}\right)-\frac{\pi}{2}\right]\label{C11the}
\end{equation}

It is interesting to extract some limits from (\ref{C11the}).
\begin{eqnarray}
D_{11}(x,z)&=&-\frac{2}{\theta(1+\theta^2)}\frac{z}{x}+O(\frac{1}{x^2})\quad x\rightarrow\infty\quad,\quad z\,  \text{given}\nonumber\\
D_{11}(x,z)&=&\frac{2}{1-\theta^4}\left[\arctan\left(\frac{2\theta}{(1-\theta^2)\sin(2\phi)}\right)-\frac{\pi}{2}\right]\quad x=r\cos\phi\, ,\, z=r\sin\phi
\end{eqnarray}
We see how the correlations are long range when we fix one of the coordinates and the other tends to infinity. However, when we follow a path in the plane $(x,z)$ such that $x=r\cos\phi$ and  $z=r\sin\phi$ the correlations are constant for a given angle $\phi$ and any $r$. Observe that in this case the  mutual distance $\vert x-z\vert=r\vert\cos\phi-\sin\phi\vert$ is proportional to $r$ for a fix $\phi$. This is a rather singular behavior of the correlations that is typical in non-equilibrium systems.

\begin{figure}[h!]
\begin{center}
\includegraphics[height=9cm]{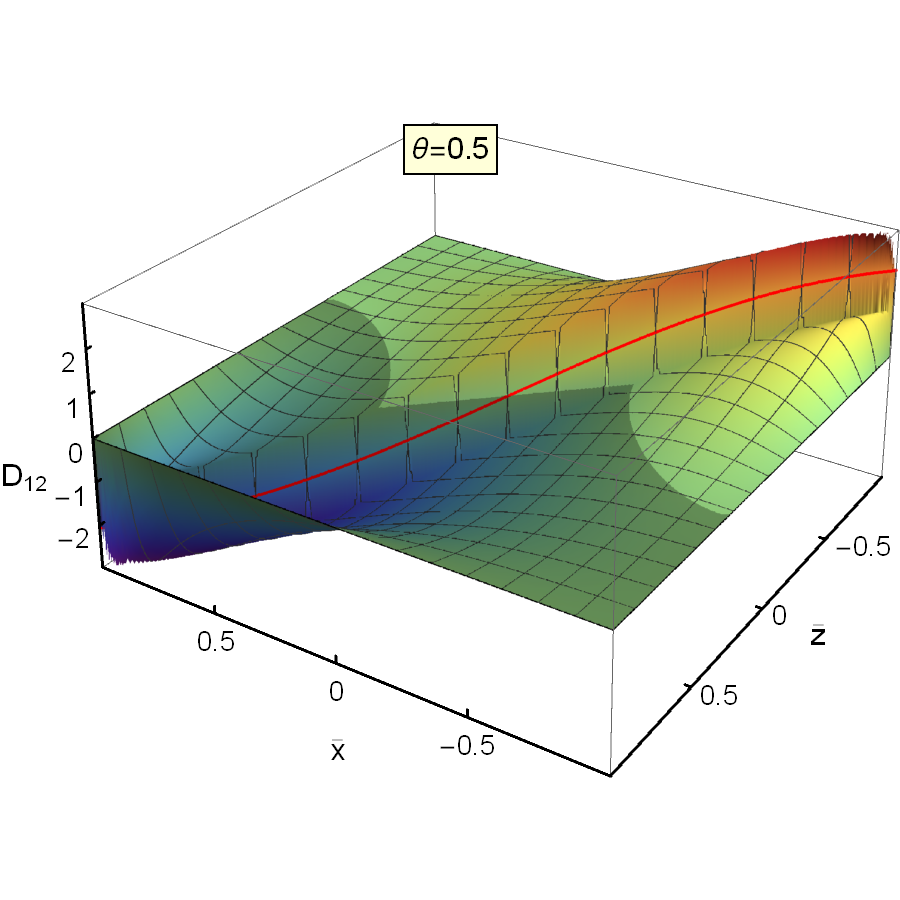}  
\end{center}
\kern -1.5cm
\caption{Numerical computation of $D_{12}=-\bar C_{12}\pi\lambda(2)^2/a_{12}$ vs. $(\bar x,\bar z)$ for dimension $d=1$ and $\theta=0.5$. The matrix $g$ is given by eq.(\ref{gex}). The red curve is $(\bar x,\bar x,D_{12}(\bar x,\bar x))$. The gray plane is the reference $(\bar x,\bar z,0)$. \label{C12fig} }
\end{figure}

The study of the $\bar C_{12}^{(1)}(x,z)$ case needs  the election of a concrete $g$. Just as an example we have chosen the matrix
\begin{equation}
g=\begin{pmatrix}
 -2&-k \\
1&1
 \end{pmatrix}\label{gex}
\end{equation}
with $k\in[2,9/4]$ to guarantee negative eigenvalues. $k$ may be expressed in function of the eigenvalues ratio $\theta=(\lambda(1)/\lambda(2))^{1/2}$, $\vert\lambda(2)\vert>\vert\lambda(1)\vert$:
\begin{equation}
k=\frac{(1+2\theta^2)(2+\theta^2)}{(1+\theta^2)^2}\label{krel}
\end{equation}
where $\theta\in[0,1]$, $\lambda(2)=-1/(1+\theta^2)$. 

We plot in figure \ref{C12fig}  $D_{12}$ for $d=1$.  We see now the discontinuity along $\bar x=\bar z$ inherited from the $F$'s behavior. We also find the  limiting values $D_{12}(L,L)=-3D_{11}(L,L)/2$ and the gap of the discontinuity: $\Delta D_{12}=\vert\lambda(2)\vert\pi/(1+\phi^2)$.

We get the analytic description of $\bar C_{12}^{(1)}$ by using the $L\rightarrow\infty$ version of $F$ given by eq. (\ref{F1d}):
\begin{eqnarray}
D_{12}^{(1)}(x,z;g)&=&\frac{1}{(1+\theta^2)^2(1-\theta^2)}\biggl[\frac{\pi}{2}(1-\theta^2)(1+\text{sign}(x-z))\nonumber\\
&+&2(1+2\theta^2)\arctan\left(\frac{x}{\theta z}\right)-2(2+\theta^2)\arctan\left(\frac{\theta x}{z}\right)
\biggr]\label{C12the}
\end{eqnarray}
where we have used (\ref{krel}). Its has the asymptotic behavior:
\begin{equation}
D_{12}(x,z)\simeq \frac{4}{\theta(1+\theta^2)}\frac{z}{x}\quad, x\rightarrow\infty\quad, z\,\text{given}
\end{equation}

\begin{figure}[h!]
\begin{center}
\includegraphics[height=11cm]{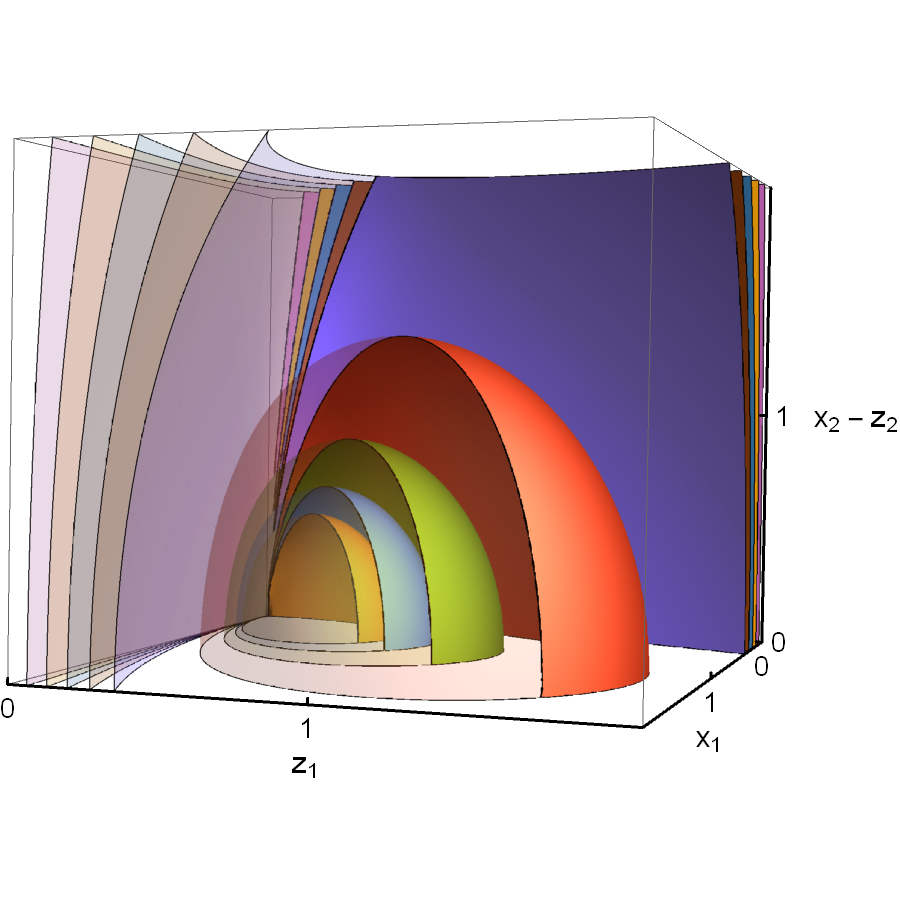}  
\includegraphics[height=7cm]{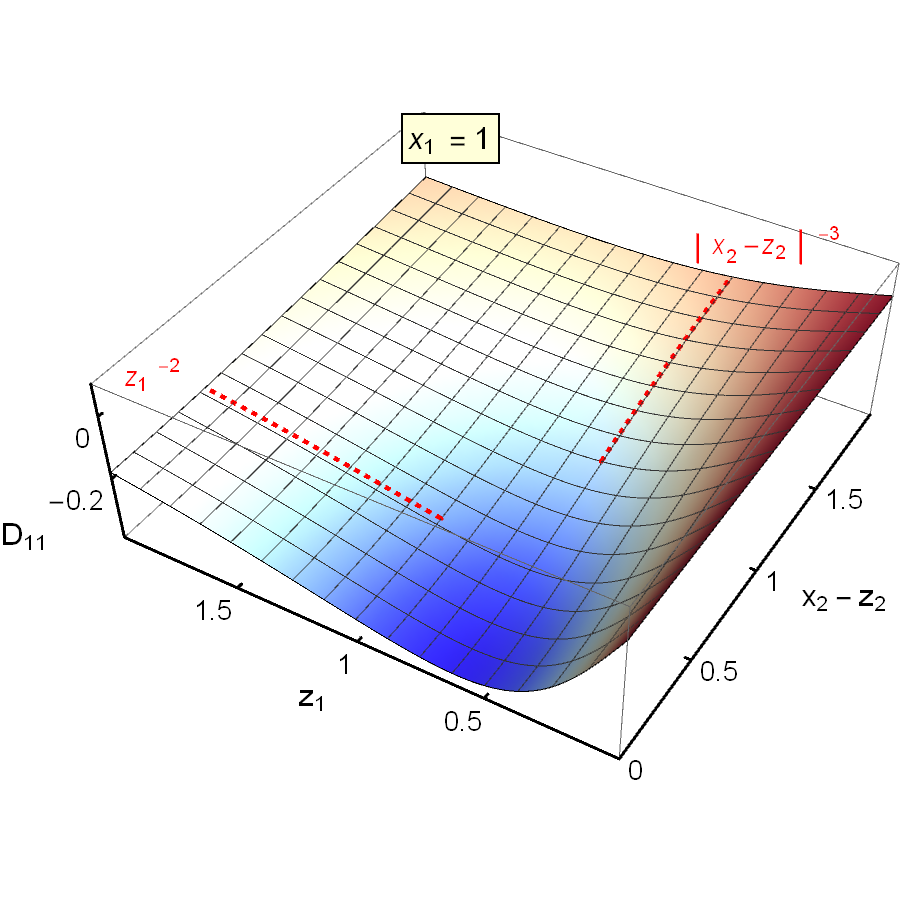} 
\includegraphics[height=7cm]{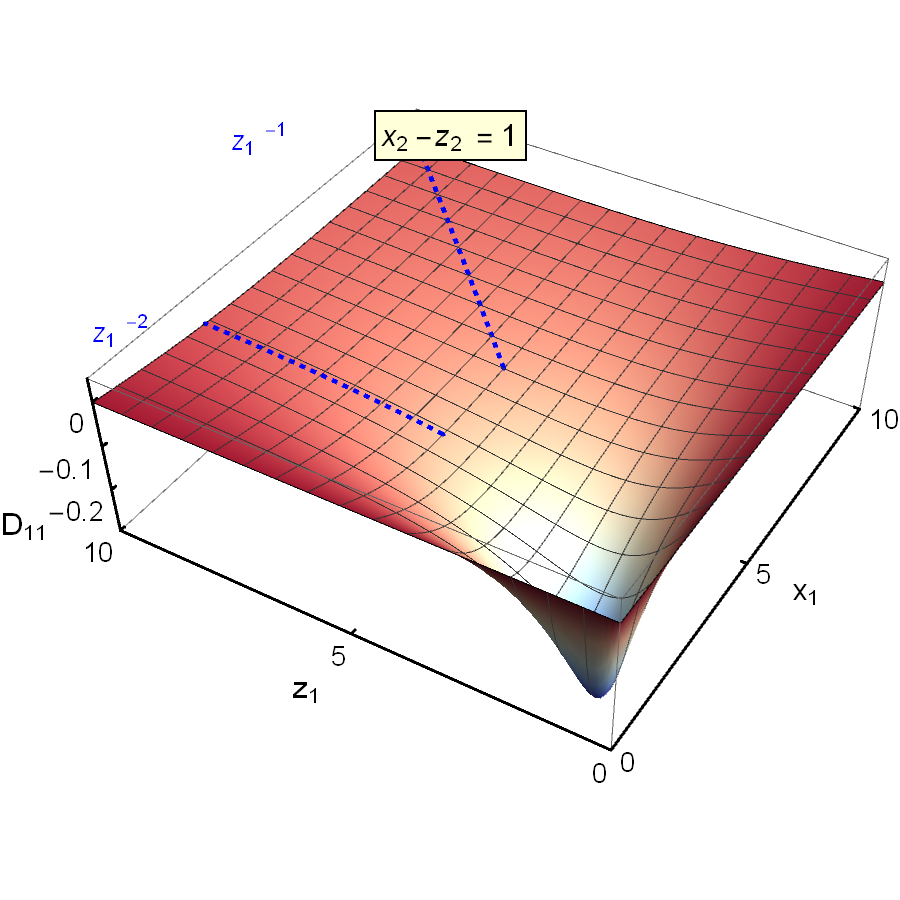} 
\end{center}
\kern -1.5cm
\caption{ $D_{11}(x,z;\theta=0.5)$ from eq.(\ref{C112d}). Top: Three-dimensional countour plot. Each surface correspond to a fix value of $D_{11}$: $-0.5$,$-0.4$,$-0.3$, $-0.2$, $-0.1$, $-0.08$, $-0.06$, $-0.04$ and $-0.02$ from inside to outside. Bottom: Behavior for fix value of $x_1=1$ (left) or fix value of $\vert x_2-z_2\vert=1$ (right). The dashed lines show the asymptotic behavior along the directions (see text).    \label{C112dfig} }
\end{figure}
\subsection{\boldmath$d=2$:}

\begin{figure}[h!]
\begin{center}
\includegraphics[height=6cm]{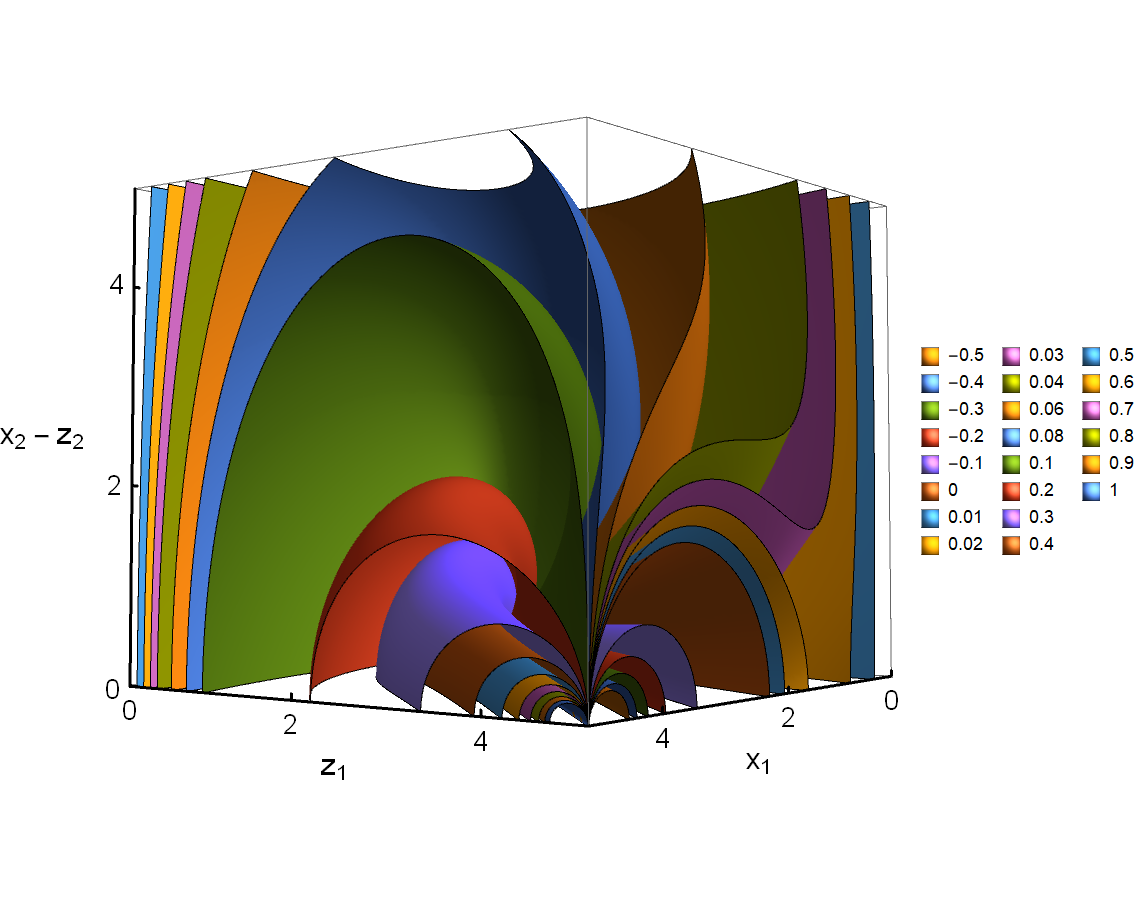}  
\includegraphics[height=6cm]{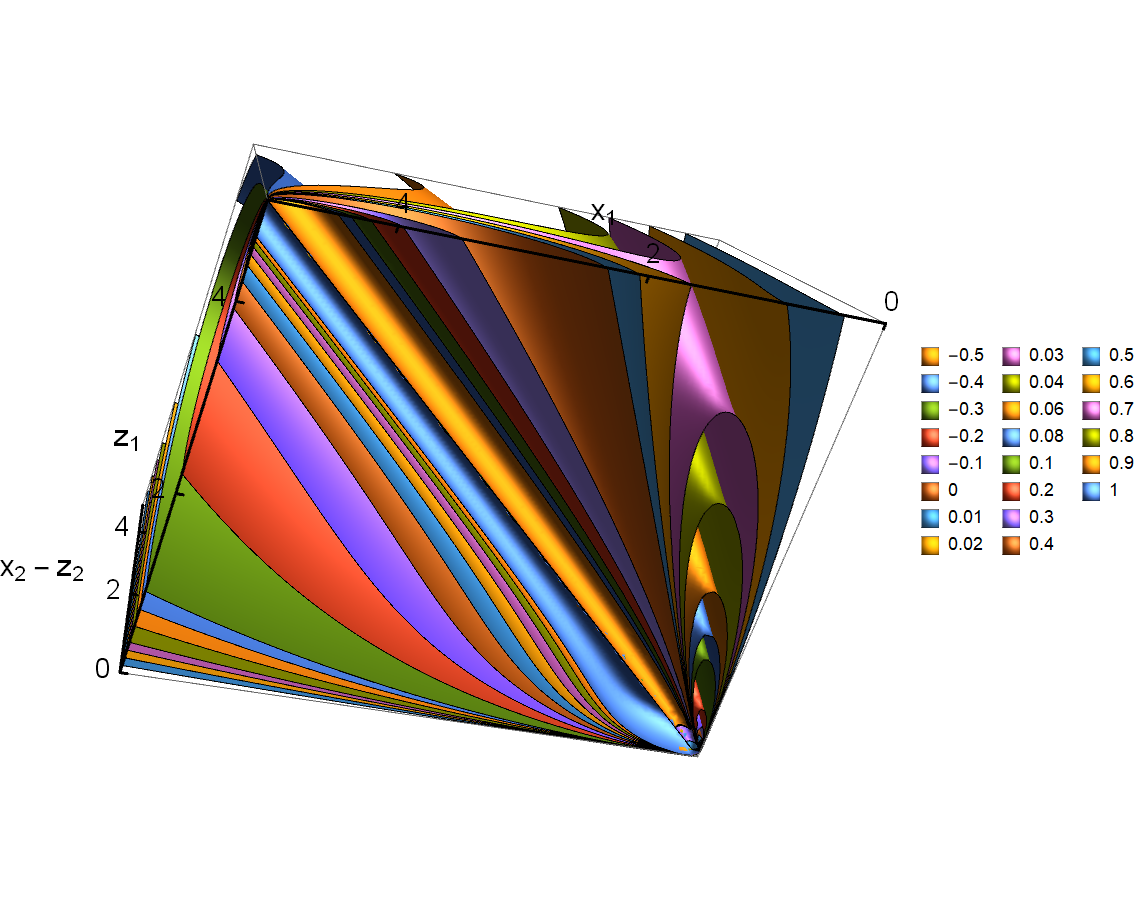}  
\end{center}
\kern -1.5cm
\caption{Three-dimensional countours-plot of $D_{12}(x,z;g)=$  for $\theta=0.5$ from eq.(\ref{C122d}).  Each surface correspond to a fix value of $D_{12}$ (see legends).   \label{C122dfig1} }
\end{figure}
The nonequilibrium correlations are obtained by using eq.(\ref{F2dtot}) into eqs. (\ref{CR1},\ref{CR2},\ref{CR3}). For $D_{11}$ we get:
\begin{eqnarray}
D_{11}^{(1)}(x,z;\theta)&=&-\frac{x_1z_1}{D_1(x,z)}\frac{1}{D_2(x,z;\theta)D_2(z,x;\theta)}\biggl[\frac{(1+\theta^2)^2(x_1^2-z_1^2)^2}{D_2(x,z;\theta)+D_2(z,x;\theta)}\nonumber\\
&+&(x_2-z_2)^2(D_2(x,z;\theta)+D_2(z,x;\theta))\biggr]\label{C112d}
\end{eqnarray}
where $D_1$ and $D_2$ are given by eqs. (\ref{D1D2}).
The asymptotic  behaviors for $D_{11}$ are:
\begin{eqnarray}
D_{11}(x,z)&\simeq&-\frac{1}{\theta(1+\theta)\sqrt{1+\theta^2}}\frac{z_1}{x_1^2}\quad,\,x_1\rightarrow\infty\nonumber\\
&\simeq&-\frac{2}{\theta(1+\theta^2)}\frac{x_1z_1}{\vert x_2-z_2\vert^3}\quad,\,\vert x_2-z_2\vert\rightarrow\infty\nonumber\\
&\simeq&-\frac{1}{2(1+\theta^2)^2}\frac{1}{y}\quad,\, x_1=z_1=y\quad, \, y\rightarrow\infty
\end{eqnarray}
We see three power-law behaviors depending on which direction we take in the space $(x_1,z_1,\vert x_2-z_2\vert)$. If we go to infinity along the transverse direction to the heat flux, the spatial decay is type $\vert x_2-z_2\vert^{-3}$. However, if we move to infinity along the heat flux direction, the correlations decay as $x_1^{-2}$. In both cases, we are assuming that the rest of the coordinates remain fixed. Finally if both coordinates $x_1=z_1=y$ are driven to infinity then, for any value of $x_2-z_2$, the correlations decay as $y^{-1}$. All these three asymptotic behaviors are paradigmatic on the complexity of correlations in these systems. Let us remind that these correlations are the first order of a perturbative expansion around the equilibrium and, thus, it is in some sense the ``simplest'' non-equilibrium case.

\begin{figure}[h!]
\begin{center}
\includegraphics[height=7cm]{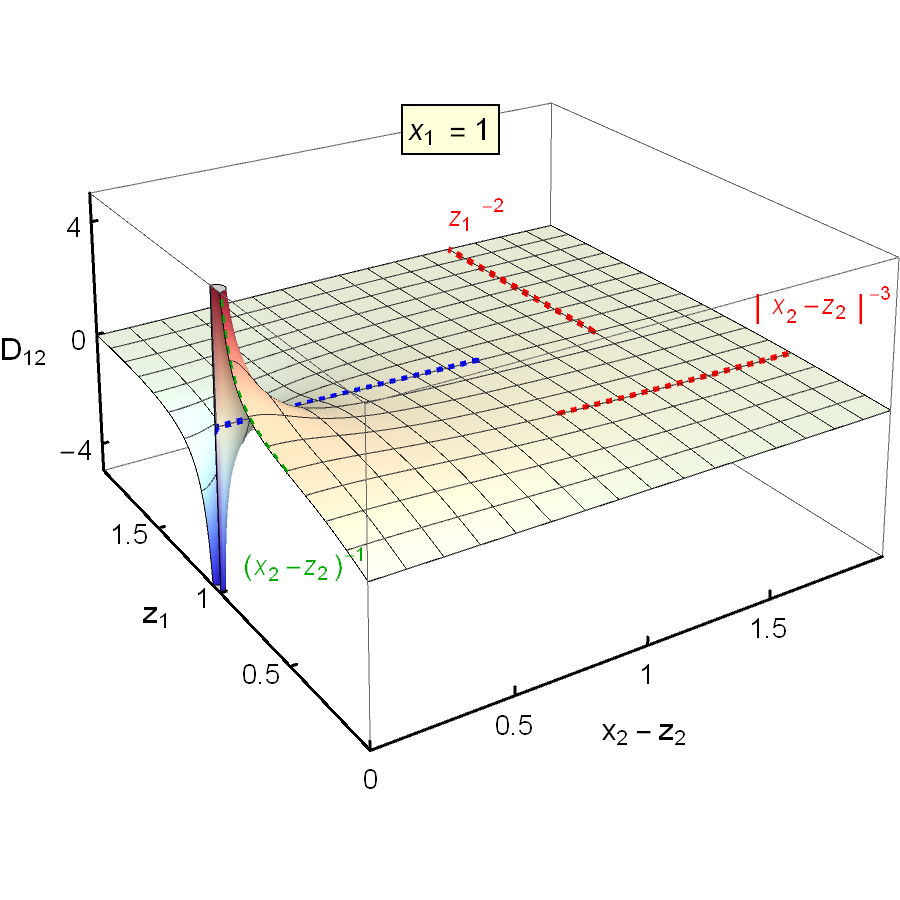} 
\includegraphics[height=7cm]{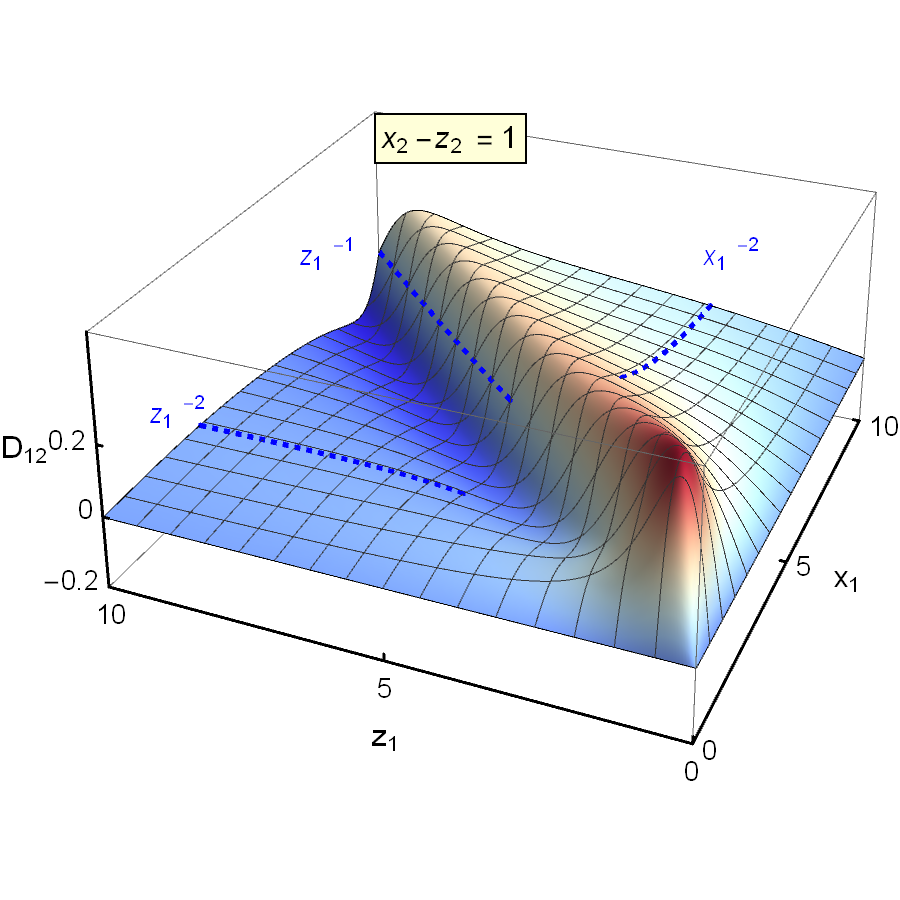} 
\includegraphics[height=7cm]{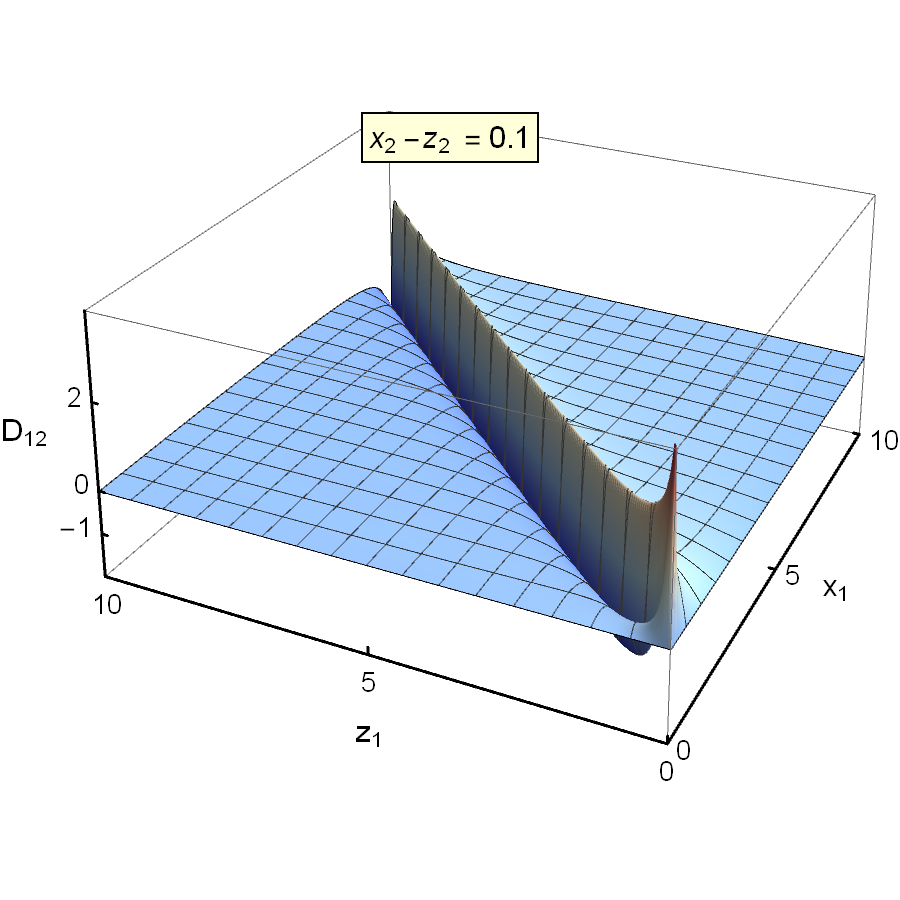}
\end{center}
\kern -1.5cm
\caption{ $D_{12}(x,z;g)$  for $\theta=0.5$ from eq.(\ref{C122d}).  Behavior for fix value of $x_1=1$ (top left), fix value of $\vert x_2-z_2\vert=1,0.1$ (top right and bottom respectively). The dashed lines show the asymptotic behavior along the directions (see text).    \label{C122dfig2} }
\end{figure}
We show in figure \ref{C112dfig} some representations of $D_{11}(x,z;\theta)$ for $\theta=0.5$. The top figure shows three dimensional contours for fixed values of $D_{11}$. We see how the correlations have a singular point at $x_1=z_1=0$. There, depending on the path we reach it, we get different limits values for $D_{11}$. For instance, when $\theta=1$,  $x_2=z_2$, $x_1=R\cos\phi$ and $z_1=R\sin\phi$ we find that $D_{11}$ diverges as  $R^{-1}$ when $R\rightarrow 0$. However, when $\phi=R\phi_0$ $D_{11}$ is finite in such limit.

Finally, in order to study  $\bar C_{12}^{(1)}$ we should fix the matrix $g$ at equilibrium. Again we choose the values given in equation (\ref{gex}) and we get:
\begin{equation}
D_{12}(x,z;g)\equiv-\frac{1}{1-\theta^4}\left[(1+2\theta^2)S(x,z;\theta)+(2+\theta^2)S(z,x;\theta)\right]\label{C122d}
\end{equation}
The limiting behaviors of $\bar C_{12}^{(1)}$ are:
\begin{eqnarray}
D_{12}(x,z)&\simeq&-\frac{2+\theta+2\theta^2}{\theta(1+\theta)(1+\theta^2)^{3/2}}\frac{z_1}{x_1^2}\quad,\,x_1\rightarrow\infty\nonumber\\
&\simeq&\frac{3x_1z_1}{\theta(1+\theta^2)}\frac{1}{\vert x_2-z_2\vert^3}\quad,\,\vert x_2-z_2\vert\rightarrow\infty\nonumber\\
&\simeq&-\frac{3}{4(1+\theta^2)^2}\frac{1}{y}\quad,\, x_1=z_1=y\quad, \, y\rightarrow\infty\nonumber\\
&\simeq&\frac{3}{4(1+\theta^2)^2x_1}\quad,\,\vert x_2-z_2\vert\rightarrow 0\nonumber\\
&\simeq&-\frac{1}{2(1+\theta^2)^2}\frac{1}{z_1-x_1}\quad,\,\vert x_2-z_2\vert=0\,,\,x_1\rightarrow z_1
\end{eqnarray}
We observe a singularity when $x_1\rightarrow z_1$ maintaining $x_2=z_2$. Observe that the  limit $x_2\rightarrow z_2$ with $x_1=z_1$ is a constant. We show in figures \ref{C122dfig1} and \ref{C122dfig2} the global behavior of $D_{12}$ for $\theta=0.5$. We see the variety of decaying behavior depending on the direction in which we do the limit. There is negative correlation whenever $x_1<z_1$ and a non-trivial equipotential structure for positive correlations. Moreover, the stronger correlation values are always near the origin.  

\section{A particle model example}

The continuum model we study in this section is based on an interactive particle system. The particles interact by a short-range potential, for instance, hard-core. Besides,  they suffer random changes in their velocities' direction during their evolution. In this way, there is no momentum conservation. In ref. \cite{Garr} we derived for dimension $d=2$ the macroscopic diffusion equations for this system starting from the Boltzmann equation.  
The system at the mesoscopic scale is characterized by only two conserved local fields: the density, $\phi_1(x,t)$ (number of particles per unit volume) and the energy, $\phi_2(x,t)$ (average kinetic energy per unit volume).
The reference equilibrium state for this model is the ideal gas. This entropy per unit volume is given by the Sakkur-Tetrode expression that for dimension $d=2$ is:
\begin{equation}
s(\phi)=\phi_1\left[\log\left(\frac{2\pi\phi_2}{\phi_1^2}\right) +1\right]\label{entropy}
\end{equation}
where we have considered $h=1$, $k_B=1$ and $m=1$. Therefore,
 the mesoscopic reference equilibrium potential (\ref{Veq}) is:
\begin{equation}
V_{eq}[\phi]=\int_\Lambda dx\,\phi_1(x)\left[2\log\frac{\phi_1(x)}{\phi_{1,eq}}-\log\frac{\phi_2(x)}{\phi_{2,eq}}+\frac{\phi_{1,eq}\phi_2(x)}{\phi_{1}(x)\phi_{2,eq}}+\frac{\phi_{1,eq}}{\phi_1(x)}-2\right]\label{Veq}
\end{equation}
where $(\phi_{1,eq},\phi_{2,eq})$ is the macroscopic equilibrium state. 

 In the BGK approximation \cite{BGK} (where the usual interaction hardcore kernel  is approximated by the local Maxwellian minus the one-particle distribution), we obtained for dimension $d=2$ the diffusion equations (\ref{dif}):
 \begin{equation}
g= \begin{pmatrix}
 0&-1\\
 2\omega\phi_2^2/\phi_1^2&-2(1+\omega)\phi_2/\phi_1
 \end{pmatrix}
 \end{equation}
where $\omega\in[0,1]$. In reference \cite{Garr} appear two parameters: $\nu$ that is related with the collision part in the BGK approximation and $\alpha$ that controls the frequency of the randomization mechanism for the particle velocities. In order to simplify computation we have assumed $\omega=2\pi\alpha/(\nu+2\pi\alpha)$ and $\alpha=1/2\pi$. Observe that the eigenvalues of $g^T$ (that we need for later computations) are: $\lambda(1,2)=-(1+\omega\mp\sqrt{1+\omega^2})\phi_2/\phi_1$.

The matrix $L$ can be obtain by using eq. (\ref{g}) once we know $S$ from (\ref{entropy}):
\begin{equation}
S=\begin{pmatrix}
-2/\phi_1&1/\phi_2\\
1/\phi_2&-\phi_1/\phi_2^2
\end{pmatrix}
\Rightarrow
L=g S^{-1}=
\begin{pmatrix}
\phi_2&2\phi_2^2/\phi_1\\
2\phi_2^2/\phi_1&2(2+\omega)\phi_2^3/\phi_1^2
\end{pmatrix}
\end{equation}

At this point, we have all the ingredients to write down the closed equations for the static two-body correlations once we detail the boundary conditions and we find the corresponding deterministic stationary state $(\phi_1^*(x),\phi_2^*(x))_{x\in\Lambda}$.

Let us assume that our system is in a strip of width unity where we impose temperatures $T_0$, $T_1$ at $x_1=0,L$ respectively. That permits only a flow of energy in the $x$ direction. We also assume that  there is not a net flow of particles through the system and the average value of the density is a given constant. The stationary state is solution of the equations (\ref{stat}) with constant currents:
\begin{equation}
J_{1,i}^D[\phi^*;x]=0\quad,\quad J_{2,i}^D[\phi^*;x]=J\delta_{i,1}\quad , i=1,2
\end{equation}
Both conditions imply:
\begin{equation}
\phi_1^*(x)=\frac{\phi_2^*}{T(x_1)}\quad,\quad T(x_1)=T_0-\Delta T \frac{x_1}{L}\quad,\quad \phi_2^*(x)=\frac{\bar n \Delta T}{\log(T_0/T_1)}\equiv\phi_2^*\label{sstate}
\end{equation}
with $\Delta T=T_0-T_1$ and $\bar n=L^{-1}\int_0^L dx\,\phi_1^*(x)$ that are our system's parameters. Let us remind here that $\phi_1^*(x)$ is the particle density, and $\phi_2^*(x)$ is the energy density. An bidimensional ideal gas at equilibrium has the equation of state $\phi_{2,eq}=\phi_{1,eq} T_{eq}$ where $\phi_{2,eq}$ is the pressure. Then, the system's stationary state has the local equilibrium property, that is, the local pressure is constant all over the system because there is no net flow of particles. 

We know from eq.(\ref{dec}) that correlations are decomposed in the sum of two terms: the local equilibrium contribution, $C_{\alpha\beta}^{LE}(x,y)$ and the correlation's excess, $\bar C_{\alpha\beta} (x,y)$ that is solution of the equation (\ref{exc}). We easily compute $C^{LE}$:
\begin{equation}
C_{\alpha\beta}^{LE}(x,y)=-(S^{-1})_{\alpha\beta}(\phi^*(x))\delta(x-y)
\end{equation}
where
\begin{equation}
S^{-1}(\phi)=-
\begin{pmatrix}
\phi_1&\phi_2\\
\phi_2&2\phi_2^2/\phi_1
\end{pmatrix}
\end{equation}

To compute $\bar C$ we need to do an expansion around the equilibrium. For this models we use $\Delta T/L$ as the $\epsilon$ parameter in the expansions we defined in Section IV. We get all necessary items there: $\vec A_{\alpha\beta}^{(1,2)}$, $g_{\alpha\beta}^{(1)}$, ... by expanding the stationary state (\ref{sstate}) for $\Delta T\simeq 0$ to obtain:
\begin{eqnarray}
h_1^{(1)}(x)&=&\frac{L\bar n}{T_1}\left(\frac{x_1}{L}-\frac{1}{2}\right)\quad ,\quad h_1^{(2)}(x)=\frac{L^2\bar n}{T_1^2}\left[\left(1-\frac{x_1}{L}\right)\left(\frac{1}{2}-\frac{x_1}{L}\right)-\frac{1}{12}\right]\nonumber\\
h_2^{(1)}(x)&=&\frac{L\bar n}{2}\quad ,\quad h_2^{(2)}(x)=-\frac{L^2\bar n}{12 T_1}
\end{eqnarray}

$\bar C_{\alpha\beta}^{(1)}$ was extensively studied in section V. However, we are going to need their explicit  expressions to study the next expansion order: $\bar C_{\alpha\beta}^{(2)}$. One can check that eq.(\ref{hatC1}) implies:
\begin{eqnarray}
\hat C_{11}^{(1)}(n,m;n_\perp)&=&-\frac{4\bar n}{\pi^2T_1}(1-\omega)\left(1-\delta_{n,m}\right)\left(1-(-1)^{n+m}\right)\hat D_{11}(n,m;n_\perp)\nonumber\\
\hat C_{12}^{(1)}(n,m;n_\perp)&=&-\frac{8\bar n}{\pi^2}(1-\omega^2)\left(1-\delta_{n,m}\right)\left(1-(-1)^{n+m}\right)\hat D_{12}(n,m;n_\perp)\nonumber\\
\hat C_{22}^{(1)}(n,m;n_\perp)&=&2\omega T_1^2\hat C_{11}^{(1)}(n,m;n_\perp)
\end{eqnarray}
where
\begin{eqnarray}
\hat D_{11}(n,m;n_\perp)&=&\frac{nm}{D(n,m;n_\perp)}\nonumber\\
\hat D_{12}(n,m;n_\perp)&=&\frac{nm}{D(n,m;n_\perp)}\frac{n^2+4n_\perp^2}{n^2-m^2}
\end{eqnarray}
and 
\begin{equation}
D(n,m;n_\perp)=\omega (n^2-m^2)^2+(1+\omega)^2(n^2+4n_\perp^2)(m^2+4 n_\perp^2)\label{DDDel}
\end{equation}

The Sinus-Fourier transform on eq.(\ref{C2}) to get a set of four linear equations whose unknowns are $\hat C_{\alpha\beta}^{(2)}$. After solving the system we get:
\begin{eqnarray}
\hat C_{11}^{(2)}&&(n,m;n_\perp)=\frac{L^2}{2\pi^2T_1^2\omega}\frac{1}{D(n,m;n_\perp)(n^2+m^2+8n_\perp^2)}\nonumber\\
&&\biggl[\frac{2}{T_1}(1+\omega)(m^2+4n_\perp^2)(n^2+4n_\perp^2)\mathcal{R}_2(n,m;n_\perp)\nonumber\\
&&+(m^2+4n_\perp^2)\left(2(1+\omega)^2(n^2+4n_\perp^2)-\omega(n^2-m^2)\right)\mathcal{R}_1(n,m;n_\perp)\nonumber\\
&&+(n^2+4n_\perp^2)\left(2(1+\omega)^2(m^2+4n_\perp^2)+\omega(n^2-m^2)\right)\mathcal{R}_1(m,n;n_\perp)\biggr]\nonumber\\
\hat C_{12}^{(2)}&&(n,m;n_\perp)=\frac{L^2}{\pi^2T_1}\frac{n^2+4n_\perp^2}{D(n,m;n_\perp)(n^2+m^2+8n_\perp^2)}\biggl[-\frac{1}{T_1}(n^2-m^2)\mathcal{R}_2(n,m;n_\perp)\nonumber\\
&&+(1+\omega)(m^2+4n_\perp^2)\mathcal{R}_1(m,n;n_\perp)-(1+\omega)(n^2+4n_\perp^2)\mathcal{R}_1(n,m;n_\perp)\biggr]\nonumber\\
\hat C_{21}^{(2)}&&(n,m;n_\perp)=-\frac{m^2+4n_\perp^2}{n^2+4n_\perp^2}\,\hat C_{12}^{(2)}(n,m;n_\perp)\nonumber\\
\hat C_{22}^{(2)}&&(n,m;n_\perp)=\frac{L^2}{\pi^2}\frac{1}{D(n,m;n_\perp)(n^2+m^2+8n_\perp^2)}\biggl[\nonumber\\
&&\frac{2}{T_1}(1+\omega)(m^2+4n_\perp^2)(n^2+4n_\perp^2)\mathcal{R}_2(n,m;n_\perp)\nonumber\\
&&+\omega (n^2-m^2)\left((m^2+4n_\perp^2)\mathcal{R}_1(m,n;n_\perp)-(n^2+4n_\perp^2)\mathcal{R}_1(n,m;n_\perp)\right)\biggr]
\end{eqnarray}
where
\begin{eqnarray}
\mathcal{R}_1&&(n,m;n_\perp)=\frac{4\bar n}{L}(1-\omega)nm\biggl[\omega(1-\delta_{n,m})\left(1-(-1)^{n+m}\right)\frac{n^2+m^2+8n_\perp^2}{D(n,m;n_\perp)}\nonumber\\
&&+\frac{16}{\pi^2}\omega\left(1+(-1)^{n+m}\right)(m^2+4n_\perp^2)\sum_{m'\neq m}\left(1-(-1)^{m'+m}\right)\frac{m'^2}{D(n,m';n_\perp)(m^2-m'^2)^2}\nonumber\\
&&+\frac{8}{\pi^2}(1+\omega)\left(1+(-1)^{n+m}\right)(n^2+4n_\perp^2)\sum_{m'\neq m}\left(1-(-1)^{m'+m}\right)\frac{m'^2}{D(n,m';n_\perp)(m^2-m'^2)^2}\nonumber\\
&&\frac{1}{n^2-m'^2}\left((1+3\omega)(m'^2-m^2)-2(1+\omega)(m'^2+4n_\perp^2)\right)
\biggr]\label{RR1}
\end{eqnarray}
\begin{eqnarray}
\mathcal{R}_2(n,m;n_\perp)&=&\frac{4\bar n T_1}{L}(1+2\omega)\delta_{n,m}-\frac{4\bar n T_1}{L}\omega(1-\omega^2)(1-\delta_{n,m})\left(1-(-1)^{n+m}\right)nm\frac{n^2+m^2+8n_\perp^2}{D(n,m;n_\perp)}\nonumber\\
&+&\frac{64\bar n T_1}{\pi^2L}\omega(1-\omega^2)\left(1+(-1)^{n+m}\right)nm\sum_{n'\neq n}\left(1-(-1)^{n'+n}\right)\frac{n'^2(n'^2+4n_\perp^2)}{(n^2-n'^2)(m^2-n'^2)}\nonumber\\
&&\left(\frac{m^2+4n_\perp^2}{D(n,n';n_\perp)(m^2-n'^2)}+\frac{n^2+4n_\perp^2}{D(n',m;n_\perp)(n^2-n'^2)} \right)\nonumber\\
&+&\frac{16\bar n T_1}{\pi^2L}\omega(1+3\omega)(1-\omega)\left(1+(-1)^{n+m}\right)nm\sum_{n'\neq n}\left(1-(-1)^{n'+n}\right)n'^2\nonumber\\
&&\left(\frac{1}{D(n,n';n_\perp)(m^2-n'^2)}+\frac{1}{D(n',m;n_\perp)(n^2-n'^2)} \right)\nonumber\\
&+&\frac{32\bar n T_1}{\pi^2L}\omega(1-\omega^2)\left(1+(-1)^{n+m}\right)nm\sum_{n'\neq n}\left(1-(-1)^{n'+n}\right)n'^2(n'^2+4n_\perp^2)\nonumber\\
&&\left(\frac{1}{D(n,n';n_\perp)(m^2-n'^2)^2}+\frac{1}{D(n',m;n_\perp)(n^2-n'^2)^2} \right)\label{RR2}
\end{eqnarray}

\begin{figure}[h!]
\begin{center}
\includegraphics[height=8cm]{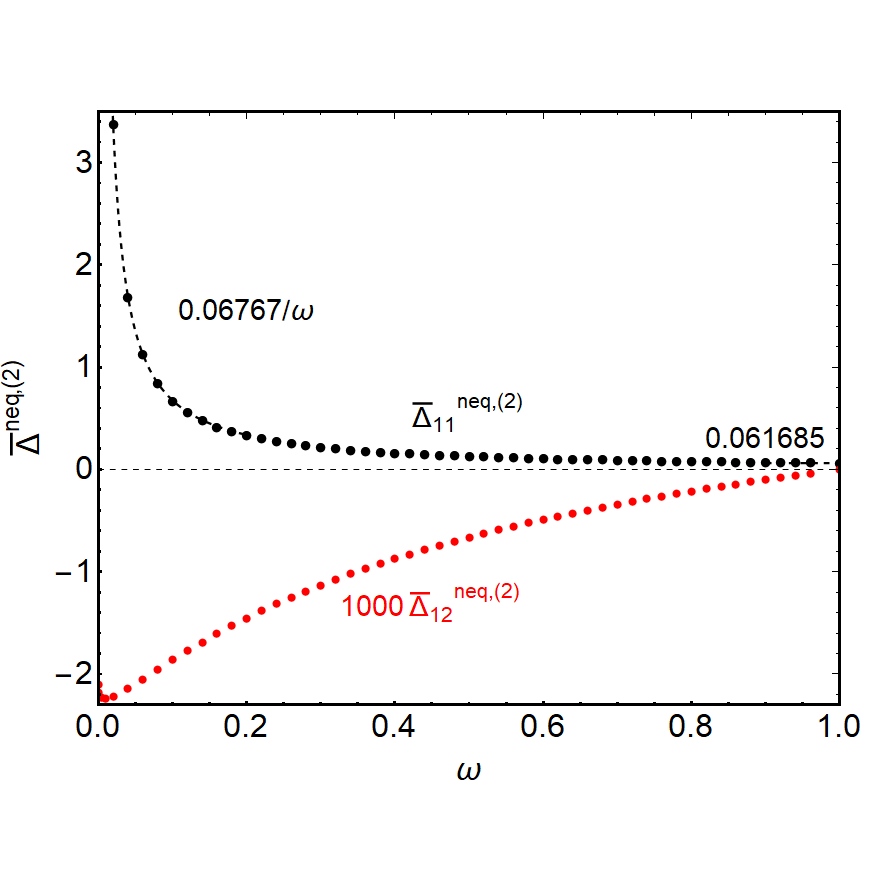} 
\end{center}
\kern -1.5cm
\caption{Second order fluctuations for the hydrodynamic two-field model computed using eq.(\ref{fluct2}). Black dots:  $\bar\Delta_{11}^{neq,(2)}(\omega)=\Delta_{11}^{neq,(2)}(\omega)T_1^2/\bar n$. Red dots: $\bar\Delta_{12}^{neq,(2)}(\omega)=\Delta_{12}^{neq,(2)}(\omega)T_1/\bar n$.  The black dashed line is the analytic asymptotic behavior for small values of $\omega$. $\bar\Delta_{11}^{neq,(2)}(1)=0.061685$ ,   $\bar\Delta_{12}^{neq,(2)}(1)=0$ and  $\bar\Delta_{12}^{neq,(2)}(0)=-0.00210306$\label{figura} }
\end{figure}

We are interested in using these solution for $\hat C_{\alpha\beta}^{(2)}$ to compute the fluctuations of the field's spatial average (\ref{fluct}) at this order:
\begin{equation}
\Delta_{\alpha\beta}^{neq,(2)}=\frac{1}{\pi^2 L}\sum_{n=1}^\infty\sum_{m=1}^\infty\frac{1}{nm}\left(1-(-1)^{n}\right)\left(1-(-1)^{m}\right)\hat C_{\alpha\beta}^{(2)}(n,m;n_\perp=0)\label{fluct2}
\end{equation}
Observe that to compute $\Delta_{\alpha\beta}$ all we need  is $\hat C_{\alpha\beta}(n,m;n_\perp)$ for $n_\perp=0$ and $n$, $m$ odd values. Therefore, some sums appearing in eqs. (\ref{RR1}) and (\ref{RR2})  just dissapear and others should be done on even values. That permits us to obtain explicitly expressions for all of them:
\begin{eqnarray} 
\hat C_{11}^{(2)}(n,m;0)&=&\frac{4L\bar n}{\pi^2T_1^2\omega}\frac{nm}{D(n,m;0)(n^2+m^2)}\biggl[(1+\omega)(1+2\omega)n^2\delta_{n,m}-\frac{16}{\pi^2}(1-\omega)A_{11}(n,m)\biggr]\nonumber\\
\hat C_{12}^{(2)}(n,m;0)&=&\frac{64 L\bar n}{\pi^4T_1}\frac{n^3m}{D(n,m;0)(n^2+m^2)}A_{12}(n,m)
\end{eqnarray}
 and
\begin{eqnarray}
A_{11}(n,m)&=&\sum_{l=0}^2\left[\alpha_{11}^{(l)}(n,m)B_l(n,m)+\alpha_{11}^{(l)}(m,n)B_l(m,n)\right]\nonumber\\
A_{12}(n,m)&=&\sum_{l=0}^2\left[\alpha_{12}^{(l)}(n,m)B_l(n,m)-\alpha_{12}^{(l)}(m,n)B_l(m,n)\right]
\end{eqnarray}
with
\begin{eqnarray}
\alpha_{11}^{(0)}(n,m)&=&m^4n^2\left(2(1+3\omega+5\omega^2+5\omega^3+3\omega^4)n^2+\omega(1+2\omega+3\omega^2)m^2\right)\nonumber\\
\alpha_{11}^{(1)}(n,m)&=&2m^2\left(\omega(1+\omega+\omega^2)n^2m^2+\omega^2m^4+(1-\omega^2)(1+\omega+\omega^2)n^4\right)\nonumber\\
\alpha_{11}^{(2)}(n,m)&=&\omega(1-\omega^2)n^2m^2
\end{eqnarray}
\begin{eqnarray}
\alpha_{12}^{(0)}&=&\omega(1-\omega)(n^2-m^2)\nonumber\\
\alpha_{12}^{(1)}&=&(1-\omega)(1+\omega+\omega^2)n^4+\omega(3+\omega)m^4\nonumber\\
\alpha_{12}^{(2)}&=&(1+2\omega+2\omega^2+3\omega^3)n^2+\omega(1+3\omega)m^2
\end{eqnarray}
where the sums
\begin{equation}
B_l(n,m)=\sum_{k=0}^{\infty}\frac{(2k)^{2+2l}}{D(n,2k;0)}\frac{1}{((2k)^2-m^2)^2(n^2-(2k)^2)}\label{sumDel}
\end{equation}
are explicitly done and they are given in Appendix IV. With all these ingredients we can compute $\Delta_{\alpha\beta}^{neq,(2)}$ given by (\ref{fluct2}). The analytic expressions are long functions of $\omega$ that we do not explicitly write here. We have plotted  their behavior in figure \ref{figura}.
We see that $\bar\Delta_{11}^{neq,(2)}(\omega)=\Delta_{11}^{neq,(2)}(\omega)T_1^2/\bar n$ behaves effectively like a $1/\omega$ function. In fact, it asymptotic behavior for small values of $\omega$ is $\simeq 0.06767/\omega$.  This singularity for $\omega=0$ is expected. When $w=0$ the microscopic model loses the velocity randomization mechanisms and its mesoscopic description changes dramatically because the momentum is locally conserved and that adds two more conserved fields. Therefore, the model with two fields breaks down.
By other hand $\bar\Delta_{11}^{neq,(2)}(1)=0.061685$. Please, observe that this value differs by only by $0.006$ the extrapolation to one of the asymptotic expression around $\omega=0$. That is, the analytic complex and long formula for $\bar\Delta_{11}^{neq,(2)}(\omega)$ only accounts for a tiny correction of the asymptotic formula. 
$\bar\Delta_{12}^{neq,(2)}(\omega)=\Delta_{12}^{neq,(2)}(\omega)T_1/\bar n$ is $10^3$ times smaller than $\bar\Delta_{11}^{neq,(2)}(\omega)$ and negative. It has a finite limit for $\omega=0$ and a minimum near it. This fluctuation does not reflects the change on the mesoscopic description when $\omega\rightarrow 0$.

\section{Conclusions}

We have studied the two body equal time correlation functions for a diffusive sytem with a reference equilibrium state (DSe) with $M$-fields. We have derived the partial differential equations they follow and studied explicitly their solutions perturbatively around the equilibrium. We show the correlation's complex and rich behavior: generic power laws that depends on the path we follow when doing the long distance limit. We show that the DSe correlations have two levels. The first level is the {\it basic correlation function} that is generic and it doesn't depend on the specific model and just on the eigenvalues's ratio of the matrix transpost  $g$ that defines the deterministic dynamics. It contains the basic spatial structure of the model. The second level is the linear combination of such basic correlation function to build the correlations. That combination strongly depends on the model's details and on the form of the stationary state. Therefore, it seems interesting to define models that focus only on the basic correlation functions in order to study the main generic properties of those non-equilibrium systems.

\section{Acknowledgements}

This work is part of the Project of I+D+i Ref. PID2020-113681GB-I00, financed by 
MICIN/AEI/10.13039/501100011033 and FEDER “A way to make Europe”.

\section*{Appendix I: Mesoscopic probability distribution of equilibrium}

Any system at equilibrium is completely determined by a small set of macroscopic variables. At mesoscopic level, such observables fluctuate around its equilibrium values.  In order to obtain the corresponding probability distribution we  can follow two equivalent strategies: the grand canonical ensemble or the Einstein fluctuation theory\cite{Einstein}. In this appendix we apply both paths to the case of a system with an equilibrium state defined by the number of particles, $N$, the volume, $\Omega$, and the energy, $E$.  

\subsection*{Grand Canonical Ensemble:}
In the Grand Canonical Ensemble the equilibrium state is determined by $(T,\mu,\Omega)$ (Temperature, Chemical potential and Volume respectively). We know that these set of variables are related with $(n_{eq},e_{eq})$ (Particle density and Energy per particle respectively) by
\begin{equation}
\beta=\frac{\partial s(n,e)}{\partial e}\biggr\vert_{eq}\quad, \quad \beta\mu=\beta e_{eq}-s(n_{eq},e_{eq})-n_{eq}\frac{\partial s(n,e)}{\partial n}\biggr\vert_{eq}
\end{equation}
where $s(n,e)$ is the entropy per particle and $\beta=1/T$.   

The probability to find the system with a given energy per particle, $e=E/N$ and density, $n=N/\Omega$, is given by
\begin{equation}
P(n,e)=\Xi^{-1}\sum_{N=1}^\infty e^{\beta\mu N}\int_\Lambda dx_N\int_{R^{Nd}}dp_N e^{-\beta H(x_N,p_N)}\delta\left(e-\frac{H(x_N,p_N)}{N}\right)\delta\left(n-\frac{N}{\Omega}\right)
\end{equation}
This relation simplifies when $\Omega\rightarrow\infty$
\begin{equation}
P(n,e)\simeq \exp\left[-\Omega(f(n_{eq},e_{eq})-f(n,e))\right]
\end{equation}
with
 \begin{equation}
 f(n,e)=n\left[\beta\mu-\beta e+s(n,e)\right]
 \end{equation}
that can be written
\begin{equation}
P(n,e)\simeq\exp\left[-\Omega V_{eq}(n,e)\right]\label{one}
\end{equation}
\begin{equation}
V_{eq}(n,e)=n\left[s(n_{eq},e_{eq})-s(n,e)-(e_{eq}-e)\frac{\partial s(n,e)}{\partial e}\biggr\vert_{eq}+\frac{n_{eq}}{n}(n-n_{eq})\frac{\partial s(n,e)}{\partial n}\biggr\vert_{eq}\right]
\end{equation}

Observe that $V_{eq}(n_{eq},e_{eq})=0$, $\partial V_{eq}(n,e)/\partial n\vert_eq=0$ and $\partial V_{eq}(n,e)/\partial e\vert_eq=0$.

\subsection*{Einstein Fluctuation Theory:}

Boltzmann proposed that the entropy of a system at an equilibrium state defined by the macroscopic variables $(A,B)$  is related with the number of compatible microstates, $\omega(A,B)$:
\begin{equation}
S(A,B)=\log w(A,B)
\end{equation}
where we have assumed $k_B=1$. When we relax the constrain that fixed $A$, the system spontaneously evolves to a new equilibrium state defined by only the observable $B$.  Then
\begin{equation}
w(B)=\sum_A w(A,B)
\end{equation}
where the sum runs over all the possible values of the observable $A$. Therefore:
\begin{equation}
e^{S(B)}=\sum_A e^{S(A,B)}
\end{equation}
We consider that $S$ is a extensive variable proportional to a volume and therefore the sum is dominated by the term of the sum that maximalizes $S(A,B)$, that is
\begin{equation}
S(B)=S(A^*,B)\quad,\quad \frac{\partial S(A,B)}{\partial A}\biggr\vert_{A=A^*}=0
\end{equation} 
Observe that $w(B)\simeq w(A^*,B)$ and the state $A^*$ is the one with larger number of microstates compared with any other value: $w(A^*,B)\geq w(A,B)$. That is, the number of microstates compatible with  the equilibrium state $B$ is overwhelm large compared to the number of microstates associated to any state $(A,B)$, $w(A,B)$ when $A$ is a macroscopic deviation from $A^*$. 

Einstein, following Boltzmann's line of reasoning, proposed that the probability that a system at equilibrium in a state $B$ to be in a fluctuating macrostate $A$ should be the ratio between the number of microstates compatibles with $A$ and the total number of microstates compatible with $B$:
\begin{equation}
P(A\vert B)=\frac{\omega(A,B)}{\omega(B)}\simeq e^{S(A,B)-S(A^*,B)}
\end{equation}
We can apply this idea to our example. The equilibrium state of a closed system $U$ is defined by the macrovariables $(N,V,E)$ (number of particles, volume and energy respectively). Let us divide the system in two disjoint subsystems $1$ and $2$. Let us assume that we have a set of constraints that can fix the equilibrium state at $1$: $(N_1,V_1,E_1)$. That fixes the equilibrium state at $2$: $(N_2,V_2,E_2)=(N-N_1,V-V_1,E-E_1)$. Therefore, the total entropy of $U$ is just the sum of the entropies of both subsystems. Therefore the $U$'s entropy per particle  in this constrained system is:
\begin{equation}
s(n_1,e_1;n,e,\alpha)=\alpha \frac{n_1}{n}s(n_1,e_1)+\left(1-\alpha\frac{n_1}{n}\right)s(n_2,e_2)\label{e1}
\end{equation}
where $n=N/V$, $e=E/N$, $\alpha=V_1/V$, $n_1=N_1/V_1$, $e_1=E_1/N_1$ and
\begin{equation}
n_2=\frac{n-\alpha n_1}{1-\alpha}\quad,\quad e_2=\frac{n e-\alpha n_1e_1}{n-\alpha n_1}
\end{equation}
We have assumed that the entropy is extensive: $S(N,V,E)=N s(n,e)$ for $N$ large enough. We see that the equilibrium state of the constrained system is defined by five variables: $(n_1,e_1;n,e,\alpha)$. Let us release the constrains over $n_1$ and $e_1$ while keeping fixed $n$, $e$ and $\alpha$. It can be checked that $n_1^*=n$ and $e_1^*=e$ are the values that make maximum the entropy (\ref{e1}). We can apply now the Einstein theory of fluctuations. The probability to observe the subsystem $1$ with values $(n_1,e_1$, while $U$ is at equilibrium state $(n,e)$ is
\begin{equation}
P(n_1,e_1;n,e,\alpha)=\exp\left[N(s(n_1,e_1;n,e,\alpha)-s(n,e))\right]
\end{equation}
It is straightforward to see that  in the limit $\alpha\rightarrow 0$ we get the same result we got using the grand canonical ensemble. 

\section*{Appendix II: Structure of the Deterministic Stationary fields for DSe Systems}

The deterministic stationary fields for DSe systems, $\phi^*$ are solutions of
\begin{equation}
\nabla_x\left(\sum_\beta g_{\alpha\beta}(\phi^*)\nabla_x\phi_\beta^*\right)=0\label{sol1}
\end{equation}
We look for the conditions to have  $\phi^*$ being solutions of 
\begin{equation}
\vec{j}_\alpha=\sum_\beta g_{\alpha\beta}(\phi^*)\nabla_x\phi_\beta^*\label{sol2}
\end{equation}
with $j$'s being constant vectors and of, at least,  $C^2$ type. Of course all $\phi^*$ solution of eq.(\ref{sol2}) is solution of eq. (\ref{sol1}) but that is not always true in the reverse case. In physics, $\vec{j}$ are the stationary currents and they contain implicit information of the form of the boundary conditions.  We focus in asking that  the cross derivatives of $\phi^*$ to be equal  ($\partial_{ij}^2\phi_\alpha^*=\partial_{ji}^2\phi_\alpha^*$) to guarantee continuity and $C^2$  differentiability. In order to study this property we invert eq.(\ref{sol2}):
\begin{equation}
\nabla_x\phi_\beta^*=\sum_\sigma\left(g^{-1}\right)_{\beta\sigma}\vec{j}_{\sigma}
\end{equation}
and by  doing the cross derivatives we get the differentiability condition:
\begin{equation}
\sum_\alpha\sum_\sigma j_{\alpha,j}j_{\sigma,i}\sum_{\gamma}\left(\frac{\partial (g^{-1})_{\beta\sigma}}{\partial\phi_{\gamma}}(g^{-1})_{\gamma\alpha}-\frac{\partial (g^{-1})_{\beta\alpha}}{\partial\phi_{\gamma}}(g^{-1})_{\gamma\sigma}\right)=0 \label{diff}
\end{equation}
We find several cases that accomplish eq.(\ref{diff}) :
\begin{itemize}
\item{(a)} $j_{\alpha,j}j_{\sigma,i}=0\quad\forall\, i\neq j$, or equivalently $j_{\alpha,i}=q_\alpha\delta_{i,k}$ for a given $k$-direccion. That is, all the currents should follow the same vector direction. That's the case when, for instance, the boundaries are two hyperplanes of $d-1$ dimensions placed one in front of the other and with homogeneous values for the fields at the boundaries.
\item{(b)} {\it Conditions on the system:} (i) $L$ is a constant matrix, (ii) $L=\tilde s(\phi) A$ where $A$ is a constant matrix and $\tilde s$ is the entropy of the reference equilibrium state, and (iii) $M=1$.
\end{itemize}

We study in this paper systems with boundary conditions as described in the (a) case. Let's take $x\in[0,L]$ as the axis perpendicular to the boundary hyper planes. Equation (\ref{sol2}) is then written:
\begin{equation}
j_\alpha=\sum_\beta g_{\alpha\beta}(\phi^*)\frac{d\phi_\beta^*(x)}{dx}\label{sol7}
\end{equation}
and we take the boundary conditions:$\phi_\alpha^*(0)=\phi_{eq}$ and $\phi_\alpha^*(L)=\phi_{eq}+\Delta\phi_\alpha$ where $\Delta\phi_\alpha$ are given constants. $j_\alpha$ are constants that are determined by the boundary conditions. We noa apply a perturbative expansion around the equilibrium solution: 
\begin{eqnarray}
\phi_\alpha^*(x)&=&\phi_{eq}+\epsilon h_{1,\alpha}(x)+\epsilon^2 h_{2,\alpha}(x)+\ldots \nonumber\\
j_\alpha&=&\epsilon j_{1,\alpha}+\epsilon^2 j_{2,\alpha}+\ldots\label{exp}
\end{eqnarray}
where $\epsilon$ is related to the distance to de equilibrium. The original boundary conditions are translated to the $h$'s functions:
\begin{equation}
h_{1,\alpha}(0)=h_{2,\alpha}(0)=0\quad,\quad \epsilon h_{1,\alpha}(L)=\Delta\phi_\alpha\quad,\quad h_{2,\alpha}(L)=0
\end{equation}
The $\epsilon$ expansion of eq.(sol7) gives the set of equations:
\begin{eqnarray}
j_{1,\alpha}&=&\sum_\beta g_{\alpha\beta}(\phi_{eq})\frac{d h_{1,\beta}(x)}{dx}\nonumber\\
j_{2,\alpha}&=&\sum_{\beta}\left[g_{\alpha\beta}(\phi_{eq})\frac{d h_{2,\beta}(x)}{dx}+\sum_{\gamma}\frac{\partial g_{\alpha\beta}}{\partial\phi_{\gamma}}\biggr\vert_{\phi=\phi_{eq}}h_{1,\gamma}(x)\frac{dh_{1,\beta}(x)}{dx}\right]\nonumber\\
&&\ldots
\end{eqnarray}
that can be solved order by order. The solutions for $h_{1,\alpha}$ and $h_{2,\alpha}(x)$  are:
\begin{equation}
\epsilon h_{1,\alpha}(x)=\frac{\Delta\phi_{\alpha}}{L}x\quad,\quad \epsilon j_{1,\alpha}=\sum_{\beta}g_{\alpha\beta}(\phi_{eq})\frac{\Delta\phi_{\beta}}{L}
\end{equation}
\begin{equation}
\epsilon^2 h_{2,\alpha}(x)=\left(\sum_{\beta}(g^{-1})_{\alpha\beta}\epsilon^2j_{2,\beta}\right)x\left(1-\frac{x}{L}\right)\, , \, \epsilon^2 j_{2,\alpha}=\frac{1}{L}\sum_{\sigma}\sum_{\gamma}\frac{\partial g_{\alpha\sigma}}{\partial\phi_{\gamma}}\biggr\vert_{\phi=\phi_{eq}}\Delta\phi_\sigma\Delta\phi_\gamma
\end{equation}
where $\Delta\phi_\alpha$ are of order $\epsilon$.

\section*{Appendix III: Correlations for systems with some strictly conserved fields}

Let us assume that the system's dynamics have a stochastic evolution that locally conserves the fields as our diffusive systems defined in the main text. Boundary conditions may or not break such conservation law. For instance, open boundary conditions introduce fluctuations on the average field that periodic boundary condition does not.
Moreover, for systems with $M>1$ fields, some of them may be strictly conserved while others are not. For example, think of a particle system where we permit open energy exchanges with the boundaries. Still, we fix the total number of particles, or the density field's average is constant during the system's evolution. This difference affects the form of the correlation functions.

 Let $P_{st}^{OB}[\phi]\simeq \exp[-\Omega V_0[\phi]]$ be the stationary distribution when $\Omega>>1$ with a set of open-boundary conditions imposed on the system for all the $M$-fields  ($OB$ stands for open boundaries). That is, it is solution of the Fokker-Planck equation (\ref{ME}) in such limit. It can be checked that the restricted distribution
\begin{equation}
P_{st}^{SC}[\phi]\simeq e^{-\Omega V_0[\phi]}\prod_{\alpha\in\tilde M}\delta\left(\int_\Lambda dx \,\phi_\alpha(x)-\vert\Lambda\vert\bar\phi_\alpha\right) 
\end{equation}
is also a stationary solution of (\ref{ME}) compatible with the boundary conditions whenever
\begin{equation}
\bar\phi_\alpha=\frac{1}{\vert\Lambda\vert}\int_\Lambda dx \,\phi_\alpha^*(x)\quad,\quad \alpha\in\tilde M
\end{equation} 
where $\phi_\alpha^*(x)$ is the deterministic stationary solution for the $\alpha$-field of the Langevin equation ($SC$ stands for strictly conserved). In this case we should think that at the microscopic level there is a strong constraint on the system that forces such strict conservation laws. 

Observe that in the limit $\Omega\rightarrow\infty$ both systems, without or with constrains on some fields, have the same macroscopic representation. In this appendix we look for the relations between the two body-correlations associated with the $SC$ and the $OB$ systems as we have defined them. Higher order correlations depend on other quasi-potential's $\Omega$ perturbative terms that may be different for the $OB$ and $SC$ cases.

We know that the two body correlation for the $OB$ case is related with the quasi-potential's second derivatives \cite{Garrido0}:
\begin{equation}
C_{\alpha\beta}^{OB}(x,y)=\left(V^{-1}\right)_{\alpha\beta}(x,y)\quad,\quad V_{\alpha\beta}(x,y)=\frac{\delta^2 V_0[\phi]}{\delta\phi_\alpha(x)\delta\phi_\beta(y)}\biggr\vert_{\phi^*}\label{cor0}
\end{equation}
Let us compute the correlations for the $SC$ case. We define the functional generator:
\begin{equation}
Z[B]=\int \mathcal{D}\phi\exp\left\{-\Omega\mathcal{F}[\phi,B]\right\}\prod_{\alpha\in\tilde M}\delta\left(\int_\Lambda dx\,\phi_\alpha(x)-\vert\Lambda\vert \bar\phi_\alpha\right)\label{gen}
\end{equation}
where
\begin{equation}
\mathcal{F}[\phi,B]=V_0[\phi]-\sum_{\alpha\in M}\int_\Lambda dx\,B_\alpha(x)\phi_\alpha(x)\label{Fgen}
\end{equation}
Observe that $M$ is the total number of fields and $\tilde M\subseteq M$ is the set of the strictly conserved fields. Then, the correlations are just derivatives of the functional  generator (\ref{gen}):
\begin{eqnarray}
C_{\alpha\beta}^{SC}(x,y)&\equiv&\lim_{\Omega\rightarrow\infty}\Omega\left[\langle\phi_\alpha(x)\phi_\beta(y)\rangle_\Omega-\langle\phi_\alpha(x)\rangle_\Omega\langle\phi_\beta(y)\rangle_\Omega \right]\nonumber\\
&=&\lim_{\Omega\rightarrow\infty}\frac{\delta}{\delta B_\alpha(x)}\frac{\delta}{\delta B_\beta(y)}\frac{1}{\Omega}\log Z[B]\biggr\vert_{B=0}\label{corSC}
\end{eqnarray}
 where $\langle\cdot\rangle_\Omega=\int \mathcal{D}\phi \cdot P_{st}^{SC}[\phi]$. We use now the Laplace representation of the Dirac's delta function to get:
 \begin{equation}
 Z[B]\simeq \left(\prod_{\alpha\in\tilde M}\int_{c-i\infty}^{c+i\infty}ds_\alpha\right)\int \mathcal{D}\phi\exp\left\{-\Omega\mathcal{G}[\phi,B,s]
\right\}
 \end{equation}
 where
 \begin{equation}
 \mathcal{G}[\phi,B,s]=\mathcal{F}[\phi,B]+\sum_{\alpha\in\tilde M}s_\alpha\int_\Lambda dx\,\left(\phi_\alpha(x)-\bar\phi_\alpha\right)
 \end{equation}
 We can get the dominant part of the integral (\ref{gen}) when $\Omega\rightarrow\infty$ by expanding $\mathcal{G}$ around the value $(\phi_0[B],s_0[B])$ that make it a minimum. That is:
  \begin{equation}
 Z[B]\simeq \exp\left\{-\Omega\,\mathcal{G}[\phi_0[B],B,s_0[B]]\right\}\quad ,\quad \Omega\rightarrow\infty\label{ZB}
 \end{equation}
 where $(\phi_0[B],s_0[B])$ are solution of the equations:
 \begin{eqnarray}
&&\frac{\delta\mathcal{G}[\phi,B,s]}{\delta\phi_\alpha(x)}\biggr\vert_{\substack{\phi=\phi_0[B]\\ s=s_0[B]}}=0\Rightarrow \frac{\delta V_0[\phi]}{\delta\phi_\alpha(x)}\biggr\vert_{\phi=\phi_0[B]}=B_\alpha(x)-\delta_{\alpha\in\tilde M} s_{0,\alpha}\nonumber\\
&&\frac{\partial\mathcal{G}[\phi,B,s]}{\partial s_\alpha}\biggr\vert_{\substack{\phi=\phi_0[B]\\s= s_0[B]}}=0\Rightarrow \int_\Lambda dx\,\phi_{0,\alpha}(x)=\vert\Lambda\vert\bar\phi\quad \alpha\in\tilde M\label{gensol}
 \end{eqnarray}
 We see that for $\phi_0(x;B=0)=\phi^*(x)$ and $s_0[B=0]=0$. Therefore we can find the solution of eqs.(\ref{gensol}) by doing a perturbative  expansion around $B=0$. We get to first order:
 \begin{equation}
 \phi_{0,\alpha}(x)=\phi_\alpha^*(x)+\Delta\phi_\alpha(x)
 \end{equation}
 where
 \begin{eqnarray}
\Delta\phi_\alpha(x)&=&\sum_{\beta\in M}\int_\Lambda dy\, C_{\alpha\beta}^{OB}(x,y)\bar B_\beta(y)\nonumber\\
\bar B_\alpha(x)&=&B_\alpha(x)-\delta_{\alpha\in\tilde M}s_0^{(1)}
 \end{eqnarray}
 and $s_0^{(1)}$ is the first order expansion in $B$. $s_0$  is solution of:
 \begin{equation}
 \int_\Lambda dx\,\Delta\phi_\alpha(x)=0\quad \alpha\in\tilde M
 \end{equation}
 Substituting the solution of $\phi_0[B]$ into eq.(\ref{ZB}) we get, 
 \begin{eqnarray}
\lim_{\Omega\rightarrow\infty} \frac{1}{\Omega}\log Z[B]&=&-V_0[\phi^*]+\sum_{\alpha\in M}\int_\lambda dx\,B_\alpha(x)\phi_\alpha^*(x)\nonumber\\
&+&\frac{1}{2}\sum_{\alpha\beta\in M}\int_\Lambda dx\,\int_\Lambda dy\,C_{\alpha\beta}^{OB}(x,y)\bar B_\alpha(x)\bar B_\beta(y)+\mathcal{O}(B^3)
 \end{eqnarray}
 Finally, form eq.(\ref{corSC}) we get the correlations for the strictly conservation case:
 \begin{equation}
 C_{\alpha\beta}^{SC}(x,y)= C_{\alpha\beta}^{OB}(x,y)-\sum_{\gamma\delta\in\tilde M}\left(A^{-1}\right)_{\gamma\delta}\int_\Lambda dz_1\,C_{\alpha\gamma}^{OB}(x,z_1)\int_\Lambda dz_2\,C_{\beta\delta}^{OB}(y,z_2)\label{corS}
 \end{equation}
 where
 \begin{equation}
 A_{\alpha\beta}=\int_\Lambda dx\int_\Lambda dy\, C_{\alpha,\beta}^{OB}(x,y)\quad\alpha,\beta\in\tilde M
 \end{equation}
 Please, observe that:
 \begin{equation}
 \int_\Lambda dx\, C_{\alpha\beta}^{SC}(x,y)=0\quad\text{if}\quad \alpha\, \text{and/or}\, \beta\in\tilde M
 \end{equation}
as we expected. 

For systems at equilibrium we know that $C_{\alpha\beta}^{OB}(x,y)=\bar C_{\alpha\beta}\delta(x-y)$. Therefore
\begin{equation}
 C_{\alpha\beta}^{SC}(x,y)= \bar C_{\alpha\beta}\delta(x-y)-\frac{1}{\vert\Lambda\vert}\sum_{\gamma\bar\gamma\in\tilde M}\bar C_{\alpha\gamma}\bar C_{\beta\bar\gamma}\left(A^{-1}\right)_{\gamma\bar\gamma}
\end{equation}
where
\begin{equation}
A_{\alpha\beta}=\bar C_{\alpha\beta}\quad\alpha,\beta\in\tilde M
\end{equation}
and they coincide in the thermodynamic limit.

\section*{Appendix IV: Fourier Transforms and sums}
We use in this paper the Sinus Fourier Transform for the $x$-axis coordinates where the functions, $f(x)$, are zero in the boundaries of the interval $f(0)=f(L)=0$:
\begin{equation}
f(x)=\sum_{n=1}^\infty\sin(\frac{n\pi x}{L})\hat f(n)\quad x\in[0,L]
\end{equation}
To use this transform we need the properties:
\begin{equation}
\frac{2}{L}\int_0^Ldx\,\sin(\frac{n\pi x}{L})\sin(\frac{m\pi x}{L})=\delta_{m,n}
\end{equation}
\begin{eqnarray}
\frac{1}{L}\int_0^Ldx\,\sin(\frac{n\pi x}{L})\cos(\frac{m\pi x}{L})&=&\frac{1}{\pi}\left(1-(-1)^{n+m}\right)\frac{n}{n^2-m^2}\quad (n\neq m) \nonumber\\
&=&0\quad (n=m)
\end{eqnarray}
The normal Fourier's Transform is used for the $x_\perp\in D\equiv [0,L]^{d-1}$ coordinates where the functions are periodic:
\begin{equation}
g(x_\perp)=\sum_{n}e^{i\frac{2\pi}{L} n\cdot x_\perp}\hat g(n)\quad n\in\mathbb{Z}^{d-1}
\end{equation}
and we have the useful property:
\begin{equation}
\frac{1}{L^{d-1}}\int_D dx_\perp\, e^{i\frac{2\pi}{L} n\cdot x_\perp}=\delta_{n,0}
\end{equation}
We needed to derive in this work some Fourier sums:
\begin{itemize}
\item $\displaystyle \begin{aligned}[t]
\sum_{m=1}^\infty\frac{2m\sin(2mx)}{(2m)^2+a^2}&=\frac{\pi}{4}\frac{\sinh\left(a(\frac{\pi}{2}-\tilde x)\right)}{\sinh\left(\frac{\pi}{2}a\right)} \quad&0<\tilde x<\pi\\
&= 0 &\tilde x=0
\end{aligned}
$

where $\tilde x=mod(x,\pi)$. 

\item $\displaystyle \begin{aligned}[t]
\sum_{m=1}^\infty\frac{(2m-1)\sin((2m-1)x)}{(2m-1)^2+a^2}
&=\text{sign}(\pi-\bar x) \frac{\pi}{4}\frac{\cosh\left(a(\frac{\pi}{2}-\tilde x)\right)}{\cosh\left(\frac{\pi}{2}a\right)}  &\quad 0<\bar x<2\pi\\
&=0 &\bar x=0\\
\end{aligned}
$

with $\bar x=mod(x,2\pi)$ and $\text{sign}(0)=0$.

and taking $a\rightarrow i a$ we also get:

\item $\displaystyle \begin{aligned}[t]
\sum_{m=1}^\infty\frac{2m\sin(2mx)}{(2m)^2-a^2}&=\frac{\pi}{4}\frac{\sin\left(a(\frac{\pi}{2}-\tilde x)\right)}{\sin\left(\frac{\pi}{2}a\right)} \quad&0<\tilde x<\pi\\
&= 0 &\tilde x=0
\end{aligned}
$

In particular, if $a=2n-1$,  $n\in\mathbb{Z}$

$$
\displaystyle \begin{aligned}[t]
 \sum_{m=1}^\infty\frac{2m\sin(2mx)}{(2m)^2-(2n-1)^2}&=\frac{\pi}{4}\cos\left((2n-1)\tilde x\right) \quad&0<\tilde x<\pi\\
&= 0 &\tilde x=0
\end{aligned}
$$

\item $\displaystyle \begin{aligned}[t]
\sum_{m=1}^\infty\frac{(2m-1)\sin((2m-1)x)}{(2m-1)^2-a^2}
&=\text{sign}(\pi-\bar x) \frac{\pi}{4}\frac{\cos\left(a(\frac{\pi}{2}-\tilde x)\right)}{\cos\left(\frac{\pi}{2}a\right)}  &0<\bar x<2\pi\\
&=0\quad &\bar x=0
\end{aligned}
$

In particular, if $a=2n$,  $n\in\mathbb{Z}$  

 $\displaystyle \begin{aligned}[t]
\sum_{m=1}^\infty\frac{(2m-1)\sin((2m-1)x)}{(2m-1)^2-(2n)^2}
&=\text{sign}(\pi-\bar x)\frac{\pi}{4}\cos(2n\tilde x)  &0<\bar x<2\pi\\
&=0\quad &\bar x=0
\end{aligned}
$

\end{itemize}

In order to show these relations we use some known result from ref. \cite{Grads}, for instance, in the first case we use  eq. 1.445.1:
\begin{equation}
I(x,a)= \sum_{m=1}^\infty\frac{m\sin(mx)}{m^2+a^2}=\frac{\pi}{2}\frac{\sinh\left(a(\pi-x)\right)}{\sinh\left(\pi a\right)}
\end{equation}
We separate the sum in two: even and odd terms: $I(x,a)=I_e(x,a)+I_o(x,a)$. But $I_e(x,a)=I(2x,a/2)/2$ and then we get the desired result: $I_o(x,a)=I(x,a)-I(2x,a/2)/2$.

Other relation that we use in the text is:
\begin{itemize}
\item $\displaystyle \sum_{n=1}^\infty\frac{\sin\left((2n-1)x\right)}{(2n-1)^2+a^2}=\frac{1}{a}\sin(x)\int_{0}^\infty d\beta\,e^{-\beta}\sin(\beta a)\frac{\displaystyle 1+e^{-2\beta}}{\displaystyle 1-2 e^{-2\beta}\cos(2x)+e^{-4\beta}} $
\end{itemize}

Finally, in Section VII we need to solve sums of the form:
\begin{equation}
B_l(n,m)=\sum_{k=0}^{\infty}\frac{(2k)^{2+2l}}{D(n,2k;0)}\frac{1}{((2k)^2-m^2)^2(n^2-(2k)^2)}\label{sumDel}
\end{equation}
where $D(n,m;n_\perp)$ is given by eq. (\ref{DDDel}) and $n$ and $m$ are odd integers. These sums are done by breaking apart the denominators and then we use some of the above relations. After some trivial algebra we get:
\begin{itemize}
\item {\boldmath $m\neq n$}:
\begin{eqnarray}
B_0(n,m)&=&\frac{\pi}{16\omega} \biggl[\frac{\pi  n^2 a_0(\omega)}{\left(n^2-m^2\right) \left(n^2 a_0(\omega)+m^2\right)^2 \left(n^2 a_1(\omega)+m^2\right)}\nonumber\\
&+&\frac{1}{\left(n^2 a_1(\omega)+m^2\right)^2}\left(\frac{4
   \sqrt{a_1(\omega)} \coth \left(\frac{1}{2} \pi  n \sqrt{a_1(\omega)}\right)}{n^3 (a_1(\omega)+1) (a_1(\omega)-a_0(\omega))}-\frac{\pi  m^2 \left(n^2
  a_1(\omega)+m^2\right)}{(m^2-n^2)\left(n^2a_0(\omega)+m^2\right)^2}\right)\nonumber\\
   &+&\frac{4 \sqrt{a_0(\omega)} \coth \left(\frac{1}{2} \pi
    n \sqrt{a_0(\omega)}\right)}{n^3 (a_0(\omega)+1) (a_0(\omega)-a_1(\omega)) \left(n^2 a_0(\omega)+m^2\right)^2}\biggr]
\end{eqnarray}
\begin{eqnarray}
B_1(n,m)&=&\frac{1}{16\omega}\biggl[-\frac{8}{(a_0(\omega)+1) (a_1(\omega)+1) \left(n^3-m^2 n\right)^2}\nonumber\\
&+&\frac{\pi ^2
   m^2-8}{\left(n^2-m^2\right) \left(n^2 a_0(\omega)+m^2\right) \left(n^2
  a_1(\omega)+m^2\right)}\nonumber\\
  &-&\frac{4 a_0(\omega) \left(\pi  n \sqrt{a_0(\omega)} \coth
   \left(\frac{1}{2} \pi  n \sqrt{a_0(\omega)}\right)-2\right)}{n^2 (a_0(\omega)+1)
   (a_0(\omega)-a_1(\omega)) \left(n^2 a_0(\omega)+m^2\right)^2}\nonumber\\
   &+&\frac{4 a_1(\omega)
   \left(\pi  n \sqrt{a_1(\omega)} \coth \left(\frac{1}{2} \pi  n
   \sqrt{a_1(\omega)}\right)-2\right)}{n^2 (a_1(\omega)+1) (a_0(\omega)-a_1(\omega))
   \left(n^2 a_1(\omega)+m^2\right)^2}\nonumber\\
   &+&\frac{8 \left(n^4 \left(a_0(\omega)
   \left(m^2-a_1(\omega) \left(m^2-2 n^2\right)\right)+m^2
   a_1(\omega)\right)+m^6\right)}{\left(m^2-n^2\right)^2 \left(n^2
   a_1(\omega)+m^2\right)^2 \left(n^2 a_1(\omega)+m^2\right)^2}\biggr]
\end{eqnarray}
\begin{eqnarray}
&&B_2(n,m)=\frac{1}{16\omega}\biggl[\frac{\left(\pi ^2 m^2-8\right) m^2}{\left(n^2-m^2\right) \left(n^2
   a_0(\omega)+m^2\right) \left(n^2 a_1(\omega)+m^2\right)}\nonumber\\
   &&-\frac{8}{(a_0(\omega)+1)
   (a_1(\omega)+1) \left(m^2-n^2\right)^2}+\frac{4 a_0(\omega)^2 \left(\pi  n
   \sqrt{a_0(\omega)} \coth \left(\frac{1}{2} \pi  n
   \sqrt{a_0(\omega)}\right)-2\right)}{(a_0(\omega)+1) (a_0(\omega)-a_1(\omega))
   \left(n^2 a_0(\omega)+m^2\right)^2}\nonumber\\
   &&+\frac{4 a_1(\omega)^2 \left(\pi  n
   \sqrt{a_1(\omega)} \coth \left(\frac{1}{2} \pi  n
   \sqrt{a_1(\omega)}\right)-2\right)}{(a_1(\omega)+1) (a_1(\omega)-a_0(\omega))
   \left(n^2 a_1(\omega)+m^2\right)^2}\\
   &&-\frac{8 m^2 n^2 \left(a_0(\omega)
   \left(a_1(\omega) \left(2 m^2 n^2-3 n^4\right)+m^4-2 m^2 n^2\right)+a_1(\omega)
   \left(m^4-2 m^2 n^2\right)-m^4\right)}{\left(m^2-n^2\right)^2 \left(n^2
   a_0(\omega)+m^2\right)^2 \left(n^2 a_1(\omega)+m^2\right)^2}\biggr]\nonumber
\end{eqnarray}
\item {\boldmath $n=m$}:
\begin{eqnarray}
B_0(n,n)&=&\frac{1}{64 n^7 w (a_0(\omega)+1)^3 (a_1(\omega)+1)^3 (a_0(\omega)-a_1(\omega))}\nonumber\\
&&\biggl[\pi  (a_1(\omega)+1) \biggl(\pi  n (a_0(\omega)+1) (a_0(\omega) (a_1(\omega)-3)-3 a_1(\omega)-7) (a_1(\omega)-a_0(\omega))\nonumber\\
&&+16 \sqrt{a_0(\omega)} (a_1(\omega)+1)^2 \coth \left(\frac{1}{2} \pi  n \sqrt{a_0(\omega)}\right)\biggr)\nonumber\\
&&-16 \pi  (a_0(\omega)+1)^3 \sqrt{a_1(\omega)} \coth \left(\frac{1}{2} \pi  n \sqrt{a_1(\omega)}\right)\biggr]
\end{eqnarray}
\begin{eqnarray}
B_1(n,n)&=&\frac{1}{64 n^5 w (a_0(\omega)+1)^3 (a_1(\omega)+1)^3 (a_0(\omega)-a_1(\omega))}\nonumber\\
&&\biggl[16 \pi  (a_0(\omega)+1)^3 a_1(\omega)^{3/2} \coth \left(\frac{1}{2} \pi  n \sqrt{a_1(\omega)}\right)\nonumber\\
&&+\pi  (a_1(\omega)+1) \biggl(\pi  n (a_0(\omega)+1) (a_1(\omega)-a_0(\omega)) (5 a_0(\omega) a_1(\omega)+a_0(\omega)+a_1(\omega)-3)\nonumber\\
&&-16 a_0(\omega)^{3/2} (a_1(\omega)+1)^2 \coth \left(\frac{1}{2} \pi  n \sqrt{a_0(\omega)}\right)\biggr)\biggr]
\end{eqnarray}
\begin{eqnarray}
B_2(n,n)&=&\frac{1}{64 n^3 w (a_0(\omega)+1)^3 (a_1(\omega)+1)^3 (a_0(\omega)-a_1(\omega))}\nonumber\\
&&\biggl[\pi  (a_1(\omega)+1) \biggl(\pi  n (a_0(\omega)+1) (a_1(\omega)-a_0(\omega)) (a_0(\omega) (9 a_1(\omega)+5)+5 a_1(\omega)+1)\nonumber\\
&&+16 a_0(\omega)^{5/2} (a_1(\omega)+1)^2 \coth \left(\frac{1}{2} \pi  n \sqrt{a_0(\omega)}\right)\biggr)\nonumber\\
&&-16 \pi  (a_0(\omega)+1)^3 a_1(\omega)^{5/2} \coth \left(\frac{1}{2} \pi  n \sqrt{a_1(\omega)}\right)\biggr]
\end{eqnarray}
\end{itemize}
where
\begin{equation}
a_{0,1}(\omega)=\frac{1}{\omega}\left(1+\omega+\omega^2\pm(1+\omega)\sqrt{1+\omega^2}\right)
\end{equation}
Plesase, we remind that these expressions only apply for $n$ and $m$ being odd-integers.

\section*{Appendix V: Computation of the basic correlation function $F(x,z;\sigma,\sigma')$ for $d=1$}
The basic correlation funtion is defined by eq. (\ref{CR2}) and (\ref{CR3}). For dimension one they reduce to:
\begin{equation} 
F(x,z;\sigma,\sigma')=\tilde F(x,z;\sigma,\sigma')-\tilde F(z,x;\sigma',\sigma) 
\end{equation}
with
\begin{eqnarray}
\tilde F(x,z;\sigma,\sigma')&=&-\frac{8}{\pi^2\lambda(\sigma')}\sum_{n=1}^\infty\sin\left(\frac{\displaystyle\pi}{L}(2n-1)x\right)\sum_{m=1}^\infty\sin\left(\frac{2\pi m}{L}z\right)\frac{(2n-1)2m}{(2n-1)^2-(2m)^2}\nonumber\\
&&\frac{1}{\theta_{\sigma\sigma'}^2(2n-1)^2+(2m)^2}
\end{eqnarray}
Now we separate the fractions:
\begin{eqnarray}
\frac{1}{(2n-1)^2-(2m)^2}&&\frac{1}{\theta_{\sigma\sigma'}^2(2n-1)^2+(2m)^2)}=\nonumber\\
\frac{1}{1+\theta_{\sigma\sigma'}^2}\frac{1}{(2n-1)^2}&&\left[\frac{1}{(2n-1)^2-(2m)^2}+\frac{1}{\theta_{\sigma\sigma'}^2(2n-1)^2+(2m)^2}\right]
\end{eqnarray}
and we get
\begin{eqnarray}
\tilde F(x,z;\sigma,\sigma')&=&-\frac{8}{\pi^2\lambda(\sigma')}\frac{1}{1+\theta_{\sigma\sigma'}^2}\sum_{n=1}^\infty\frac{\displaystyle\sin\left(\frac{\pi}{L}(2n-1)x\right)}{2n-1}\sum_{m=1}^\infty\sin\left(\frac{2\pi m}{L}z\right)2m\nonumber\\
&&\biggl[\frac{1}{(2n-1)^2-(2m)^2}+\frac{1}{\theta_{\sigma\sigma'}^2(2n-1)^2+(2m)^2}\biggr]
\end{eqnarray}
At this point we can use the formulas in Appendix IV to do explicitly the sum over $m$'s. We find:
\begin{eqnarray}
 \tilde F(x,z;\sigma,\sigma')&=&-\frac{2}{\pi\lambda(\sigma')}\frac{1}{1+\theta_{\sigma\sigma'}^2}\sum_{n=1}^\infty\frac{\displaystyle\sin\left(\frac{\pi}{L}(2 n-1)x\right)}{2n-1}\biggl[-\cos\left(\frac{\pi}{L}(2n-1)z\right)\nonumber\\
&+&\frac{\displaystyle\sinh\left(\frac{\pi}{2}\theta_{\sigma\sigma'}(2n-1)\left(1-\frac{2z}{L}\right)\right)}{\displaystyle\sinh\left(\frac{\pi}{2}\theta_{\sigma\sigma'}(2n-1)\right)}\biggr]\label{Ftil}
\end{eqnarray}
The first sum can be done by converting the sinus cosinus product into a sum of sinus. Then, we use the Gradsteyn's formula GR.1.442.1\cite{Grads} to get:
\begin{equation}
\sum_{n=1}^\infty\frac{\displaystyle\sin\left(\frac{\pi}{L}(2 n-1)x\right)}{2n-1}\cos\left(\frac{\pi}{L}(2n-1)z\right)=\frac{\pi}{8}\left[sgn(x-z)+sgn(L-(x+z))\right]
\end{equation}
The second sum in eq. (\ref{Ftil}) needs more work to get a simple version. First we use Gradsteyn's GR.3.743.1 that converts an hyperbolic sinus ratio into an integral:
\begin{equation}
\frac{\sinh(a\beta)}{\sinh(b\beta)}=\frac{2\beta}{\pi}\int_{0}^{\infty}dy\,\frac{\sin(ay)}{\sin(by)}\frac{1}{y^2+\beta^2}
\end{equation}
in our case we choose $b=1$, $a=1-2z/L$ and $\beta=\theta_{\sigma\sigma'}(2n-1)\pi/2$. Therefore we can write:
\begin{eqnarray}
I=\sum_{n=1}^\infty\frac{\displaystyle\sin\left(\frac{\pi}{L}(2 n-1)x\right)}{2n-1}&&\frac{\displaystyle\sinh\left(\frac{\pi}{2}\theta_{\sigma\sigma'}(2n-1)\left(1-\frac{2z}{L}\right)\right)}{\displaystyle\sinh\left(\frac{\pi}{2}\theta_{\sigma\sigma'}(2n-1)\right)}\nonumber\\
&=&\theta_{\sigma\sigma'}\int_0^\infty dy\,\frac{\sin(y\bar z)}{\sin y}\sum_{n=1}^\infty\frac{\displaystyle\sin\left(\frac{\pi}{L}(2 n-1)x\right)}{\displaystyle y^2+\left(\frac{\pi}{2}\theta_{\sigma\sigma'}(2n-1)\right)^2}
\end{eqnarray}
We can convert the last sum into another integral (see Appendix IV) and  we get:
\begin{equation}
I=-\frac{2}{\pi}\cos\left(\frac{\pi}{2}\bar x\right)\int_0^\infty d\beta\,e^{-\beta}\frac{\displaystyle 1+e^{-2\beta}}{\displaystyle 1+2 e^{-2\beta}\cos(\pi\bar x)+e^{-4\beta}}\int_0^\infty dy\,\frac{\displaystyle\sin(y\bar z)\sin(\frac{2}{\pi}\theta_{\sigma'\sigma}\beta y)}{y\sin(y)}
\end{equation}
where $\bar x=2x/L-1$ and $\bar z=2z/L-1$. We substitute the last integral with the relation that we derive in Appendix VI:
\begin{equation}
\int_0^\infty dy \frac{\displaystyle\sin(ay)\sin(by)}{y\sin(y)}=\frac{\pi}{2}\text{sign}(ab)\sum_{n=0}^\infty\chi\left[ 2n+1-\vert a\vert<\vert b\vert<2n+1+\vert a\vert\right]\quad, \vert a\vert<1
\end{equation}
where $\chi[\text{condition}]=1$ whenever the condition holds and $0$ otherwise. Therefore we get
\begin{equation}
I=-\cos\left(\frac{\pi}{2}\bar x\right)\sum_{n=0}^\infty\int_{\pi\theta_{\sigma\sigma'}(2n+1-\bar z)/2}^{\pi\theta_{\sigma\sigma'}(2n+1+\bar z)/2} d\beta\,e^{-\beta}\frac{\displaystyle 1+e^{-2\beta}}{\displaystyle 1+2 e^{-2\beta}\cos(\pi\bar x)+e^{-4\beta}}
\end{equation}
Finally, the last integral can be done explicitly:
\begin{equation}
\int d\beta \,e^{-\beta}\frac{\displaystyle 1+e^{-2\beta}}{\displaystyle 1+2 e^{-2\beta}\cos(\pi\bar x)+e^{-4\beta}}=-\frac{1}{\displaystyle 2\cos\left(\frac{\pi}{2}\bar x\right)}\arctan\left[2\cos\left(\frac{\pi}{2}\bar x\right)\frac{e^{-\beta}}{1-e^{-2\beta}} \right]
\end{equation} 
and, after some straight-ahead algebra, we get eq.(\ref{F}).

\section*{Appendix VI: Math relations}

We show in this Appendix some formulas we have derived and used along the paper.
\subsection*{1. The integral:}
\kern-1cm
\begin{eqnarray}
\boldsymbol{I(y,z)}&\boldsymbol{\equiv}&\boldsymbol{\int_0^\infty d\alpha\,\frac{\sin(y\alpha)\sin(z\alpha)}{\alpha\sin\alpha}}\nonumber\\
&\boldsymbol{=}&\boldsymbol{\frac{\pi}{2}\text{\bf sign}(yz)\sum_{n=0}^\infty \chi\left[2n+1-\vert z\vert<\vert y\vert<2m+1+\vert z\vert\right]\quad \vert z\vert<1}\label{rel2}
\end{eqnarray}
with $\chi[\text{condition}]=1$ whenever the condition holds and zero otherwise.

\begin{figure}[h!]
\kern-1cm
\begin{center}
\includegraphics[height=9cm,clip]{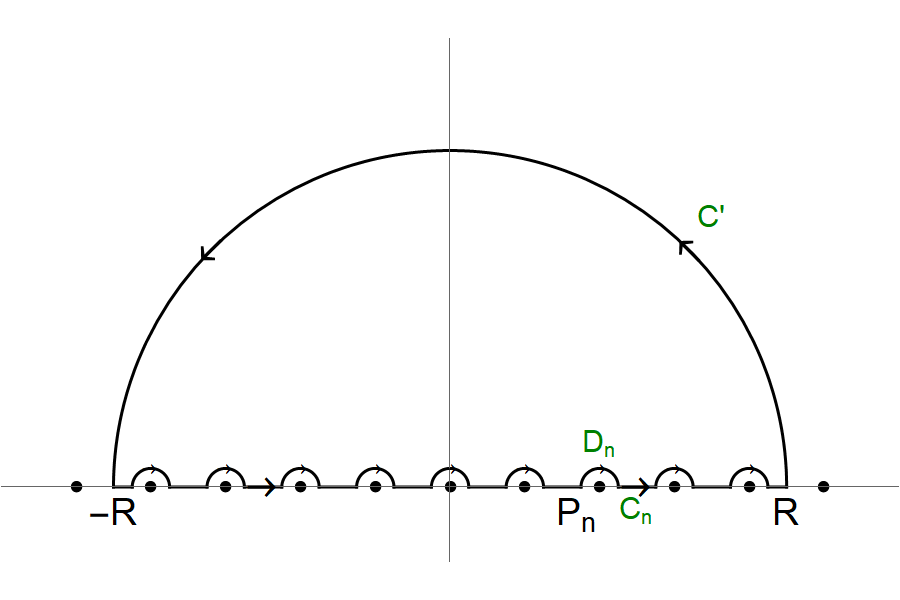}  
\end{center}
\kern -1.5cm
\caption{Contour $C=(\bigcup\limits_{n}( C_n\bigcup D_n )\bigcup C'$ used to evaluate integral (\ref{int2}). $P_{n}$ are the simple  poles at $w(P_{n})= n \pi$.  }\label{figap2}
\end{figure}

We prepare the integral $I(y,z)$ to be analyzed in the complex plane:
\begin{equation}
I(y,z)=\frac{1}{4}\left[J(a)-J(b)\right]\quad,\quad a=\vert y-z\vert\quad b=\vert y+z\vert
\end{equation}
where
\begin{equation}
J(a)=\int_{-\infty}^\infty d\alpha\,f(\alpha;a)\quad,\quad f(\alpha;a)=\frac{\displaystyle e^{i\alpha a}-1}{\displaystyle\alpha\sin\alpha}\quad,\quad a\ge 0
\end{equation}
Therefore, we study in the complex plane the integral
\begin{equation}
J(C)=\int_C dw f(w;a)\label{int2}
\end{equation}
We see that $f(w,a)$ have an infinitely number of simple poles located at $w(P_n)=n\pi$ $\forall n\in\mathbb{Z}$ . Then we choose  the contour shown in figure \ref{figap2}. That implies:
\begin{equation}
J(C)=J(C')+\sum_n J(C_n)+\sum_n J(D_n)=0
\end{equation}

We get for each piece of the contour:
\begin{itemize}
\item{$C'$:} $w=R e^{i\phi}$, $\phi\in[0,\pi]$. The term $\exp[iwa]\propto\exp[-Ra\sin\phi]$  tends to zero when $R\rightarrow\infty$ and therefore at such limit $J(C')=0$.
\item{$C_n$:} $w\in [n+\epsilon,n+1-\epsilon]$ in the limit $\epsilon\rightarrow 0$ and $R\rightarrow\infty$ we have
\begin{equation}
\sum_{n=-\infty}^\infty J(C_n)=J(a)
\end{equation}
in the sense that $J(a)$ is the Cauchy's Principal Part of the integral.
\item{$D_n$:} $w=n\pi+\epsilon e^{i\phi}$, $\phi\in [\pi,0]$. We find when $\epsilon\rightarrow 0$:
\begin{equation}
J(D_0)=a\pi\quad,\quad J(D_n)=\frac{(-1)}{n}^n i\left(1-e^{ian\pi}\right)\quad \forall n\neq 0
\end{equation}
\end{itemize}
Therefore 
\begin{equation}
J(a)=-\sum_{n=-\infty}^\infty J(D_n)=-\pi a-2\sum_{n=1}^{\infty}(-1)^n\frac{\sin(\pi m a)}{m}
\end{equation}
Finally we use GR.1.441.3 to do the last sum and we get:
\begin{eqnarray}
J(a)&=&0\qquad\quad 0\leq a<1\nonumber\\
&=&-2\pi k\qquad 2k-1<a<2k+1\quad (k=1,2,\ldots)
\end{eqnarray}
From this result, it follows a simple algebra to show the initial statement.

\subsection*{2. The integral relation:}
\kern-1cm
\begin{equation}
\boldsymbol{W(x,z;a,\theta)+W(z,x;a,\frac{1}{\theta})=\frac{\pi}{2}e^{-a(x+z)}\quad,\quad x,z>0}\label{rel1}
\end{equation}
{\bf where}
\begin{equation}
\boldsymbol{W(x,z;a,\theta)=\int_0^{\infty}dq\, \frac{q \sin(qz)}{q^2+a^2}e^{-x\sqrt{\theta q^2+(1+\theta)a^2}}}
\end{equation}
This relation includes the well known result: 
\begin{equation}
W(x,z;0,1)+W(z,x;0,1)=\frac{\pi}{2}\quad\Rightarrow \quad W(x,x;0,1)=\frac{\pi}{4} 
\end{equation}
We first prepare the integral $W(x,z;a,\theta)$ to be suitable to a complex variable integration:
\begin{equation}
W(x,z;a,\theta)=\frac{1}{2i}\int_{-\infty}^{\infty}f(q;x,z;a,\theta)\quad, \quad f(q;x,z;a,\theta)\equiv\frac{q e^{iqz}}{q^2+a^2}e^{-x\sqrt{\theta q^2+(1+\theta)a^2}}
\end{equation} 
by using the integrant symmetry $q\rightarrow -q$. We do the integral on the complex plane, $w\in\mathbb{C}$:
\begin{equation}
J(C)\equiv\int_C dw f(w;x,z;a,\theta)\label{int}
\end{equation}
We choose the closed integration contour $C$ shown in figure \ref{figap1} taking into account that there are two poles at $w_{P_{1,2}}=\pm ia$ and two branch lines due to the square root: $[w_{B_{1,2}},\pm\infty)$ where  $w_{B_{1,2}}=\pm ia\sqrt{1+1/\theta}$.  
\begin{figure}[h!]
\begin{center}
\includegraphics[height=9cm,clip]{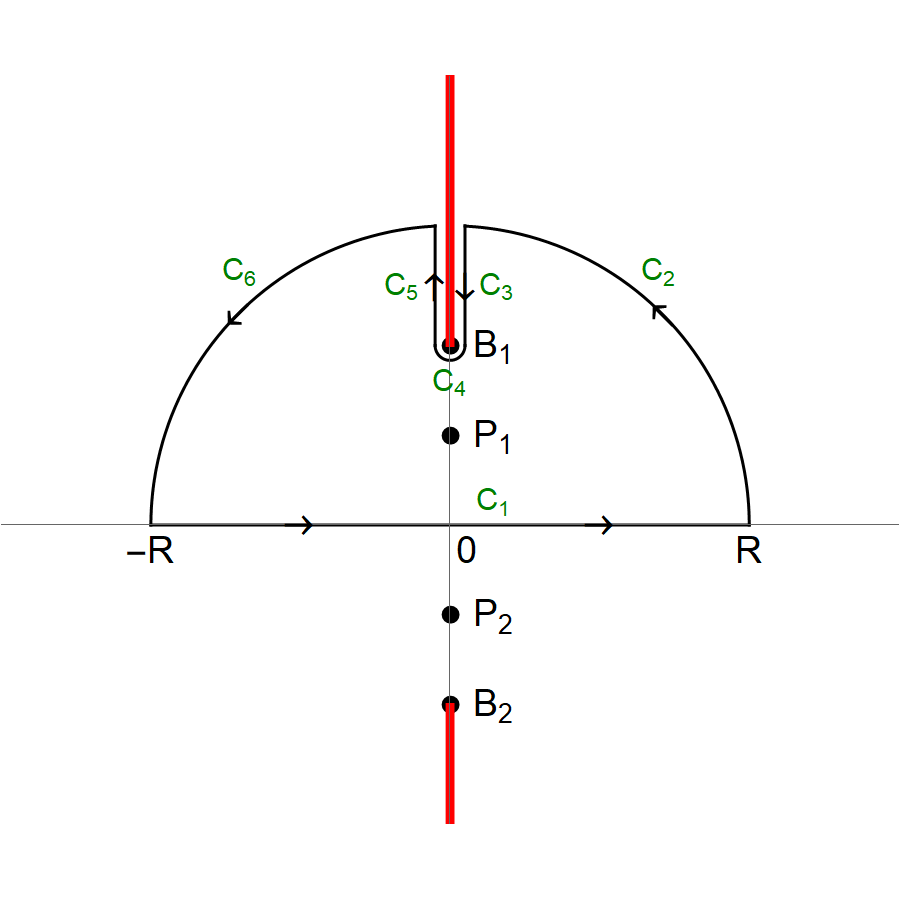}  
\end{center}
\kern -1.5cm
\caption{Contour $C=\bigcup\limits_{i=1}^6 C_i$ used to evaluate integral (\ref{int}). $P_{1,2}$ are the poles at $w_{P_{1,2}}=\pm a i$ and $B_{1,2}$ are the branch points at $w_{B_{1,2}}=\pm ia\sqrt{1+1/\theta}$. The branch lines are shown in red. }\label{count1}
\end{figure}
We get from each contour piece $C_i$:
\begin{itemize}
\item{$C_1$:} $w=q \in [-R,R]$,
\begin{equation}
J(C_1)=\int_{-\infty}^\infty dq\,f(q;x,z;a,\theta)
\end{equation}
after doing the limit $R\rightarrow\infty$.
\item{$C_2$ and $C_6$:} $w=R e^{i\phi}$, $\phi\in[0,\pi/2]$. We observe that in this path $\exp[iwz]=\exp[iRz\cos\phi-Rz\sin\phi]$ that tends to zero when $R\rightarrow\infty$ (Jordan's Lemma) and therefore $J(C_2)=J(C_6)=0$.
\item{$C_3$:} $w=w_{B_1}+q e^{i\pi/2}$, $q\in[R,0]$,
\begin{equation}
J(C_3)=-\int_{0}^\infty dk\,f(k;z,-x;a,1/\theta)
\end{equation}
where we have done the change of variables $q=-a\sqrt{1+1/\theta}+\sqrt{k^2/\theta+a^2(1+1/\theta)}$.
\item{$C_4$:}  $w=w_{B_1}+\epsilon e^{i\phi}$, $\phi\in[0,-\pi]$. This path goes to zero when $\epsilon\rightarrow 0$ and therefore $J(C_4)=0$.
\item{$C_5$:} $w=w_{B_1}+q e^{i5\pi/2}$, $q\in[0,R]$ and similarly to the $C_3$ path we get:
\begin{equation}
J(C_5)=\int_0^\infty dk\,f(k;z,x;a,1/\theta)
\end{equation}  
\end{itemize}
Then, we apply the residue theorem:
\begin{equation}
J(C)=J(C_1)+J(C_3)+J(C_5)=2\pi i Res(w_{P_1}=ia)
\end{equation}
where in our case $Res(ia)=\exp[-(x+z)a]/2$. Finally,
\begin{equation}
W(x,z;a,\theta)=\frac{1}{2i}J(C_1)=\frac{\pi}{2}e^{-(x+z)a} -\frac{1}{2i}(J(C_3)+J(C_5))
\end{equation}
that is the result desired.

\subsection*{3. The integral:}
\kern-1cm
\begin{eqnarray}
&&\boldsymbol{\int_0^\infty dq_2\, \cos(q_2 (x_2-z_2))\int_0^\infty dq\, \frac{q\sin(x_1 q)}{q^2+q_2^2}e^{-z_1 \sqrt{\theta q^2+(1+\theta^2)q_2^2}}}\nonumber\\
&\boldsymbol{=}&\boldsymbol{\frac{\pi x_1}{2D_1(x,z)}\biggl[x_1^2-z_1^2+(x_2-z_2)^2}\nonumber\\
&\boldsymbol{+}&\boldsymbol{\frac{z_1}{D_2(x,z)}\left((1+\theta^2)(z_1^2-x_1^2)+(1-\theta^2)(x_2-z_2)^2\right)\biggr]}\label{exp7}
\end{eqnarray}
{\bf where}
\kern-1cm
\begin{eqnarray}
\boldsymbol{D_1(x,z)}&\boldsymbol{=}&\boldsymbol{[(x_1-z_1)^2+(x_2-z_2)^2][(x_1+z_1)^2+(x_2-z_2)^2]}\nonumber\\
\boldsymbol{D_2(x,z;\theta)}&\boldsymbol{=}&\boldsymbol{[(1+\theta^2)x_1^2+\theta^2((1+\theta^2)z_1^2+(x_2-z_2)^2)]^{1/2}}
\end{eqnarray}

First we  transform the square root in the exponential to a single variable by means of an elliptic change of variables $(q,q_2)\rightarrow(u,v)$: $q=u\cos v/\theta$, $q=u\sin v/\sqrt{1+\theta^2}$ whose Jacobian is $J=u/(\theta\sqrt{1+\theta^2})$ and the domain of integration is $u\in[0,\infty]$ and $v\in[0,\pi]$. The integration over $u$ can be explicitly done using GR.3.893.1 \cite{Grads}. The two remaining integrals over the $v$-variable are of the form:
\begin{equation}
\int_0^{\pi/2}dv\,\frac{\cos v (\alpha\cos v+\beta\sin v)}{(\alpha\cos v+\beta\sin v)(b+(\alpha\cos v+\beta\sin v)^2}
\end{equation}
We do the change of variables $u=w/2$ to convert such integral in one of the form GR.2.559.2 \cite{Grads} and after we apply the limits and we sum the two integrals we obtain the expression (\ref{exp7}).


\begin{thebibliography}{99}
\bibitem{water}Eisenberg, D. and Kauzmann W. {\it The Structure and Properties of Water}. Oxford Classic Texts in the Physical Sciences. ISBN: 9780198570264 (2005); Stillinger, F.H. and Rahman, A. {\it Improved simulation of liquid water by molecular dynamics}. Journal of  Chemical Physics  {\bf 60}, 1545 (1974) https://doi.org/10.1063/1.1681229;  Cisneros et al. {\it Modeling Molecular Interactions in Water: From Pairwise to Many-
Body Potential Energy Functions}. Chemical Reviews, {\bf 116}, 7501 (2016) https://doi.org/10.1021/acs.chemrev.5b00644.
\bibitem{Bat} Batchelor, G.K., {\it An introduction to Fluid Dynamics}, Cambridge University Press, (2000); Gallavotti, G. ,{\it Foundations of fluid dynamics}, Springer, (2003).
\bibitem{eco} Maynard-Smith, J., {\it Models in Ecology}, Cambridge University Press (1978) ISBN: 9780521294409;  Gotelli, N.J. {\it A Primer of Ecology}. Sinauer Associates, Inc.  (Oxford University Press) (2008) ISBN:9780878933181.
\bibitem{Onsager} Onsager, L. and Machlup, S., {\it Fluctuations and Irreversible Processes}, Physical Review {\bf 91}, 1505 (1953).
\bibitem{Bertini2} Bertini, L., de Sole, A., Gabrielli, D., Jona Lasinio, G. and Landim, C., {\it Macroscopic fluctuation theory}, Reviews of Modern Physics, {\bf 87}, 593 (2015).
\bibitem{seminal} Graham,R. and Tel T., {\it On the weak-noise
limit of Fokker–Planck models}. Journal of  Statistical Physics {\bf} 35,729–748 (1984); Graham, R.,
Tel, T., {\it Weak-noise limit of Fokker-Planck models and non-differentiable potentials
for dissipative systems}. Phys. Rev. A {\bf 31}, 1109–1122 (1985);
 Graham, R., Roekaerts, D. and T{\'e}l, T., {\it Integrability of Hamiltonians associated with Fokker-Planck equations}, Physical Review A, {\bf 31}, 3364 (1985);
 Graham, R. and T{\'e}l, T., {\it Nonequilibrium potential for coexisting attractors}, Physical Review A, {\bf  33}, 1322 (1986).
\bibitem{Garrido0} Garrido, P.L. {\it Notes about the Macroscopic Fluctuating Theory}, Journal of Statistical Mechanics, 024001 (2021).
\bibitem{Derrida} Derrida, B., Lebowitz, J.L. and Speer, E.R., {\it Large Deviation of the Density Profile in the Steady State of the Open Symmetric Simple Exclusion Process}, Journal of Statistical Physics, {\bf 107}, 599 (2001).
\bibitem{Bertini} Bertini, L., Gabrielli, D. and  Lebowitz, J.L., {\it Large Deviations for a Stochastic Model of Heat Flow}, Journal of Statistical Physics, {\bf 121}, 843 (2005).
\bibitem{Garrido1} Garrido, P.L., {\it Nonequilibrium quasi-potentials} arXiv:2103.16121 (2021).
\bibitem{Groot} de Groot, S.R. and Mazur, P., {\it Non-Equlibrium Thermodynamics}, Dover Publications (2011).
\bibitem{Fox} Landau, L. D., Lifshitz, E. M. {\it Statistical Physics. Part I}. Pergamon, London (1958);
Landau, L. D., Lifshitz, E. M. {\it  Fluid Mechanics}. Pergamon, London (1959);
Fox, R.F., {\it Gaussian Stochastic Processes in Physics}, Physics Reports {\bf 48} 179 (1978); Schmitz R., {\it Fluctuations in Nonequilibrium fluids}, Phys. Rep. {\bf 171}1 (1988). Ortiz de Zarate, J.M. and Sengers J. V. {\it Hydrodynamic Fluctuations in Fluids and Fluid Mixtures}. Elsevier (2006).
\bibitem{Tremblay} Tremblay, A.M.S., Arai, M. and Siggia, E.D. {\it Fluctuations about simple nonequilibrium steady states}. Physical Review A, {\bf 23} 1451 (1981); Schmitz, R. and Cohen, E.G.D. {\it Fluctuations in a Fluid under a Stationary Heat Flux. I. General Theory}. Journal of Statistical Physics {\bf 38} 285 (1984).
\bibitem{Mansour} Mansour, M.M., Turner, J.W. and Garcia, A.J. {\it Correlation Functions for Simple Fluids in a Finite System under Nonequilibrium Constraints}. Journal of Statistical Physics, {\bf 48} 1157 (1987); Mansour, M.M., Garcia, A.L. and Lie, G.C. {\it Fluctuating Hydrodynamics in a Dilute Gas}. Physical Review Letters, {\bf 58}, 874 (1987).
\bibitem{Garr}  Garrido, P.L. and Lebowitz, J.L., {\it Diffusion equations from kinetic models with non-conserved momentum}, Nonlinearity {\bf 31} 5441 (2018). 
\bibitem{Einstein} Einstein, A., {\it The Theory of the Opalescence of homogeneous fluids and liquid mixtures near the critical states}, Annalen der Physik {\bf 33} 1275-1298 (1910).
\bibitem{Grads} Gradshteyn, I.S. and Ryzhik , {\it Table of Integrals, Series and Products}, Elsevier (2007).
\bibitem{Riem}  Butzer, P.L. and  Stens, R.L., {\it The Euler-MacLaurin Summation Formula, the Sampling Theorem,
and Approximate Integration over the Real Axis}, Linear Algebra and its Applications {\bf 52/53}, 141-155 (1983).

\bibitem{BGK}  Bhatnagar, P.L.,  Gross, E.P. and  AND M. Krook, M., {\it A Model for Collision Processes in Gases. I. Small Amplitude Processes in Charged and Neutral One-Component Systems}, Physical Review {\bf 94}, 511 (1954). For an application with boundary conditions: Bassanini, P, Cercignani, C. and Pagani C.D., {\it Comparison of Kinetic Theory Analyses of Linearized Heat Transfer Between Parallel Plates}, International Journal of Heat Transfer {\bf 10}, 447 (1967).


\end{thebibliography}
\end{document}